\documentclass[twocolumn]{aastex63}

\usepackage{xcolor}
\usepackage{newtxtext,newtxmath,amsmath}

\usepackage[version=3]{mhchem}
\hypersetup{linkcolor=red,citecolor=green,filecolor=cyan,urlcolor=magenta}

\received{}
\revised{}
\accepted{}
\submitjournal{ApJ}

\shorttitle{Ru Isotopes from NuGrid Project}
\shortauthors{Kim, Sung, and Kwak}

\begin{document}

\title{Isotopic Compositions of Ruthenium Predicted from the NuGrid Project}

\author{Seonho Kim}
\affiliation{Department of Physics, Ulsan National Institute of Science and Technology, Ulsan 44919, Republic of Korea}
\email{shkim0707@unist.ac.kr}

\author{Kwang Hyun Sung}
\affiliation{Department of Physics, Ulsan National Institute of Science and Technology, Ulsan 44919, Republic of Korea}

\author[0000-0002-2304-7798]{Kyujin Kwak}
\affil{Department of Physics, Ulsan National Institute of Science and Technology, Ulsan 44919, Republic of Korea}
\email{kkwak@unist.ac.kr}










\begin{abstract}

The isotopic compositions of ruthenium (Ru) are measured from presolar silicon carbide (SiC) grains. In a popular scenario, the presolar SiC grains formed in the outskirt of an asymptotic giant branch (AGB) star, left the star as a stellar wind, and joined the presolar molecular cloud from which the solar system formed. 
The Ru isotopes formed inside the star,  moved to the stellar surface during the AGB phase, and were
locked into the SiC grains. 
Following this scenario, we analyze the NuGrid data which provide the abundances of the Ru isotopes in the stellar wind for a set of stars in a wide range of initial masses and metallicities. We apply the C>O (carbon abundance larger than the oxygen abundance) condition which is commonly adopted for the condition of the SiC formation in the stellar wind. The NuGrid data confirm that SiC grains do not form in the winds of massive stars. The isotopic compositions of Ru in the winds of low-mass stars can explain measurements. We find that lower-mass stars ($1.65~M_\sun$ and $2~M_\sun$) with low metallicity (Z=0.0001) can explain most of the measured isotopic compositions of Ru. We confirm that the abundance of ${^{99}}$Ru inside the presolar grain includes the contribution from the in-situ decay of ${^{99}}$Tc. 
We also verify our conclusion by comparing the isotopic compositions of Ru integrated over all the pulses with those calculated at individual pulses.


\end{abstract}



\section{Introduction} \label{sec:intro}

\begin{figure*}
\epsscale{1.17}
\plottwo{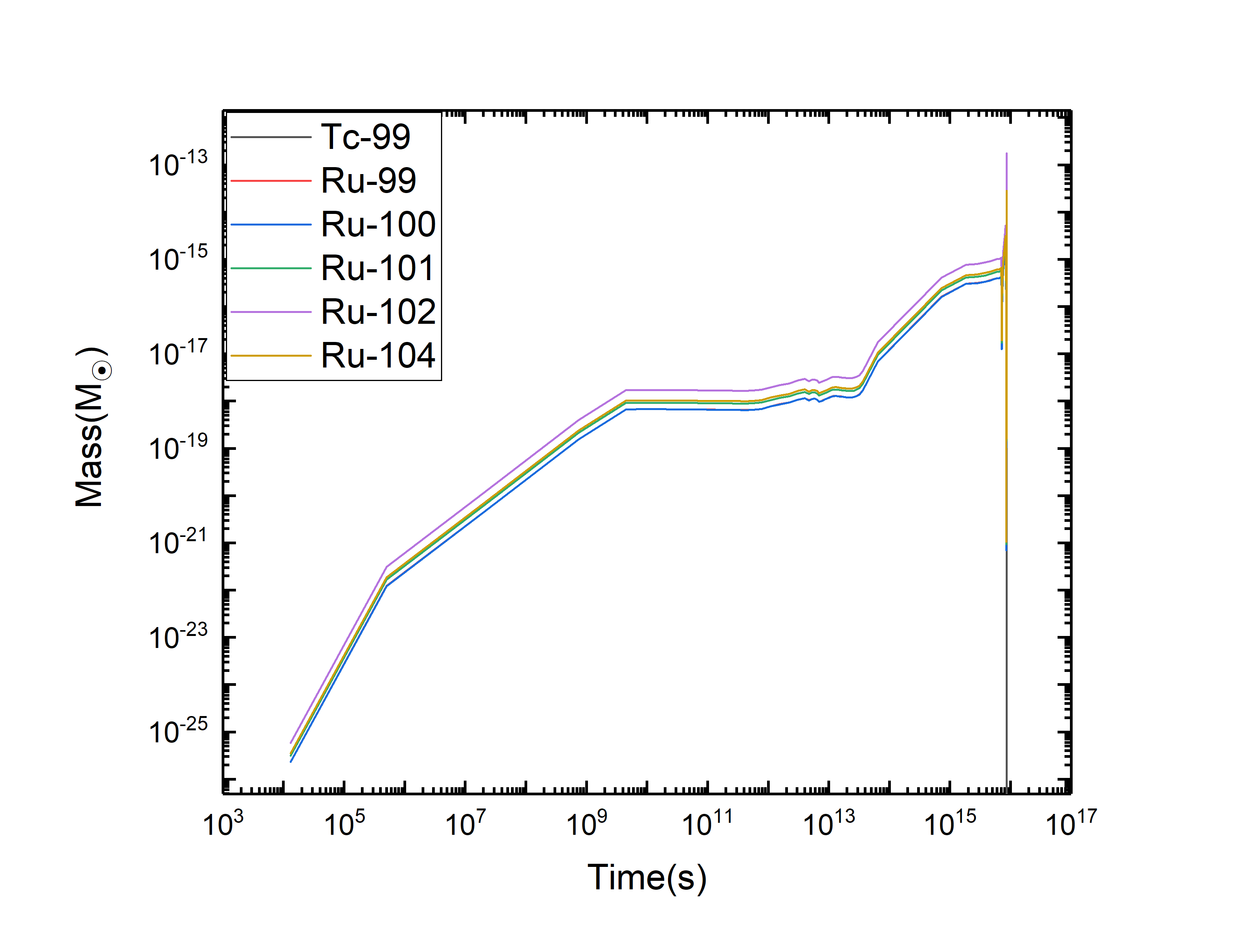}{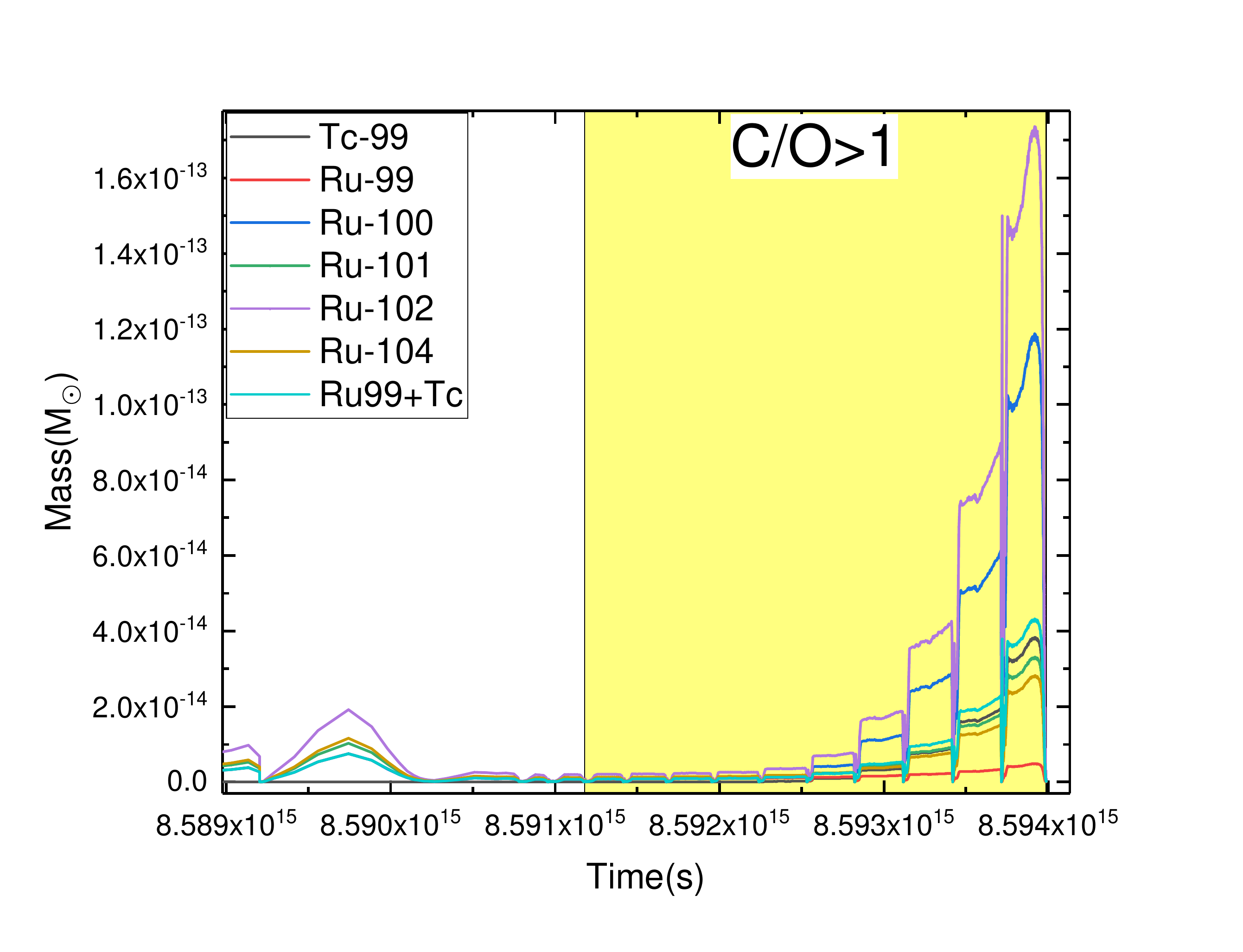}
\caption{Abundances of Ru isotopes and \ce{^{99}Tc} in the stellar wind as a function of stellar age for a low-mass star with $M=3~M_{\sun}$ and $Z=0.0001$. The abundances in the y-axes are presented in the unit of solar mass. Left panel shows a full evolution in a logarithmic scale while right panel is zoomed-in near the end of the evolution in a linear scale. 
The yellow region in the right panel corresponds to an epoch in which the C>O condition is satisfied. The step-like increase in the abundance indicates the TP features. 
%
}
\label{fig:evol}
\end{figure*}

Our solar system or a planetary system in general formed from a molecular cloud that had collapsed due to its own gravity. The molecular cloud from which the solar system formed was under the influence of its environments. In particular, its grain (or dust) contents were affected by the stardusts which had been produced in the nearby stellar winds. The grains in the molecular could affect the evolution of the solar system such as formation of meteorites and terrestrial planets \citep{1988EM&P...40..165H}. Some of the grains that existed when the solar system formed can be found even at the present time because they have been able to survive the destruction process during the history of the solar system. These survived grains, often called presolar grains, are found within meteorites \citep{2005ChEG...65...93L} and provide information on the early times of the solar system by answering questions like ``Under what environment had the solar system formed?".

More accurately, the elemental abundances that presolar grains contain provide constraints on the environments of the presolar system. In particular, relative abundances of isotopes, commonly called isotopic compositions, can be used to differentiate the effects of various environments such as stellar winds at various stages during the stellar evolution, novae, and supernovae (SNe). The isotopic compositions are also known to be affected by different types of nuclear reactions such as s-, r-, and p-process, depending on the environments. 
From the isotopic compositions measured within presolar grains, 
previous studies revealed that various types of presolar grains could form in various environments. For example, \citet{1998AREPS..26..147Z} summarized that diamond, silicon carbide (SiC), graphite, corundum, and silicon nitride showed a stellar origin such as red giant and asymptotic giant branch (AGB) stars while evidence for a SN origin could be found in diamond, low-density graphite and silicon nitride, and a subgroup of SiC. 
Novae were also found to a source of presolar grains \citep{2001ApJ...551.1065A}, but this discovery was disputed by \citet{2005ApJ...631L..89N} who claimed that SNe were the source.


In this work, we re-visit the isotopic compositions of ruthenium (Ru) measured within the presolar SiC grains \citep{2004Sci...303..649S} by using the publicly available abundance data of the NuGrid project \citep{2016ApJS..225...24P,2018MNRAS.480..538R}. In the previous work of \citet{2004Sci...303..649S}, the measured isotopic compositions of Ru were compared with those calculated with the FRANEC (Frascati Raphson-Newton Evolutionary Code) and the Torino s-process post processing code \citep{1997ApJ...478..332S,1998ApJ...497..388G} which were also used to calculate the isotopic compositions of other s-process elements such as Sr, Zr, Mo, and Ba \citep{2003ApJ...593..486L}. The previous work confirmed that the s-process isotopes of Ru (\ce{^{99}Ru}, \ce{^{100}Ru}, \ce{^{101}Ru}, and \ce{^{102}Ru}) found inside the SiC grains had originated from AGB stars although the in-situ decay of \ce{^{99}Tc}, a short-lived radionuclide (SLR), must be taken into account for the abundance of \ce{^{99}Ru}. Since the NuGrid project covers a wider range of initial stellar masses and metallicities than the models considered in the previous study, it is worth checking whether the same conclusion can be drawn from a larger number of samples. 
Additional advantage of using the NuGrid data comes from the fact that the relevant physics for the stellar evolution, which is often determined by choosing parameters in the code, can be handled more systematically. 
We note that in the models of the previous study, the amounts of Ru isotopes varied with the size of the \ce{^{13}C} pocket, which was chosen as a free parameter while the mass and the metallicity were fixed.

Previous works assumed a common scenario in order to explain the isotopic compositions of Ru (and other s-process elements) contained within presolar SiC grains. In this scenario, 
%
SiC grains form during the AGB phase, often including the later thermal pulse (TP) phase, and the isotopes of Ru (and other s-process elements) are locked up into the SiC grains~\citep[e.g.,][and references therein]{2003ApJ...593..486L}. During the AGB phase, convection occurring on top of the carbon-oxygen (CO) core delivers carbon to stellar surface, which is often called third dredge-up (TDU). The increase in the carbon abundance near the surface provides better environments for the formation of SiC (and possibly other grains containing carbon) because molecules and grains are likely to form in the outskirt of an AGB star, i.e., an expanding surface/envelope during the (TP) AGB phase. In our work presented in this paper, we also take into account the same, currently accepted, scenario and apply the C>O constraint (i.e., the carbon abundance larger than the oxygen abundance), which is commonly adopted for the condition of the SiC formation. 

This paper is organized as follows. In the next section, we briefly describe the NuGrid project. Section \ref{sec:results} presents the isotopic compositions of Ru calculated from the selected NuGrid data sets. Conclusion is in Section \ref{sec:conclusion}. 

\begin{deluxetable*}{ccccccc}

\tablecaption{Low-Mass Stars in NuGrid Data Sets \label{tab1}}

\tabletypesize{\footnotesize}
\tablehead{
\colhead{Metallicity} & \colhead{~~~~~Mass~~~~~} & \colhead{~~~~C>O~~~~} & \colhead{~~~~~~Pulse~~~~~~} & \colhead{Number} & \colhead{Number} & \colhead{Location} \\
\colhead{($Z$)} & \colhead{($M_{\sun}$)} & \colhead{(Y/N)} & \colhead{Type} 
& \colhead{~~~of Pulses \tablenotemark{a}}~~~ & 
\colhead{~~of 90\% Pulses \tablenotemark{b}~~} &
\colhead{~~of 90\% Pulses \tablenotemark{c}~~}
}
\startdata
\hline
0.02 & 1 & N & & & & \\
& 1.65 & N & & & & \\
& 2 & Y & I & 5 & 3 & $3-5$ \\
& 3 & Y & I &11 & 7 & $5-11$ \\
& 4 & Y & I & 5 & 4 & $2-5$ \\
& 5 & N & & & & \\
& 6 & N & & & & \\
& 7 & N & & & & \\
\hline
0.01 & 1 & N & & & & \\
& 1.65 & Y & I & 4 & 2 & $3-4$\\
& 2 & Y & I & 7 & 4 & $4-7$ \\
& 3 & Y & I & 9 & 5 & $5-9$ \\
& 4 & Y & I & 10 & 5 & $6-10$ \\
& 5 & N & & & & \\
& 6 & N & & & & \\
& 7 & N & & & & \\
\hline
0.006 & 1 & N & & & & \\
& 1.65 & Y & I & 4 & 3 & $2-4$ \\
& 2 & Y & I & 9 & 4 & $5-9$\\
& 3 & Y & I & 6 & 2 & $5-6$\\
& 4 & Y & I & 1 & 1 & 1\\
& 5 & N & & & & \\
& 6 & N & & & & \\
& 7 & N & & & & \\
\hline
0.001 & 1 & N & & & & \\
& 1.65 & Y & I & 13 & 4 & $10-13$ \\
& 2 & Y & I & 12 & 5 & $8-12$ \\
& 3 & Y & I & 7 & 2 & $6-7$ \\
& 4 & Y & I & 5 & 2 & $4-5$ \\
& 5 & Y & I & 4 & 3 & $2-4$ \\
& 6 & N & & & & \\
& 7 & Y & II & 1 & 1 & 1 \\
\hline
0.0001 & 1 & Y & I & 2 & 1 & 2 \\
& 1.65 & Y & I & 11 & 4 & $8-11$ \\
& 2 & Y & I & 11 & 3 & $9-11$ \\
& 3 & Y & I & 10 & 3 & $8-10$ \\
& 4 & Y & I & 8 & 1 & 8 \\
& 5 & Y & I & 12 & 6 & $7-12$ \\
& 6 & Y & I & 20 & 12 & $9-20$ \\
& 7 & Y & II & 1 & 1 & 1 \\
\enddata

\tablenotetext{a}{Number of pulses that satisfy the C>O condition}
\tablenotetext{b}{Number of pulses within which $90\%$ of the total abundances are contained. The $90\%$ of the total abundances of individual Ru isotopes vary from isotope to isotope as their total abundances in the total pulses that satisfy the C>O condition do. We counted the number of pulses that contain $90\%$ of the total abundances of all Ru isotopes considered in this work. As a result, more than $90\%$ of the total abundances of some Ru isotopes are contained in these pulses.}
\tablenotetext{c}{Location of pulses that contain $90\%$ of the total abundances. We located each pulse that satisfies the C>O condition by assigning a set of increasing integer numbers to pulses in chronological order. See the right panels of Figures \ref{fig:Z002M2} and \ref{fig:Z0001M7}.}

\end{deluxetable*}

\section{Method: NuGrid Project} \label{sec:method}

Stellar yields play important roles in many astronomical phenomena like the galactic chemical evolution, and 
the NuGrid project aimed to estimate stellar yields for elemental abundances, in particular, those of heavy nuclei, which are produced by stars~\citep{2018MNRAS.480..538R}. Various sources for the stellar yields; stellar winds, planetary nebulae (PNe), and supernova (SN) explosions were included in the NuGrid project. Another goal of the NuGrid project was to identify uncertainties in nuclear reactions that affect the stellar yields by analyzing simulation data in comparison with measurements. In order to achieve these goals, the NuGrid project used a post-processing nucleosynthesis method based upon a 1D stellar evolution code, Modules for Experiments in Stellar Astrophysics \citep[MESA,][in particular, revision 3709]{2011ApJS..192....3P}. MESA first calculated stellar evolution by using so called basic nuclear reaction networks (e.g., abb.net, agbtomassive.net, and approx21.net) which included the reactions contributing to most thermonuclear energy inside a star such as p-p chain, CNO cycles, triple-$\alpha$, and $\alpha$-capture. Then, a post-processing code, called "mppnp", calculated the updates of elemental abundances by performing all relevant nuclear reactions at each time cycle which was determined in the first step~\citep{2016ApJS..225...24P}. Because the nuclear reactions relevant to the elemental abundances of heavy elements do not contribute to the total thermonuclear energy as much as those in the basic networks do, the errors for the elemental abundances obtained by this post-processing calculation are small when compared with the results obtained with the fully self-consistent calculation which evolves a star with the full reactions during the whole lifetime. The advantage of using this post-processing nucleosynthesis method is obvious: it saves a large amount of computing time. The mppnp code traces the elemental abundances up to Bi.

\begin{figure*}
\plotone{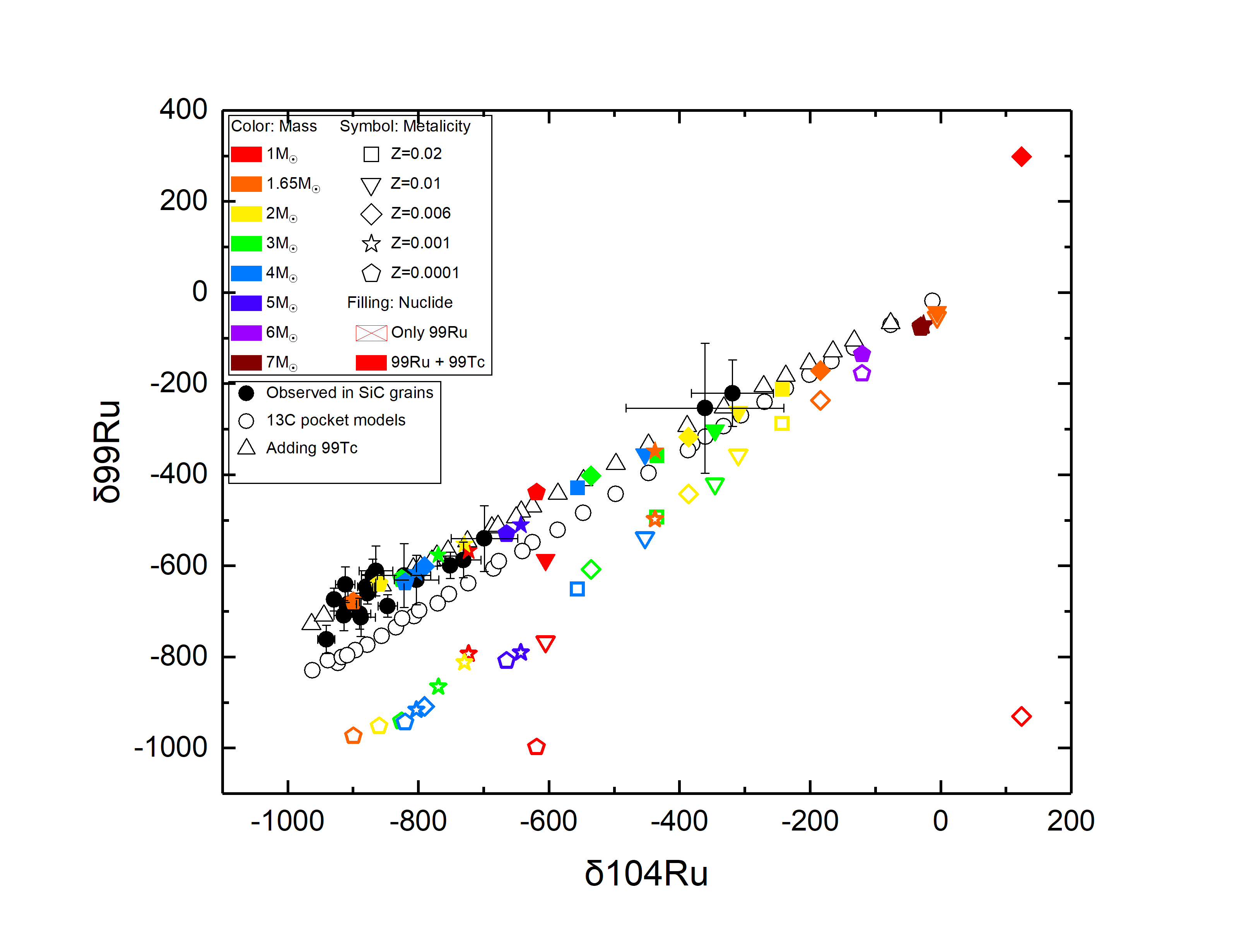}
\caption{$\delta\ce{^{99}Ru}$ versus $\delta\ce{^{104}Ru}$. 
$\delta\ce{^{x}Ru}$ is defined in equation (\ref{eq:delta}). Black filled circles with error bars are from the measured abundances of Ru isotopes within the presolar SiC grains \citep{2004Sci...303..649S}. Open black circles are predictions from the FRANEC model with the Torino s-process code \citep{1997ApJ...478..332S,1998ApJ...497..388G} in which the size of the $\ce{^{13}C}$ pocket varies inside the same low-mass star with $1.5~M_\sun$. 
The size of the $\ce{^{13}C}$ pocket increases from top right to bottom left in all three panels. 
Detailed description on the comparison between the measurements and the predictions of the $\ce{^{13}C}$ pocket model can be found in \citet{2004Sci...303..649S}. 
The results obtained from the NuGrid data are presented with different colors and symbols that correspond to different initial masses and metallicities, respectively. We apply the C>O condition to the NuGrid data during the entire lifetime. See the text for more details on how this condition applies. 
Open black triangles are the predictions from the $\ce{^{13}C}$ pocket model, including the in-situ decay of $\ce{^{99}Tc}$ to $\ce{^{99}Ru}$. We present the NuGrid results in a similar fashion by using open color symbols only for $\ce{^{99}Ru}$ and filled color symbols for the results that include the contribution of the in-situ decay of $\ce{^{99}Tc}$ to $\ce{^{99}Ru}$.
}
\label{fig:wCO1}
\end{figure*}

\begin{figure*}
\plotone{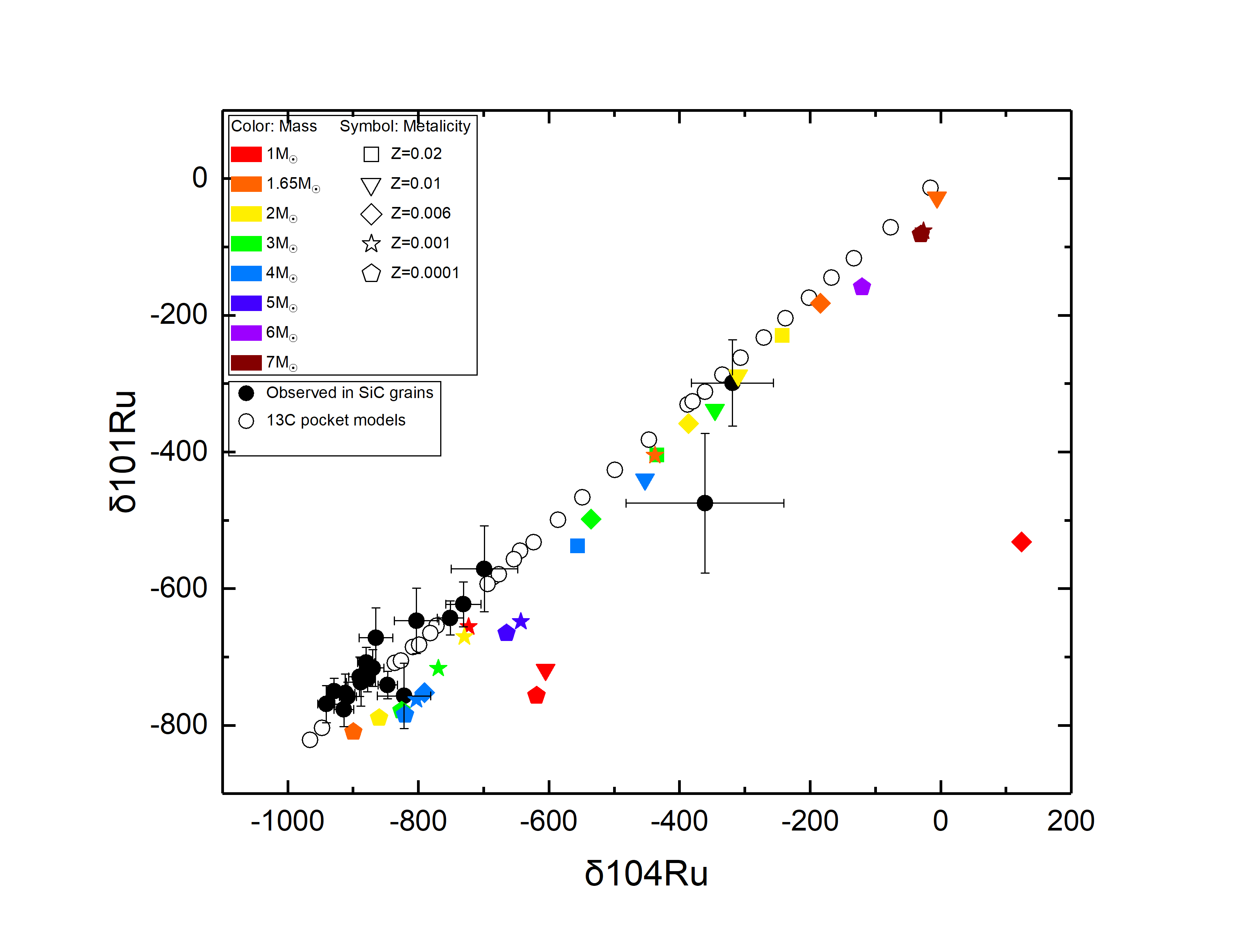}
\caption{$\delta\ce{^{101}Ru}$ versus $\delta\ce{^{104}Ru}$. Symbols and colors are the same as in Figure \ref{fig:wCO1} except the contribution of $\ce{^{99}Tc}$, which is relevant only for $\ce{^{99}Ru}$.} 
\label{fig:wCO2}
\end{figure*}

\begin{figure*}
\plotone{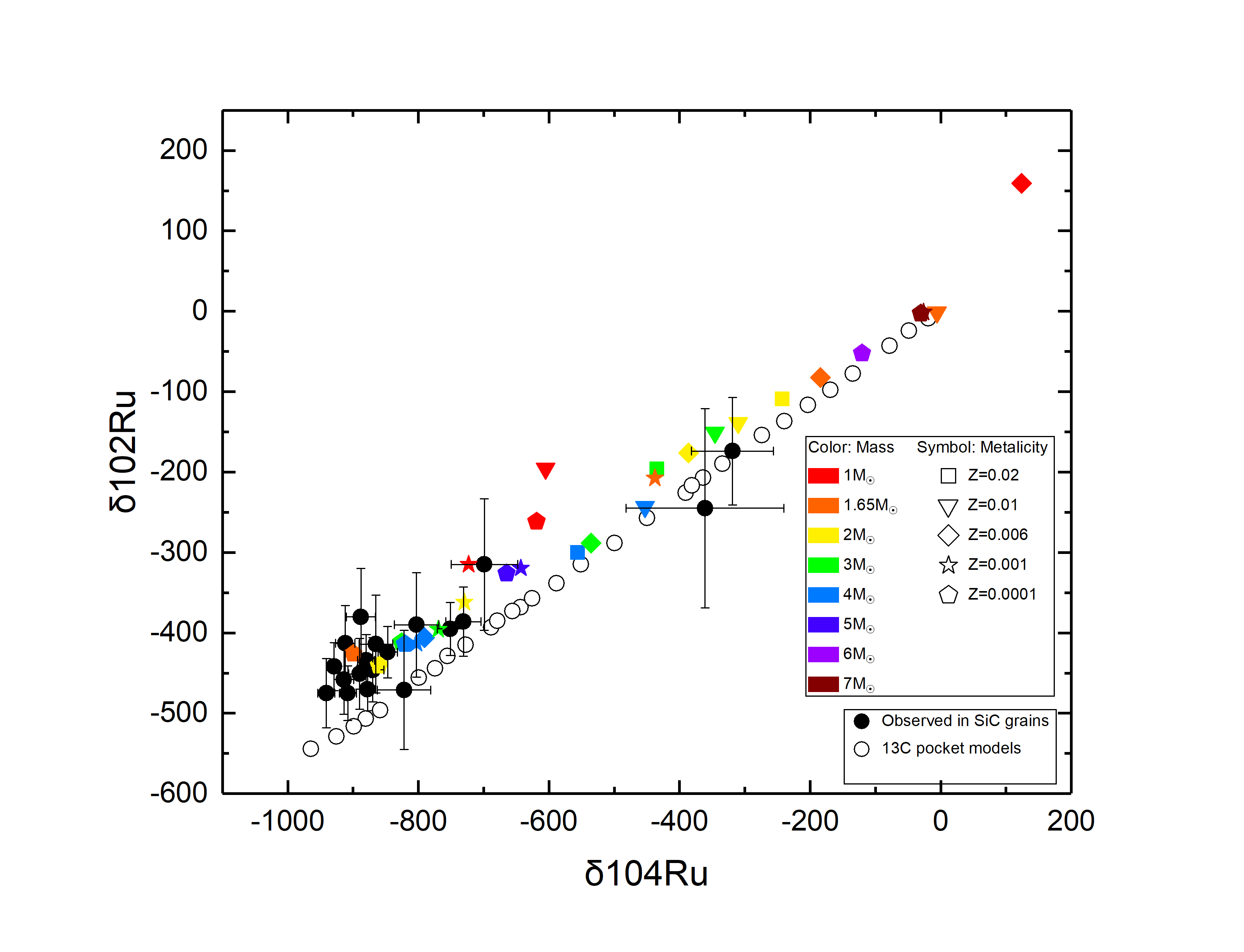}
\caption{$\delta\ce{^{102}Ru}$ versus $\delta\ce{^{104}Ru}$. Symbols and colors are the same as in Figure \ref{fig:wCO1} except the contribution of $\ce{^{99}Tc}$, which is relevant only for $\ce{^{99}Ru}$.}
\label{fig:wCO3}
\end{figure*}

For our current work, we use the data in the NuGrid data release set1 and set1ext \citep{2016ApJS..225...24P,2018MNRAS.480..538R} which are publicly available at the NuGrid website (\url{https://nugrid.github.io})\footnote{A more recent data set of the NuGrid project is also available \citep{2019MNRAS.489.1082B,2021Univ....7...25B}. The updated results were obtained by using an updated 
convective-boundary-mixing model which lead to the formation of a \ce{^{13}C}-pocket three times wider in comparison with the one formed in the previous set of models. However, because the updated results were calculated for a smaller set of initial masses and metallicities ($M_{\mbox{ZAMS}}$ = $2~M_{\sun}$, $3~M_{\sun}$ and Z = 0.03, 0.02, 0.01, 0.001, 0.002), we do not include them in our current work.}. 
In these data sets, the results for the elemental abundances are available with the following initial conditions of mass and metallicity: $M_{\mbox{ZAMS}}$ = $1~M_{\sun}$, $1.65~M_{\sun}$, $2~M_{\sun}$, $3~M_{\sun}$, $4~M_{\sun}$, $5~M_{\sun}$, $6~M_{\sun}$, $7~M_{\sun}$, $12~M_{\sun}$, $15~M_{\sun}$, $20~M_{\sun}$, $25~M_{\sun}$ and $Z = 0.02$, 0.01, 0.006, 0.001, 0.0001. Following convention, stars up to $7~M_{\sun}$ and those from $12~M_{\sun}$ and larger are categorized as low-mass stars and massive stars, respectively. 

%
%
%
%
In the NuGrid project, stellar yields are calculated from stellar winds (both for low-mass and massive stars), PNe (only for low-mass stars), and SN explosions (only for massive stars). 
However, the current work focuses on the stellar yields produced only by the stellar winds because Ru isotopes inside presolar SiC grains are assumed to be produced in the scenario mentioned in the previous section. 
%
Therefore, we search for the C>O condition in the stellar wind under which SiC grains are assumed to form. We obtain the isotopic compositions of Ru under the C>O condition by analyzing the raw data in the NuGrid database which are provided in the format of h5 files. In these data, all elemental abundances, including those of C, O, and Ru isotopes, which are lost from the star as a part of the stellar wind, are given as a function of the stellar age. 
Figure \ref{fig:evol} shows an example for a low-mass star. In this figure, we indicate an epoch during which the C>O condition is satisfied, and this epoch corresponds to a part of the TP/TDU phase. 
In low-mass stars, TP is caused by the explosive He burning in the He-intershell, which exists between C/O core and H-envelope. TDU is driven by TP. During TDU, the materials contained in the He-intershell, such as C and Ru isotopes, are delivered to the stellar surface through the convective H-envelope, and then they are lost as stellar wind. In the He-intershell, s-process elements form by taking neutrons which are produced via the $^{13}C(\alpha,n)^{16}O$ reaction. The $\ce{^{13}C}$ pocket, which is required to produce s-process elements like Ru isotopes in consideration, also forms in the He-intershell. 
A very steep spike near the end of the evolution in the left panel of Figure \ref{fig:evol} corresponds the beginning of TP/TDU.
Note that production of different Ru isotopes (and $\ce{^{99}Tc}$) varies pulse to pulse even when the same C>O condition is satisfied, which is seen in the right panel of Figure \ref{fig:evol}. 
%
%

From the NuGrid data, the stellar yields produced by the stellar winds are generally calculated as below. 
\begin{equation}
Y_{i}^{wind} = \int_{0}^{\tau}\dot{M}(t) X_i(t) dt ~,
\end{equation}
where $\tau$ is the final age of the star, i.e., the time just before PN for the low-mass star and the SN explosion (also called presupernova) for the massive star, and $\dot{M}(t)$ and $X_i (t)$ are the mass loss rate and the mass fraction of a specific isotope (indicated as the $i$-th isotope), respectively, at time $t$. The subscript "$i$" in $Y$ and $X$ indicates the same $i$-th isotope. Because we apply the C>O condition in this work, we calculate the yield of each Ru isotope in consideration only during the epoch in which the C>O condition is satisfied. 
As mentioned above, in low-mass stars, the relative amounts of Ru isotopes (and $\ce{^{99}Tc}$) change pulse to pulse during the TP/TDU phase, which can affect the isotopic composition of Ru. For this reason, we calculated two stellar yields for low-mass stars: one integrated over all the pulses and the other at each pulse, having the C>O condition still applied. As a result, two different isotopic compositions of Ru were calculated from these two yields, respectively. We present and compare both isotopic compositions of Ru for the low-mass stars in the next section.

\section{Results} \label{sec:results}

We compare the isotopic compositions of Ru obtained from the NuGrid project with the measured ones by using the $\delta$-notation which is commonly used in this field of study. Because \ce{^{100}Ru} is produced only by the s-process, it is useful to apply the $\delta$-notation to other Ru isotopes with respect to \ce{^{100}Ru} as follows. 
\begin{equation}
\delta\ce{^{x}Ru} (\text{\textperthousand}) = 1000 \times \left[ \frac{ \left( \frac{\ce{^{x}Ru}}{\ce{^{100}Ru}} \right)_{grain/model} } { \left( \frac{\ce{^{x}Ru}}{\ce{^{100}Ru}} \right)_{solar} } -1 \right]~, \label{eq:delta}
\end{equation}
where grain/model and solar indicate that the abundance 
of each isotope is measured or obtained from the grain/model and the Sun, respectively. 
%
Note that the same $\delta$-notation was used in \citet{2004Sci...303..649S} who compared their measured isotopic compositions of Ru within the presolar SiC grains (i.e., the grain $\delta \ce{^{x}Ru}$) with predictions of a model which changes the size of the $\ce{^{13}C}$ pocket within the same low-mass star. We use the same $\delta \ce{^{x}Ru}$ for the grain (i.e., measured) part as in \citet{2004Sci...303..649S} and compare it with the model $\delta \ce{^{x}Ru}$ obtained from the NuGrid data. More specifically, the comparison is made on the basis of correlation between two different $\delta \ce{^{x}Ru}$'s, which are calculated with two different isotopes indicated with two different $x$'s. 

As mentioned in the previous section, the NuGrid $\delta \ce{^{x}Ru}$ is calculated from two different stellar yields of Ru isotopes: one from the entire lifetime and the other from individual pulses during the TP/TDU phase for low-mass stars. In both cases, we applied the same C>O condition.
We first blindly searched for the C>O condition in the stellar winds in all NuGrid data sets, including massive stars. As expected, we did not find any case to satisfy the condition in the massive stars, regardless of the metallicity, for the entire lifetime. As a result, we present the results obtained only from low-mass stars in this section. In case of low-mass stars, we found that all the epochs that satisfy the C>O condition were a fraction of the TP/TDU phase. The next two subsections present the NuGrid $\delta \ce{^{x}Ru}$ calculated for the entire lifetime (or integrated over all the pulses) and at individual pulses during TP/TDU, respectively. We also compare these two results and give some discussions on them.

\begin{figure*}
\epsscale{1.17}
\plottwo{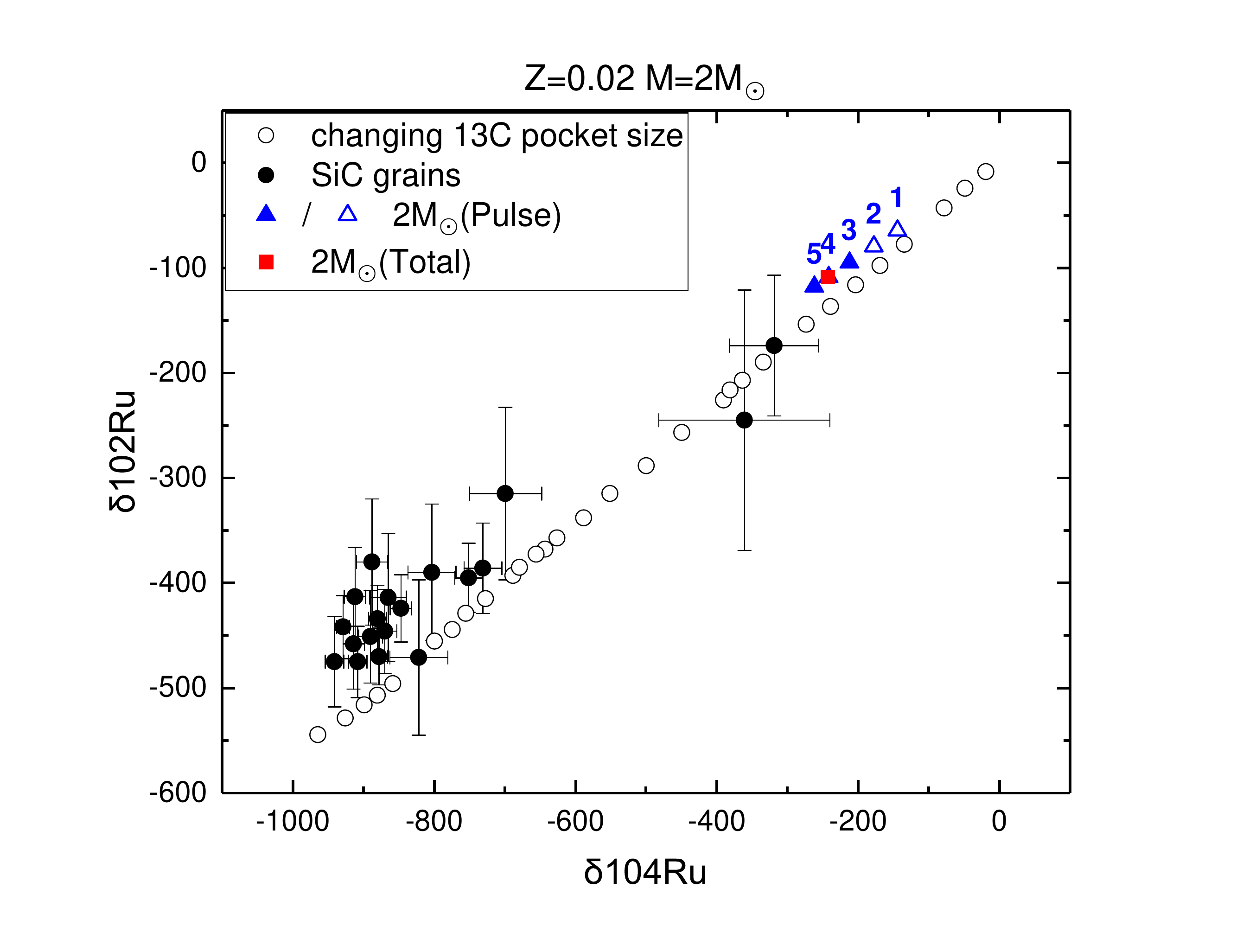}{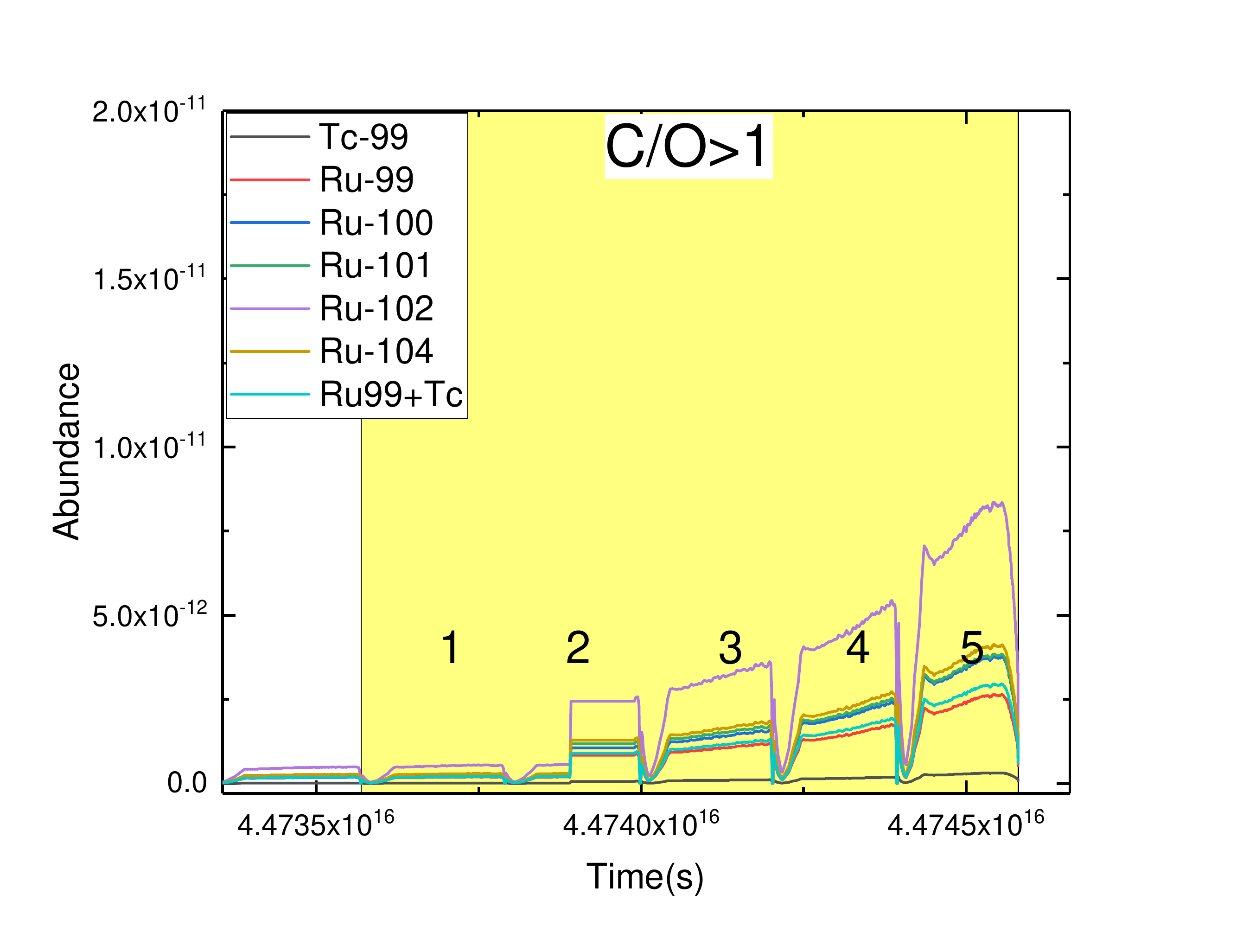}
\caption{Left panel: $\delta\ce{^{102}Ru}$ versus $\delta\ce{^{104}Ru}$ for a low-mass star with $Z=0.02$ and $M=2~M_{\sun}$. Filled and open circles represent the same quantities as in Figure \ref{fig:wCO3}. Triangles are calculated from the abundances at individual pulses. The number on each triangle corresponds to each pulse in the right panel. Filled triangles, combined, contain $90\%$ of the total abundances, more accurately, at least $90\%$ of the total abundances of \ce{^{99}Ru}, \ce{^{100}Ru}, \ce{^{101}Ru}, \ce{^{102}Ru}, and \ce{^{104}Ru}, which are contained in all the pulses under the C>O condition (see the caption in Table \ref{tab1}). Red square is calculated from the total abundances, summed over all the pulses that satisfy the C>O condition. Note that this data point (red square) is identical to yellow square in Figure \ref{fig:wCO3}. Right panel: Abundances of Ru isotopes as a function of time. As in Figure \ref{fig:evol}, the region that satisfies the C>O condition is shaded with yellow and the y-axis is in the unit of solar mass. We assign the integer numbers to pulses in chronological order. (Similar figures for the remaining 22 sets that belong to Type I can be found online.)} 
\label{fig:Z002M2}
\end{figure*}

\subsection{Entire Lifetime}

The results obtained for the entire lifetime are shown in Figures \ref{fig:wCO1} to \ref{fig:wCO3}. 
First of all, we need to mention that some low-mass stars in the NuGrid data sets do not satisfy the C>O condition even when the condition is searched for during the entire lifetime. 
It has been generally known that low-mass stars within a mass range of $1~M_{\sun}$ to $8~M_{\sun}$ go through the AGB phase, including TP and TDU, and the abundance patterns of the Ru isotopes in all of the 40 sets confirm this, i.e., that all the low-mass stars in the NuGrid data show the AGB features. 
However, as shown in Table \ref{tab1}, out of 40 sets of low-mass stars (8 different initial masses times 5 different initial metallicities), only 25 sets satisfy the C>O condition.
For lower-mass stars such as $1~M_\sun$ and $1.65~M_\sun$, the C>O condition is not met for high metallicity ($Z=0.02$). Similarly, for upper-mass stars such as $5~M_\sun$, $6~M_\sun$, and $7~M_\sun$, the C>O condition is met only for low metallicities: $Z=0.001$ and $0.0001$ for $5~M_\sun$ and $7~M_\sun$, and $Z=0.0001$ for $6~M_\sun$. In case of $2~M_\sun$, $3~M_\sun$, and $4~M_\sun$, the C>O condition is satisfied for all values of metallicity. 
We emphasize that the C>O condition is satisfied during the TP/TDU phase in all of 25 sets. We also note that during TP/TDU, abundances of Ru isotopes in the stellar wind as a function of time also show pulse-like shapes as shown in Figure \ref{fig:evol}.

From Figures \ref{fig:wCO1} to \ref{fig:wCO3}, we find that although some outliers exist, the NuGrid results generally show linear correlations between $\delta \ce{^{99,101,102}Ru}$ and $\delta \ce{^{104}Ru}$, which were also seen in the predictions from the $\ce{^{13}C}$ pocket model \citep{1997ApJ...478..332S,1998ApJ...497..388G} in the previous study \citep{2004Sci...303..649S}. We emphasize that the NuGrid results are obtained with a wider range of masses and metallicities for low-mass stars while the predictions of the $\ce{^{13}C}$ pocket model come from a single mass and metallicity (with the size of the $\ce{^{13}C}$ pocket varied). Because similar linear correlations seem to exist between $\delta \ce{^{99,101,102}Ru}$ and $\delta \ce{^{104}Ru}$ which are calculated from the measured Ru isotopes within the presolar SiC grains, both the NuGrid results obtained from the low-mass stars and the predictions from the $\ce{^{13}C}$ pocket model can explain the measurements.
However, the NuGrid results are more capable than the $\ce{^{13}C}$ pocket model. Because each NuGrid data point in the $\delta\ce{^{99,101,102}Ru}$--$\delta\ce{^{104}Ru}$ plot in Figures \ref{fig:wCO1} to \ref{fig:wCO3} is calculated from a specific set of mass and metallicity, the environment for the measurement data point can be differentiated with the NuGrid results. For example, in all three plots in Figures \ref{fig:wCO1} to \ref{fig:wCO3}, two NuGrid data points, orange and yellow pentagons, which are obtained from $Z=0.0001$ with $1.65~M_\sun$ and $2~M_\sun$, respectively, are located closely to the region where most of measurement data points are clustered. This could imply that population II stars affected the environment of the presolar system. 

From the NuGrid data which cover a wide range of metallicities, a trend of $\delta \ce{^{x}Ru}$ can be claimed to exist as a function of metallicity as follows. For a given (same) initial mass, the absolute value of $\delta \ce{^{x}Ru}$ increases as metallicity decreases; for example, in case of $M=2 M_{\sun}$ (yellow), symbols are put in order of square ($Z=0.02$), inverted-triangle ($Z=0.01$), diamond ($Z=0.006$), star ($Z=0.001$), and pentagon ($Z=0.0001$), from upper right to lower left, in Figures \ref{fig:wCO1} to \ref{fig:wCO3}. The exception for this trend is found between $Z=0.02$ (green square) and $0.01$ (green inverted-triangle) for $3~M_\sun$ and between the same metallicities (blue square and blue inverted-triangle) for $4~M_\sun$. Note that the results for $1~M_\sun$ do not show any feature of this trend; four red symbols (no red square) are randomly spread out in the plots. 

Figure \ref{fig:wCO1} confirms the previous result of \citet{2004Sci...303..649S} for $\ce{^{99}Ru}$ with the NuGrid data; the contribution of $\ce{^{99}Tc}$, an SLR, which decays to $\ce{^{99}Ru}$ in situ, is required to explain the discrepancy between the measurements and the predictions for $\ce{^{99}Ru}$. 
%
%
However, the contribution of $\ce{^{99}Tc}$ seems more prominent in the NuGrid results than in the $\ce{^{13}C}$ pocket model; the gaps between open and filled color symbols, which represent the NuGrid results, are larger than those between open black triangles and circles, which represent the results of the $\ce{^{13}C}$ pocket model, although the gaps in both results have a similar trend such that they increase from the origin to the bottom-left corner. 
%
Closer examination of the NuGrid results, which cover a wider range of initial masses and metallicities than the $\ce{^{13}C}$ pocket model, reveals that the contribution of $\ce{^{99}Tc}$ decreases as the metallicity increases, regardless of the mass, although exception for this trend is found for $1~M_\sun$. 
Because the results of the $\ce{^{13}C}$ pocket model also show that the contribution of $\ce{^{99}Tc}$ decreases as the size of the $\ce{^{13}C}$ pocket decreases, one can speculate that metallicity affects the size of the $\ce{^{13}C}$ pocket in such a way that the pocket size decreases as the metallicity increases for the same stellar mass. This speculation could be verified by more detailed examination of the stellar evolution results which is beyond the scope of the current work. 
%

Finally, we make a comment on overshooting which may affect the size of the $\ce{^{13}C}$ pocket and the corresponding elemental abundances. 
%
%
%
MESA adopts a convective-boundary-mixing model which is related to overshooting \citep{1997A&A...324L..81H}. The presence of overshooting can change the hydrogen distribution around the He-intershell \citep{1976A&A....47..389M} and the amount of He-shell \citep{2013ApJ...766..118M}. The TDU can be enhanced with the presence of overshooting \citep{2000A&A...360..952H}. 
Because the size of the \ce{^{13}C} pocket depends on TDU, the presence of overshooting can also affect the chemical composition in the stellar evolution, in particular, the s-process nuclei. 
As mentioned earlier, new NuGrid data sets, which were obtained by using an updated convective-boundary-mixing model, were released recently, and the size of the \ce{^{13}C} pocket was found to increase with the updated treatment of overshooting~\citep{2019MNRAS.489.1082B,2021Univ....7...25B}. Although the new NuGrid data covered a relatively smaller range of initial masses and metallicities than the data that we use for the current study, the new data were compared with the isotopic compositions of four s-process elements (Sr, Ba, Zr, and Mo) measured from the presolar SiC grains. We plan to compare the isotopic compositions of Ru obtained from the new NuGrid data with those measured from the SiC grains. It is also useful to compare the isotopic compositions of s-process elements obtained from different NuGrid data sets, the new data and the data that we use. The comparison will be able to further reveal the effect of convective-boundary-mixing (or overshooting) on the abundances of s-process elements.

\begin{figure*}
\epsscale{1.17}
\plottwo{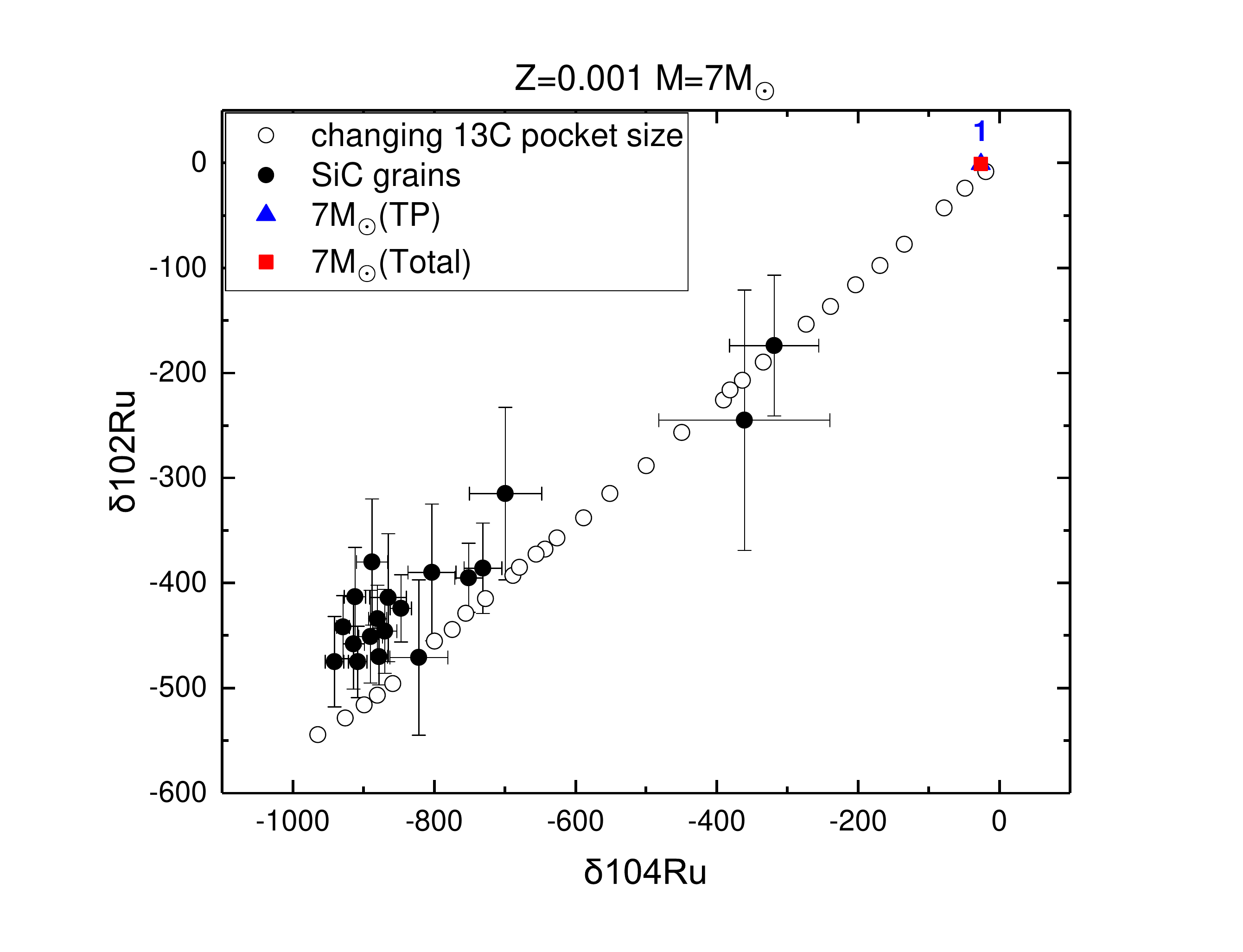}{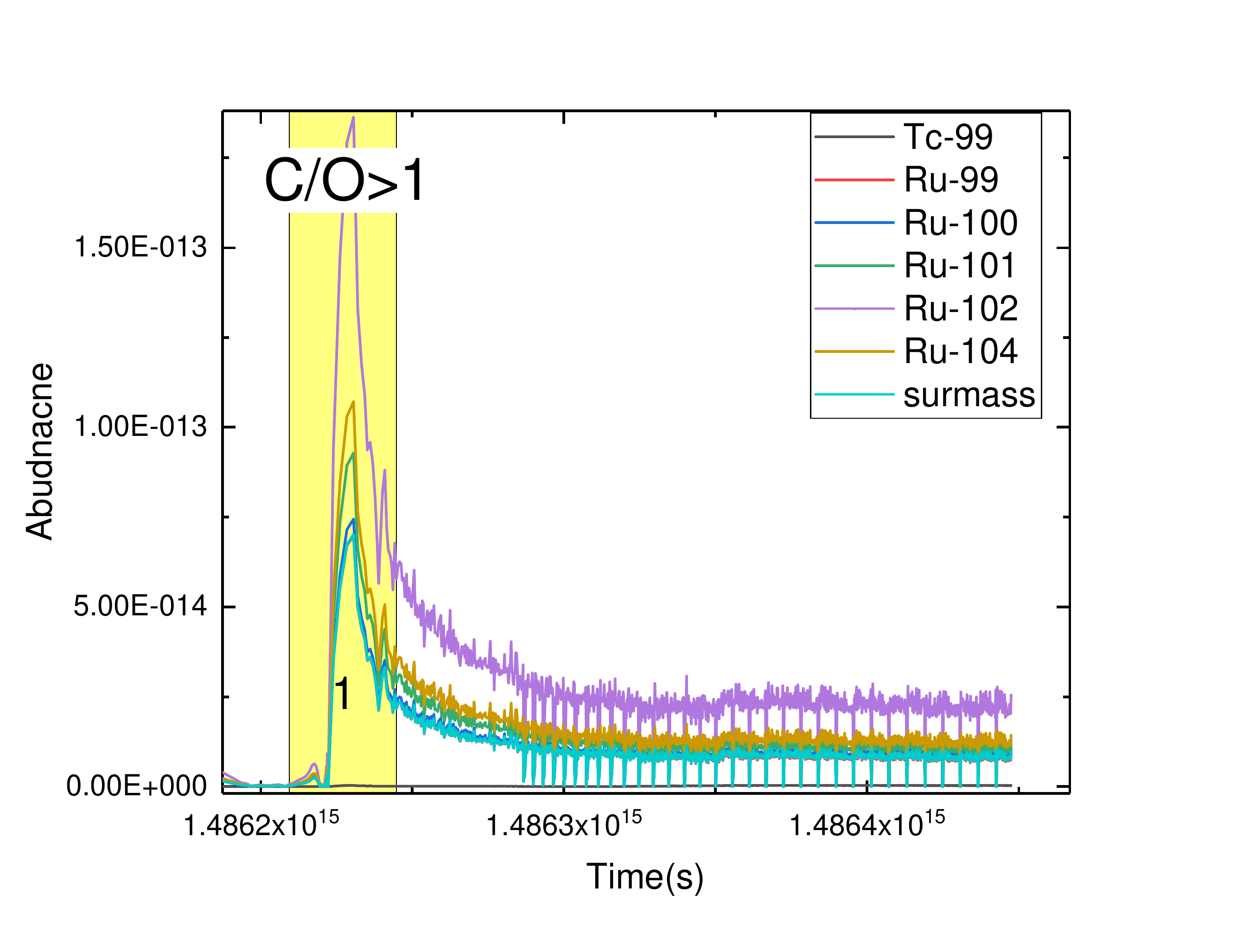}
\caption{Left panel: $\delta\ce{^{102}Ru}$ versus $\delta\ce{^{104}Ru}$ for a low-mass star with $Z=0.001$ and $M=7~M_{\sun}$. Colors and symbols are the same as in Figure \ref{fig:Z002M2}, except that the red square here is identical to brown star in Figure \ref{fig:wCO3}, where brown star is overlapped with brown pentagon. Blue triangle and red square are identical to each other in this plot because a single peak exists. Right panel is the same as in Figure \ref{fig:Z002M2}. (A similar figure for the other set that belongs to Type II, a low-mass star with $Z=0.0001$ and $M=7~M_{\sun}$, can be found online.)} 
\label{fig:Z0001M7}
\end{figure*}

\subsection{Individual Pulses}

The results presented in the previous subsection reveal limited pieces of information on the isotopic compositions of Ru in such a way that the yields are obtained first by summing over many pulses and then $\delta \ce{^{x}Ru}$'s are calculated over the integrated yields. The values of $\delta \ce{^{x}Ru}$ calculated in this way represent weighted averages of all grains, with weights corresponding to the amount of grain production occurring during any particular pulse. 
However, individual grains form during individual pulses. For example, as mentioned before, the TP/TDU phase for the star in Figure \ref{fig:evol} lasts for $\sim 10^5$ years, but grains form quasi-continuously during that period and the Ru isotopic composition is changing throughout the period. Thus, one expects the total output of a star to be an ensemble of grains with varying isotopic compositions corresponding to their times of formation along the evolutionary track of the star.
%
For this reason, in this subsection, we present the isotopic compositions of Ru at individual pulses. In particular,  we investigate the correlations between $\delta \ce{^{x}Ru}$'s with different $x$'s which are calculated at individual pulses. We also compare the results obtained at individual pulses to those in the previous section which were obtained by integrating over all the pulses.

As Table \ref{tab1} shows, among 40 sets of low-mass stars in the NuGrid data, 25 sets have at least one pulse that satisfies the C>O condition during TP/TDU. First of all, we examine the overall shape of pulses for the 25 sets, in particular, the abundances of individual Ru isotopes as a function of time as shown in the right panel of Figure \ref{fig:evol}. The examination reveals that two different types, Type I or II, can be distinguished. In Type I, the abundances generally increase until the end of TP/TDU, while a large peak, which satisfies the C>O condition, exists in early times in Type II. 
Figures \ref{fig:Z002M2} and \ref{fig:Z0001M7} show an example of Type I and II, respectively\footnote{Similar figures for all 25 sets can be found online.}. Table \ref{tab1} shows which type each of 25 sets belongs to. Note that only two out of 25 sets show the Type II pulse shape. We also find that regardless of the metallicity, the abundance patterns of the Ru isotopes in all the sets with $M=7~M_{\sun}$ look similar to Type II. This implies that the largest mass ($M=7~M_{\sun}$) among the low-mass stars in the NuGrid data goes through a qualitatively different evolution during the AGB phase, which may be caused by the carbon burning in the core. 
We identify each pulse in each set by assigning integer numbers to pulses as shown in the right panels of Figures \ref{fig:Z002M2} and \ref{fig:Z0001M7}. For each pulse, we calculate $\delta \ce{^{102}Ru}$ and $\delta \ce{^{104}Ru}$, and plot them together in the left panels of Figures \ref{fig:Z002M2} and \ref{fig:Z0001M7}. 

In case of Type II, the results obtained in this subsection are identical to those in the previous section simply because only a single pulse at early times can satisfy the C>O condition. In contrast, in case of Type I, isotopic compositions of Ru vary pulse to pulse, depending on the pulse shape, although variation is not random. First of all, we can observe that the linear correlation between two $\delta \ce{^{x}Ru}$'s is still valid even if the isotopic compositions of Ru are calculated at individual pulses (for a given low-mass star). Furthermore, the absolute value of $\delta \ce{^{x}Ru}$ increases with time, i.e., with the pulse number. Because the size of the \ce{^{13}C} pocket increases as the absolute value of $\delta \ce{^{x}Ru}$ increases, it can be speculated that the size of the \ce{^{13}C} pocket also increases with time during the TP/TDU phase in the MESA simulations from which the NuGrid data were calculated. Detailed investigation of this speculation, however, is beyond the scope of the current work. 

In almost all the cases of Type I (the only exception is the set with $Z=0.0001$ and $M=6~M_{\sun}$), the amplitude of pulse (or the abundance of each Ru isotope) increases with time. Due to this time behavior of pulses, the isotopic compositions of Ru calculated over the entire pulses do not deviate much from those calculated at individual pulses at later times, which contain more abundant Ru isotopes. As shown in Figure \ref{fig:Z002M2} and other online figures of the Type I sets, red squares (data points calculated from the entire lifetime) are close to filled triangles (data points calculated from the pulses containing $90\%$ of the total abundances). This finding implies that the results in the previous section can be "representative" isotopic compositions of Ru predicted from the NuGrid data. It is because predicted isotopic compositions of Ru are more likely to be these values under the assumption that isotopic compositions of Ru calculated with more abundant Ru isotopes are more likely to be found or observed. In particular, for the two special sets of the NuGrid data, which are closer to most measurements as mentioned in the previous section, the isotopic compositions of Ru calculated at individual pulses that contain $90\%$ of the total abundances are very close to those summed over the entire pulses; see two online figures (or Figures \ref{special1} and \ref{special2}) with $Z=0.0001$ and $M=1.65~M_{\sun}$ and with $Z=0.0001$ and $M=2~M_{\sun}$, respectively.

\section{Conclusion}\label{sec:conclusion}

We obtain the isotopic compositions of Ru by analyzing the NuGrid data and compare them with those measured from the presolar SiC grains. In this work, we assume a scenario which is commonly adopted to explain the isotopes of Ru and other s-process elements found inside the presolar SiC grains. In this scenario, SiC grains form in the outskirt of a star when the carbon abundance is larger than the oxygen abundance (C>O) there, leave the star as a part of the stellar wind, and reach the presolar molecular cloud. 
The Ru isotopes form in the He-intershell between the C/O core and the H envelope, move to the surface of the star during TP/TDU, and then are locked up into the SiC grains.
Following this scenario, we first identify when the C>O condition is satisfied in the stellar wind in the NuGrid data sets which cover a wide range of initial stellar masses and metallicities. The NuGrid data confirm that the C>O condition required for the SiC grain formation is not satisfied in the wind of any massive star, regardless of the metallicity. In contrast, the C>O condition is satisfied in the stellar winds of most of low-mass stars in the NuGrid data. 
We calculate the stellar yield of each Ru isotope produced by the stellar winds of the low-mass stars by applying the C>O condition. From the stellar yield of each Ru isotope calculated in this way, the isotopic compositions of Ru are obtained. We find that the isotopic compositions of Ru calculated from the NuGrid data can explain those measured from the presolar SiC grains. Because the NuGrid data cover a wide range of initial stellar masses and metallicities, the measured isotopic compositions of Ru can be differentiated with the initial masses and the metallicities. We find that most of the measured isotopic compositions of Ru can be explained better with lower-mass stars ($1.65~M_\sun$ and $2~M_\sun$) with low metallicity (Z=0.0001) than with other low-mass stars having different masses and metallicities. The NuGrid data also confirm that the in-situ decay of $\ce{^{99}Tc}$, an SLR, contributed to the abundance of $\ce{^{99}Ru}$, which was found in the previous study~\citep{2004Sci...303..649S}. 

The conclusion mentioned above was drawn from the isotopic compositions of Ru integrated over the entire lifetime (or all the pulses during TP/TDU). Because some isotopic compositions of Ru at individual pulses seem to vary significantly from pulse to pulse, especially when multiple pulses satisfy the C>O condition, it is necessary to confirm whether the conclusion above is valid or not after investigating the isotopic compositions of Ru calculated at individual pulses. We find that the isotopic compositions of Ru integrated over all the pulses, i.e., the ensemble-average isotopic compositions are a reasonable representation which can be compared with measurements because most of Ru isotopes are contained in the last few pulses and the isotopic compositions of Ru in these pulses are close to those integrated over all the pulses.

Alternative scenarios may also provide possible explanations for the isotopic compositions of Ru and other elements measured from the presolar grains. PNe and SNe have been considered as those scenarios. Because the NuGrid data also provide stellar yields produced by PNe and SNe, these scenarios can be investigated with the NuGrid data. 
Furthermore, the NuGrid data that we use in the current study can be compared with the isotopic compositions of other s-process elements, such as Sr, Zr, Mo, and Ba, measured from the presolar SiC grains although the comparison was made with the new NuGrid data. 

\acknowledgements 
We thank the referee for giving us constructive comments and suggestions which improved the original manuscript significantly. We would like to thank all the developers and the contributors of the NuGrid project the results of which this work is based upon. 
This work was supported by a National Research Foundation (NRF) of Korea grant funded 
by the Ministry of Science and ICT of the Korean government (No. 2016R1A5A1013277). 
KHS was also supported by a NRF grant funded by the Ministry of Education of the Korean government (NRF-2015H1A2A1031629-Global Ph.D. Fellowship Program).
KK was also supported by a NRF grant funded by the Ministry of Education of the Korean government (No. 2019R1F1A1062276). 
We acknowledge the hospitality at the Asia Pacific Center for Theoretical Physics (APCTP) where part of this work was done.


\appendix

We include the online figures here. Figures are listed in the same order as in Table \ref{tab1}: Type I sets are presented first and then comes a Type II set. 

\bibliography{RuNuGrid}{}
\bibliographystyle{aasjournal}

\begin{figure*}
\figurenum{A1}
\epsscale{1.17}
\plottwo{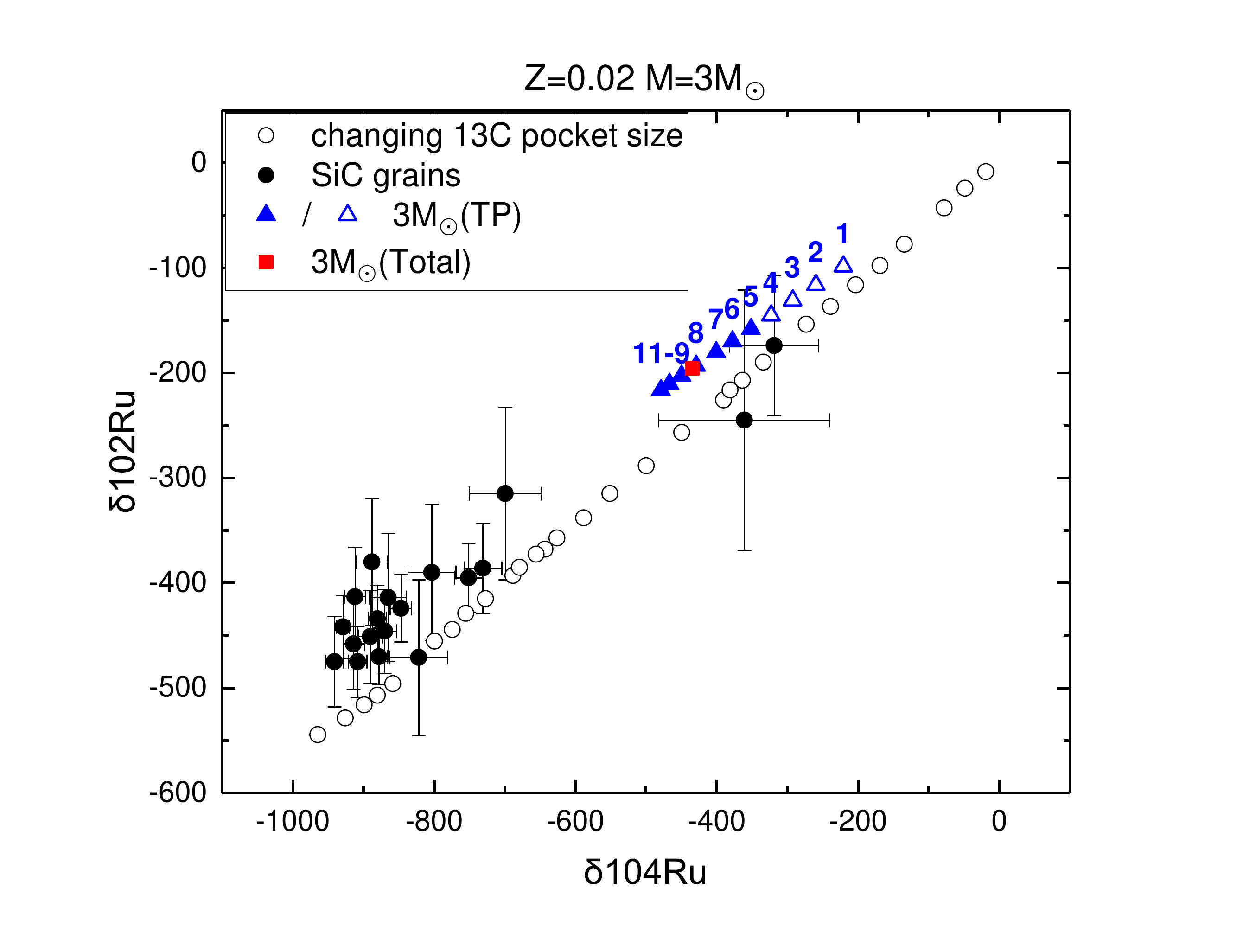}{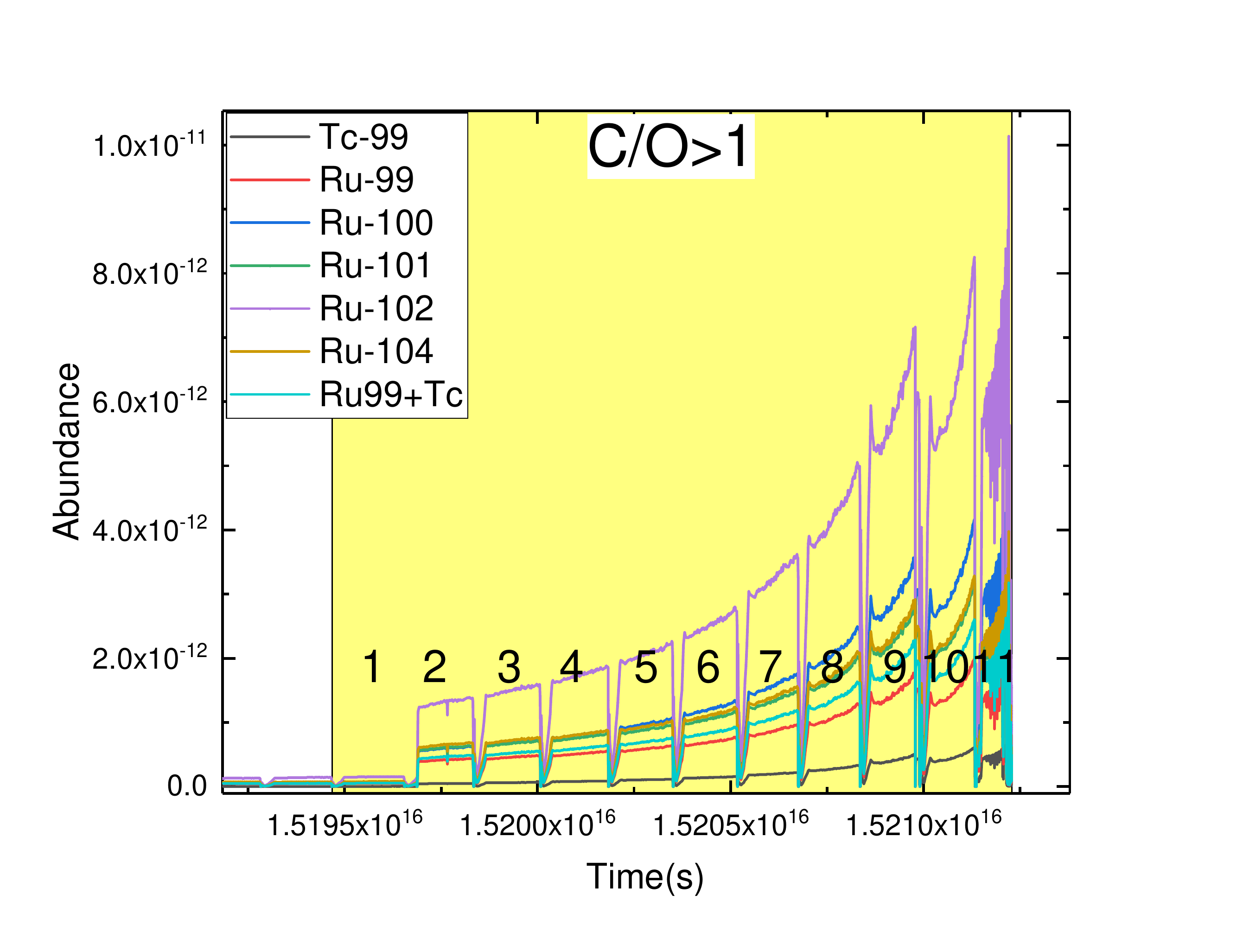}
\caption{A low-mass star with $Z=0.02$ and $M=3~M_{\sun}$}
\end{figure*}

\begin{figure*}
\figurenum{A2}
\epsscale{1.17}
\plottwo{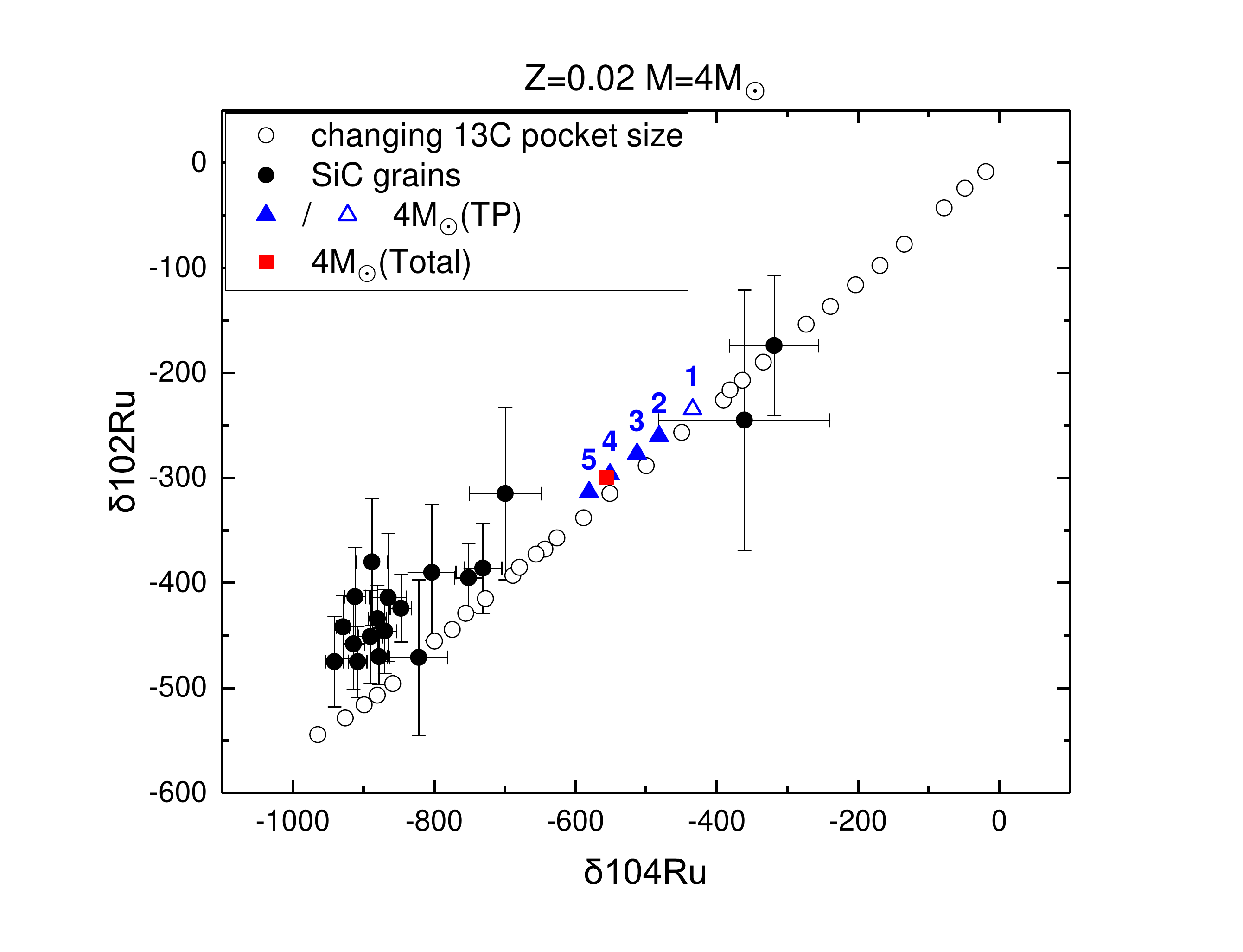}{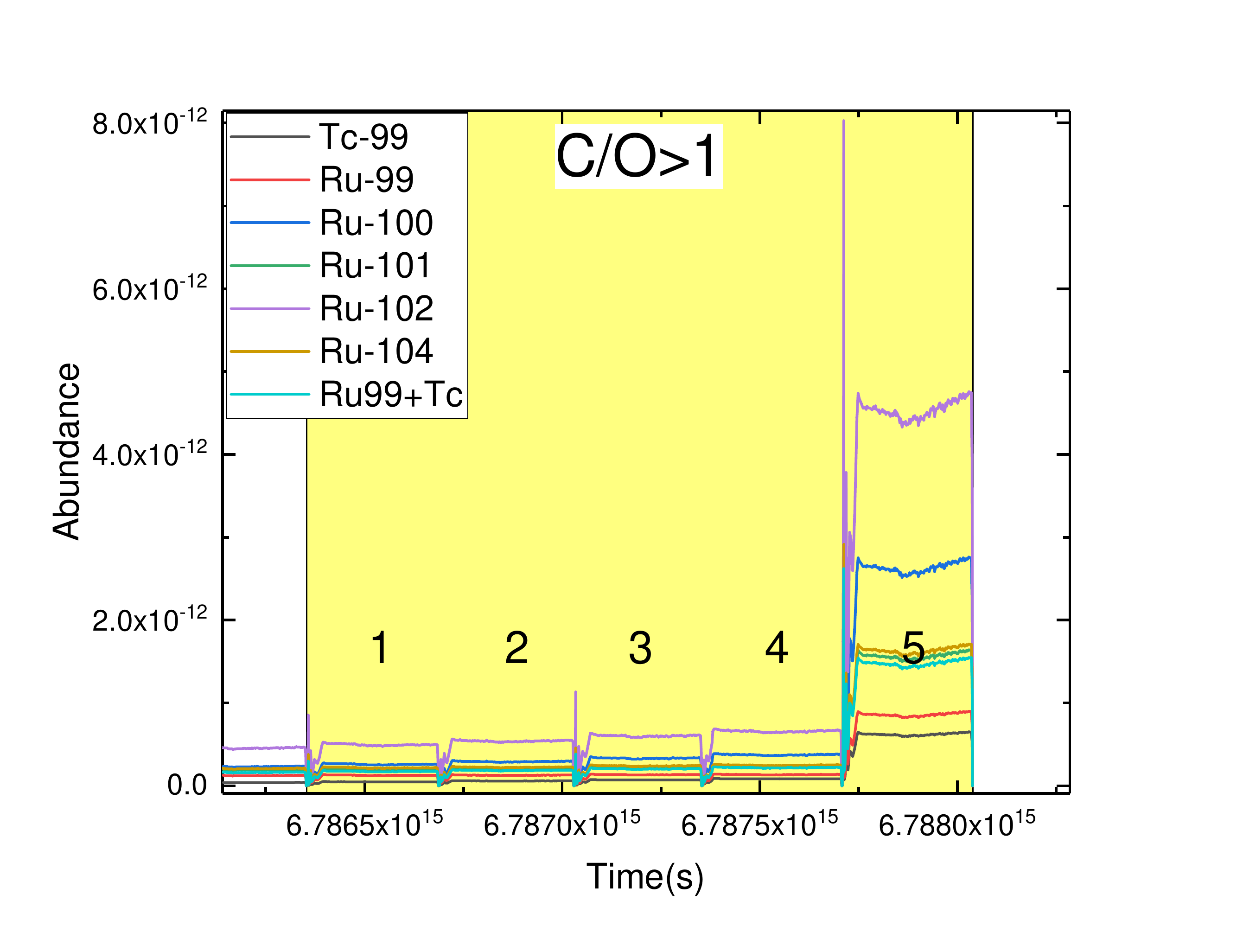}
\caption{A low-mass star with $Z=0.02$ and $M=4~M_{\sun}$}
\end{figure*}

\begin{figure*}
\figurenum{A3}
\epsscale{1.17}
\plottwo{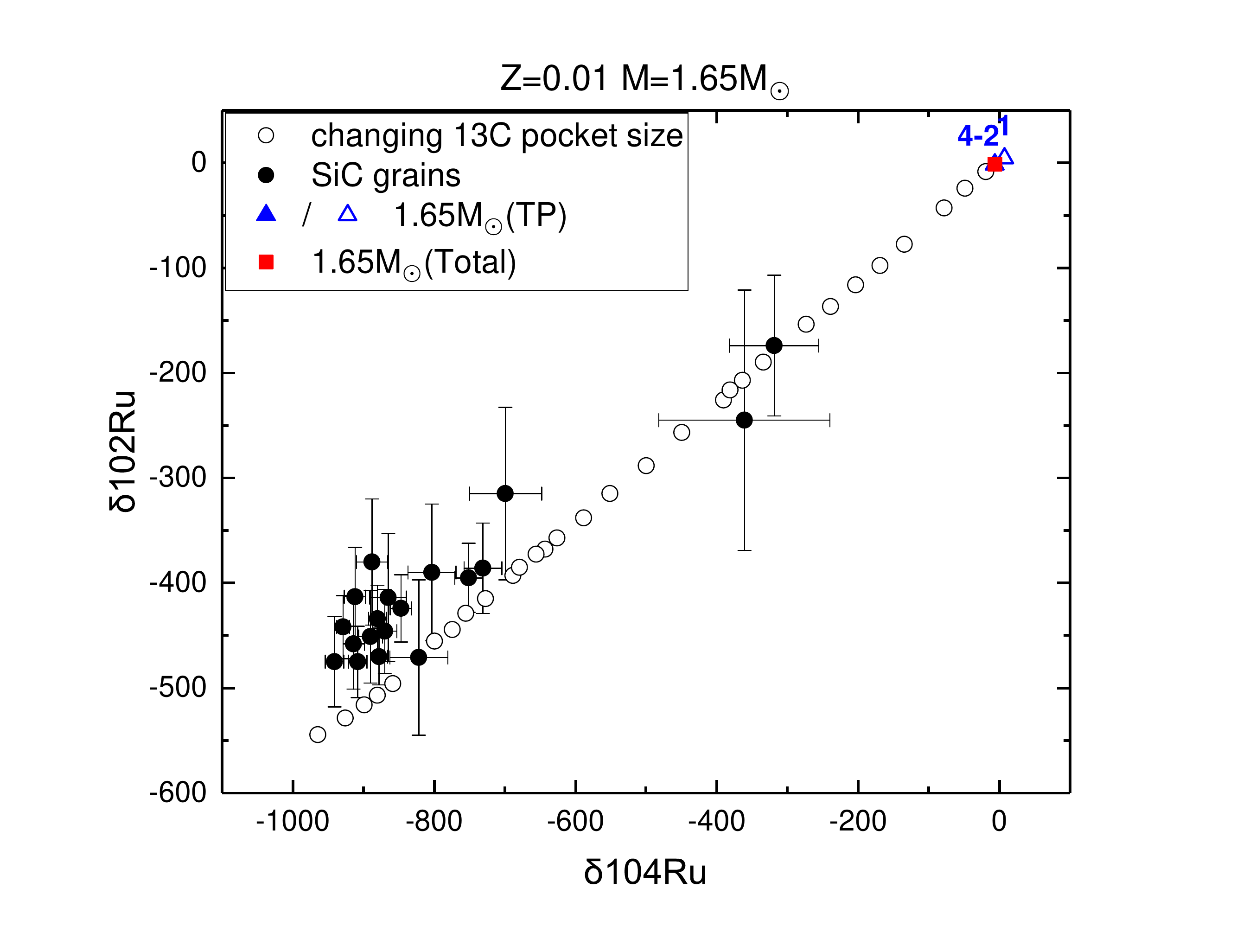}{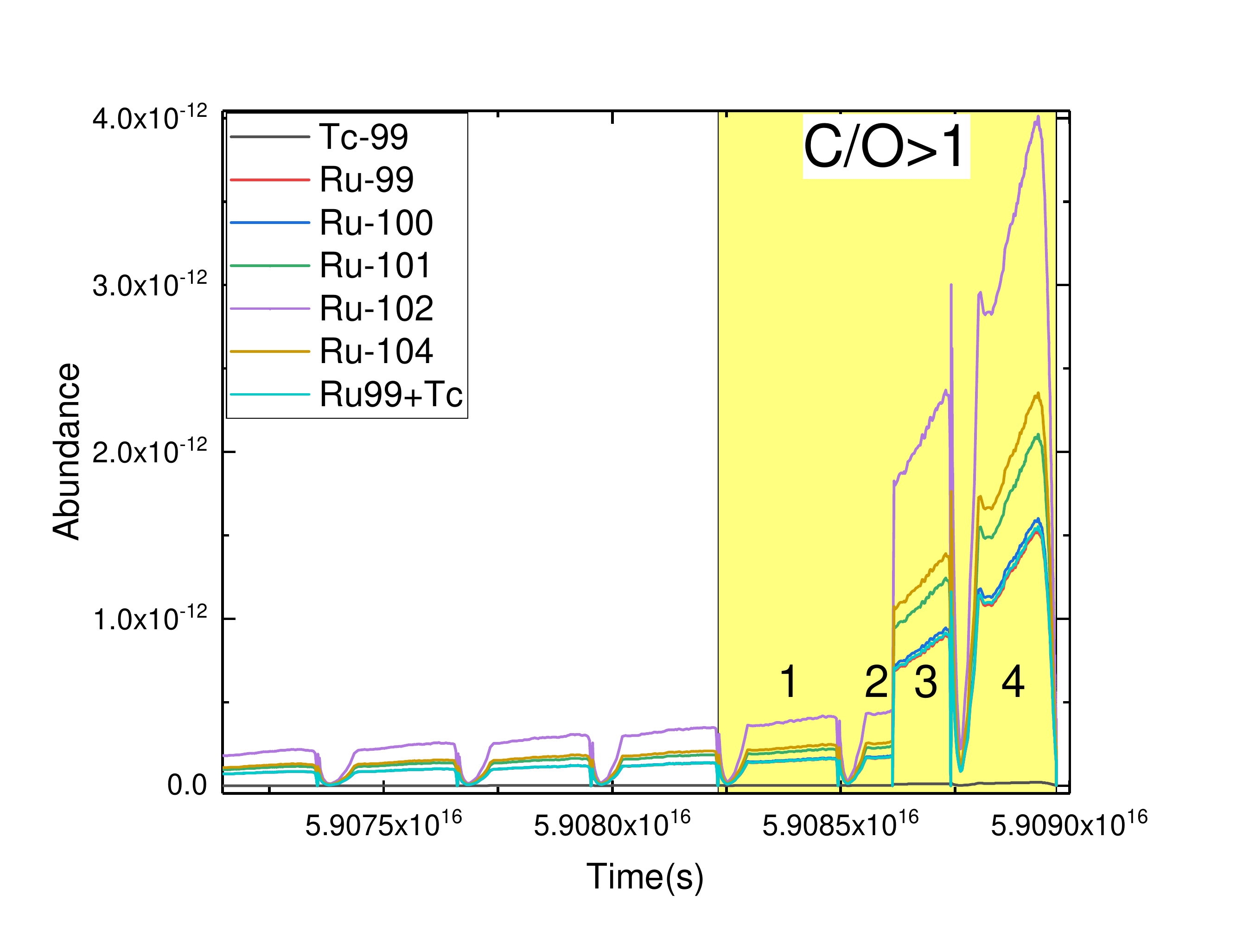}
\caption{A low-mass star with $Z=0.01$ and $M=1.65~M_{\sun}$} 
\end{figure*}

\begin{figure*}
\figurenum{A4}
\epsscale{1.17}
\plottwo{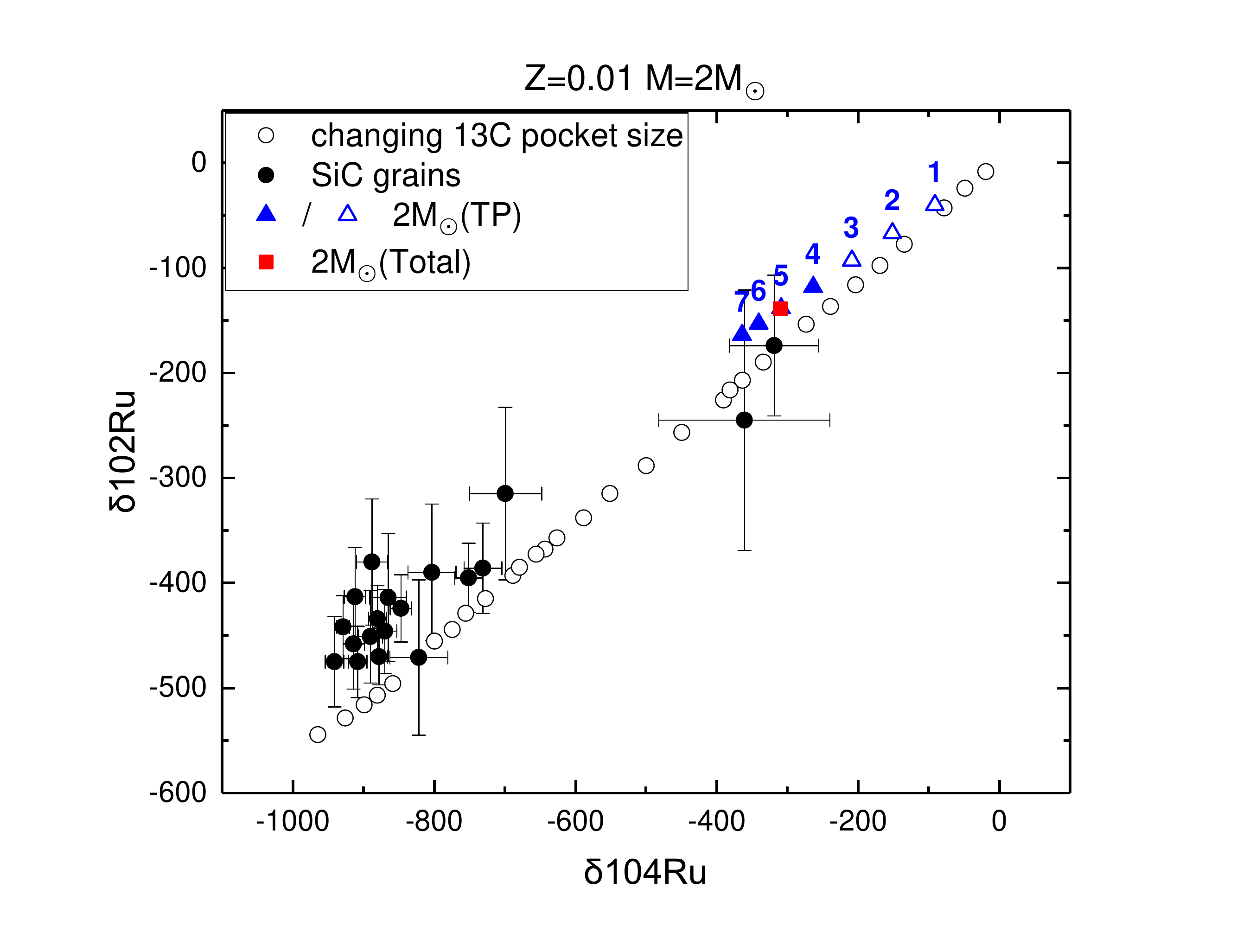}{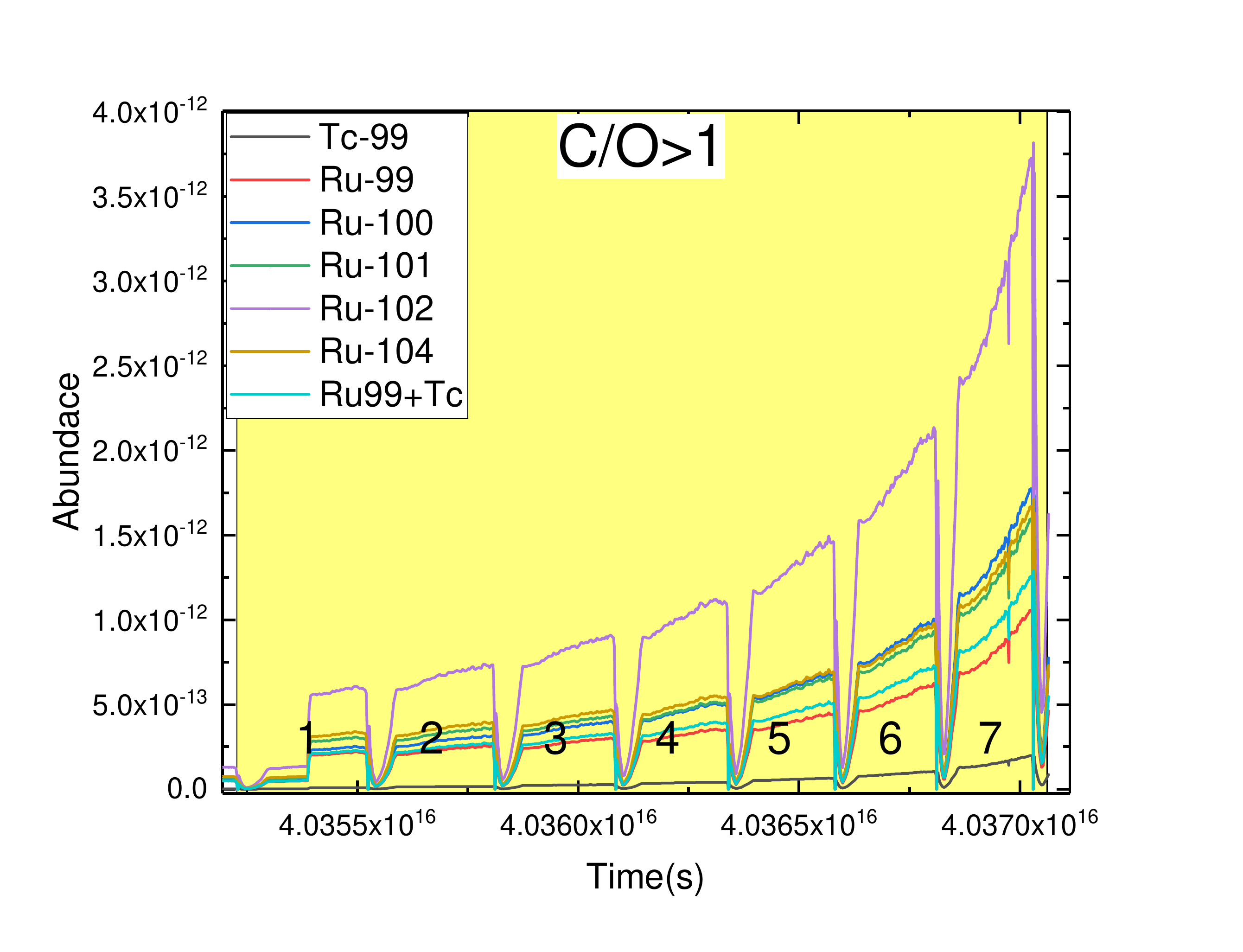}
\caption{A low-mass star with $Z=0.01$ and $M=2~M_{\sun}$}
\end{figure*}

\begin{figure*}
\figurenum{A5}
\epsscale{1.17}
\plottwo{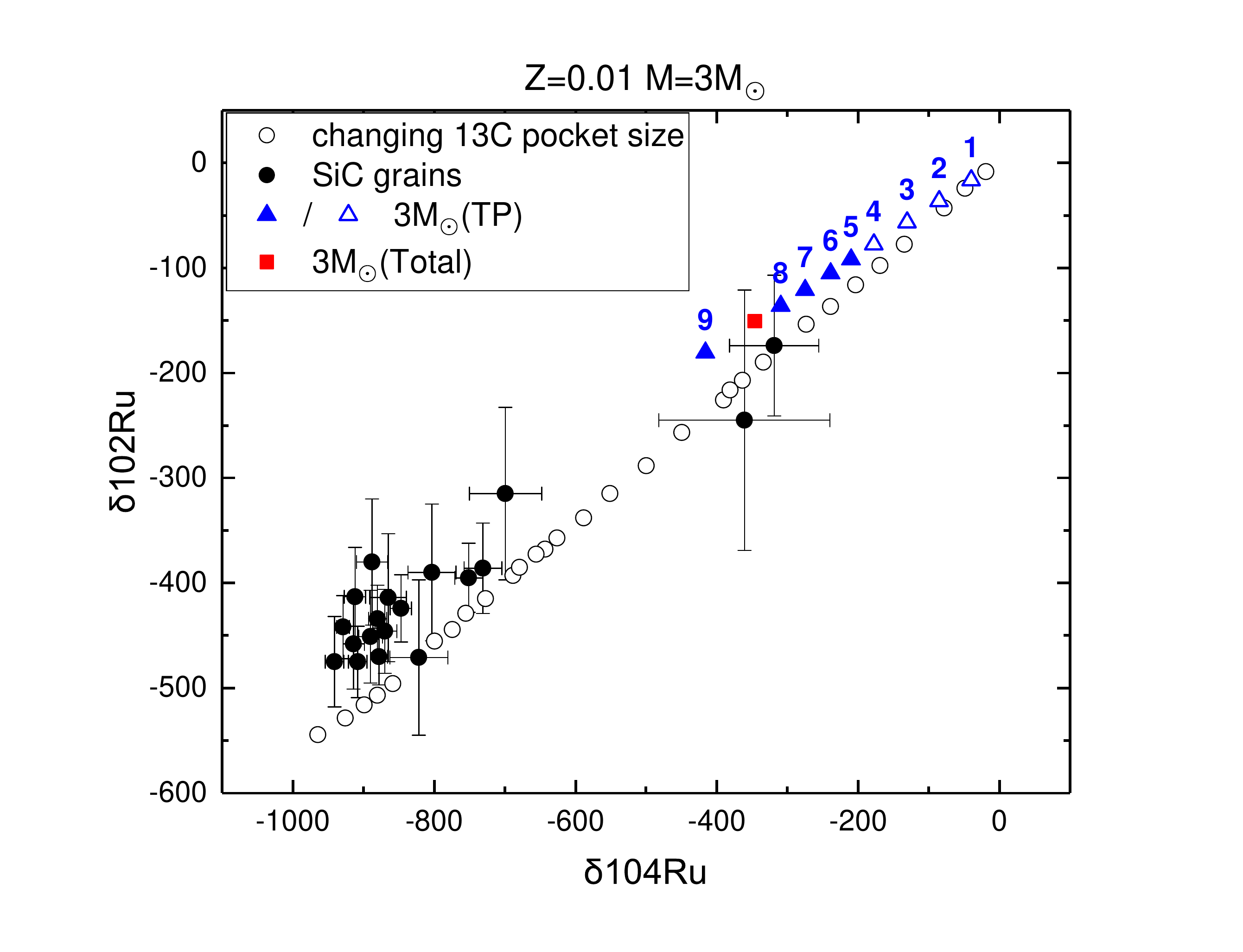}{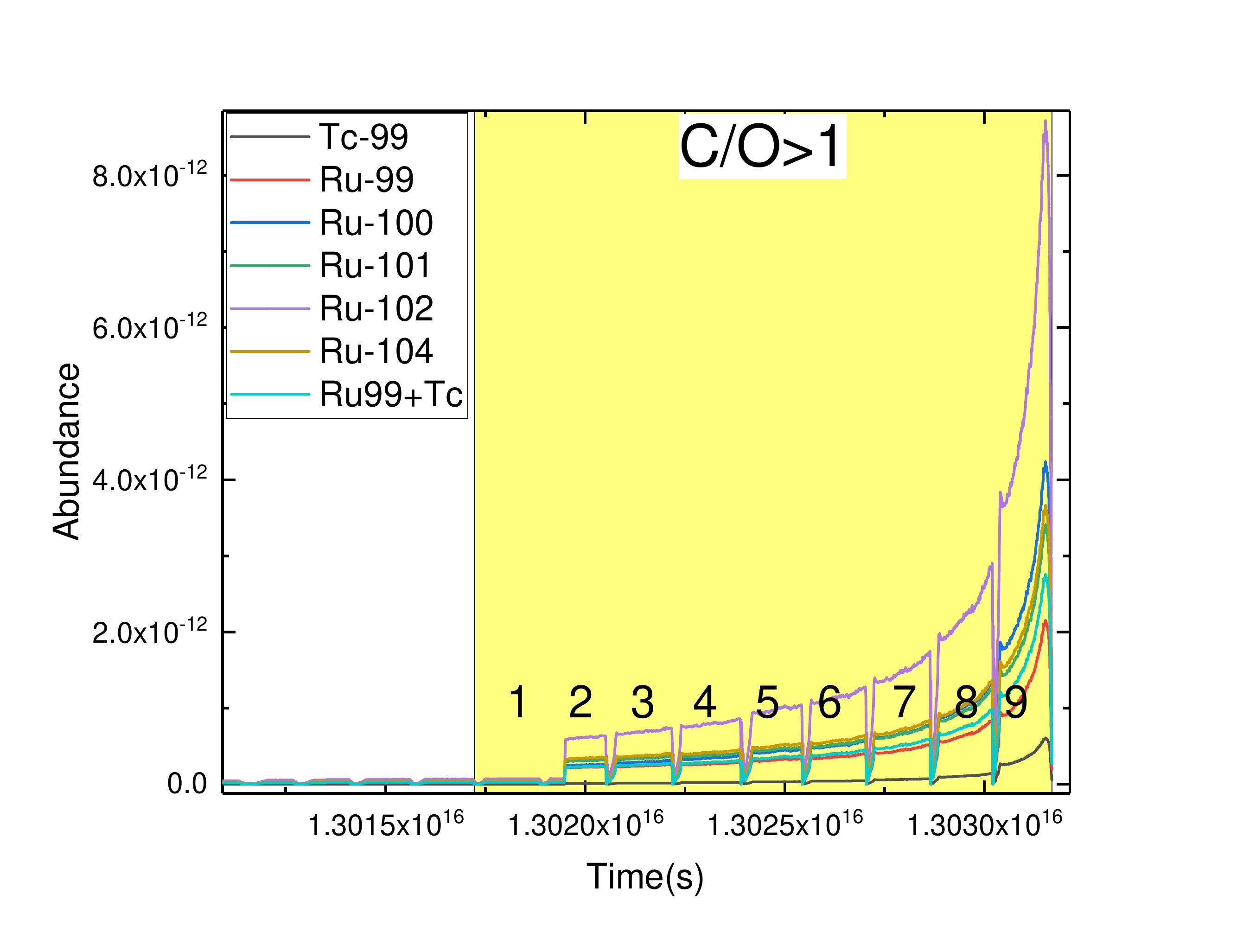}
\caption{A low-mass star with $Z=0.01$ and $M=3~M_{\sun}$} 
\end{figure*}

\begin{figure*}
\figurenum{A6}
\epsscale{1.17}
\plottwo{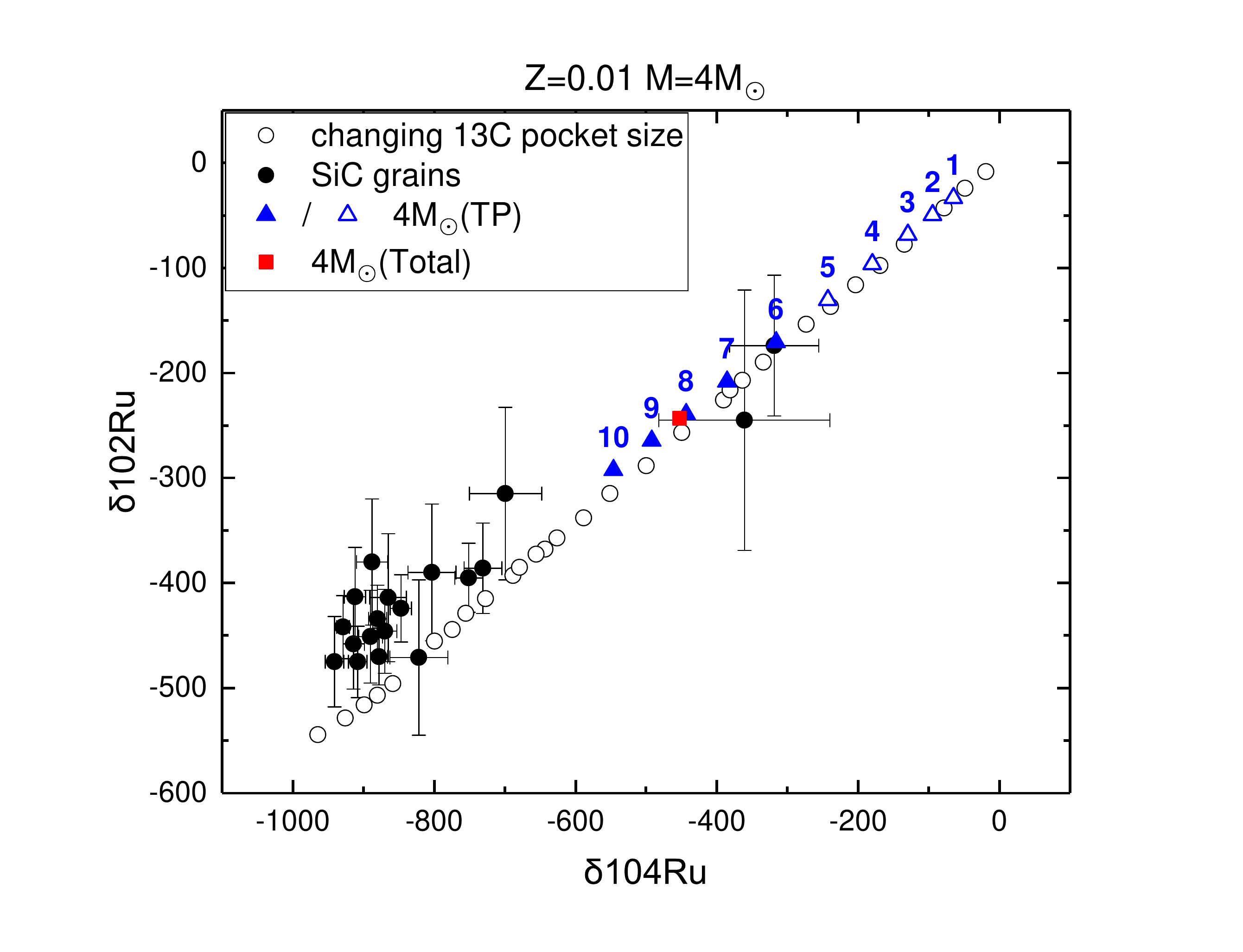}{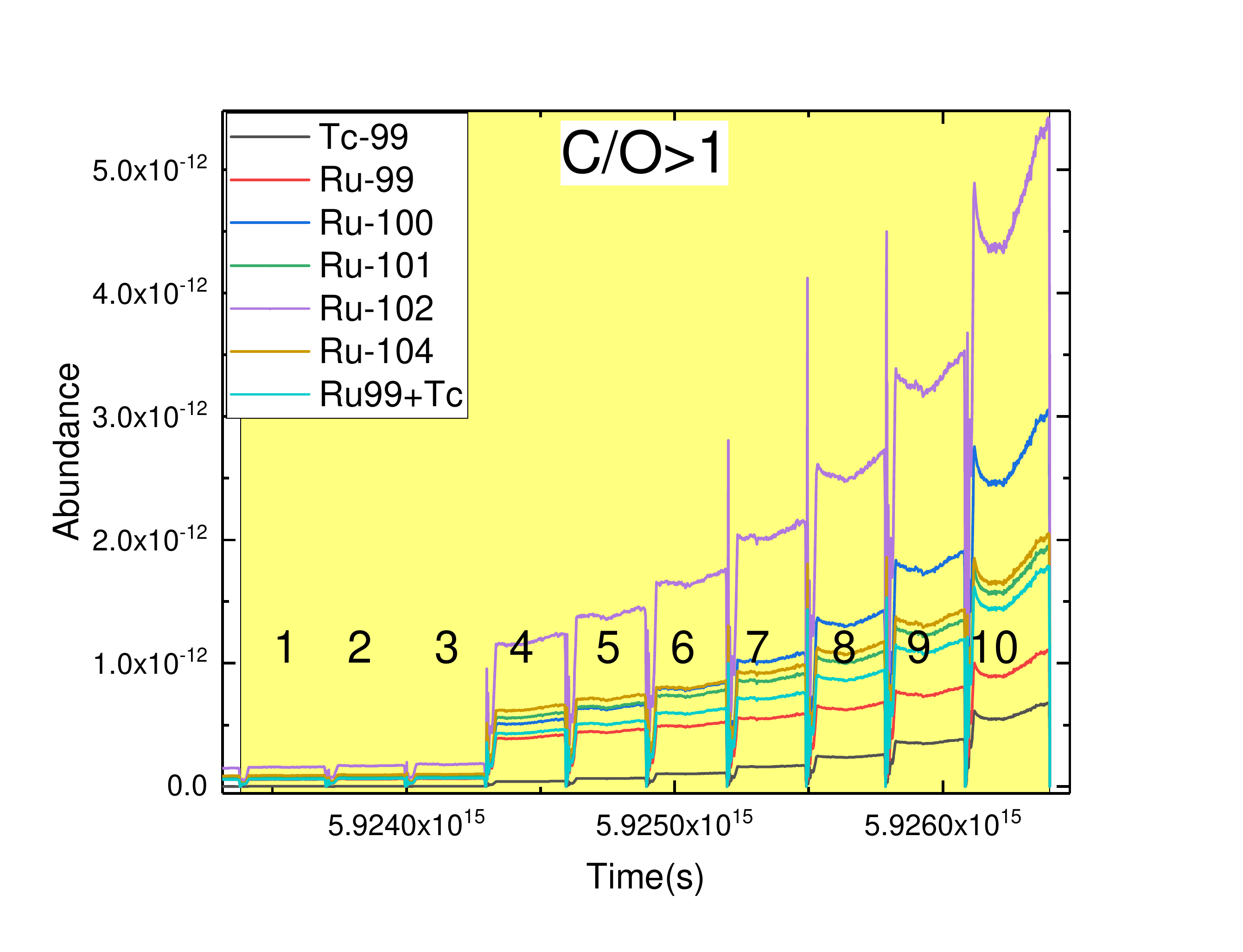}
\caption{A low-mass star with $Z=0.01$ and $M=4~M_{\sun}$}
\end{figure*}

\begin{figure*}
\figurenum{A7}
\epsscale{1.17}
\plottwo{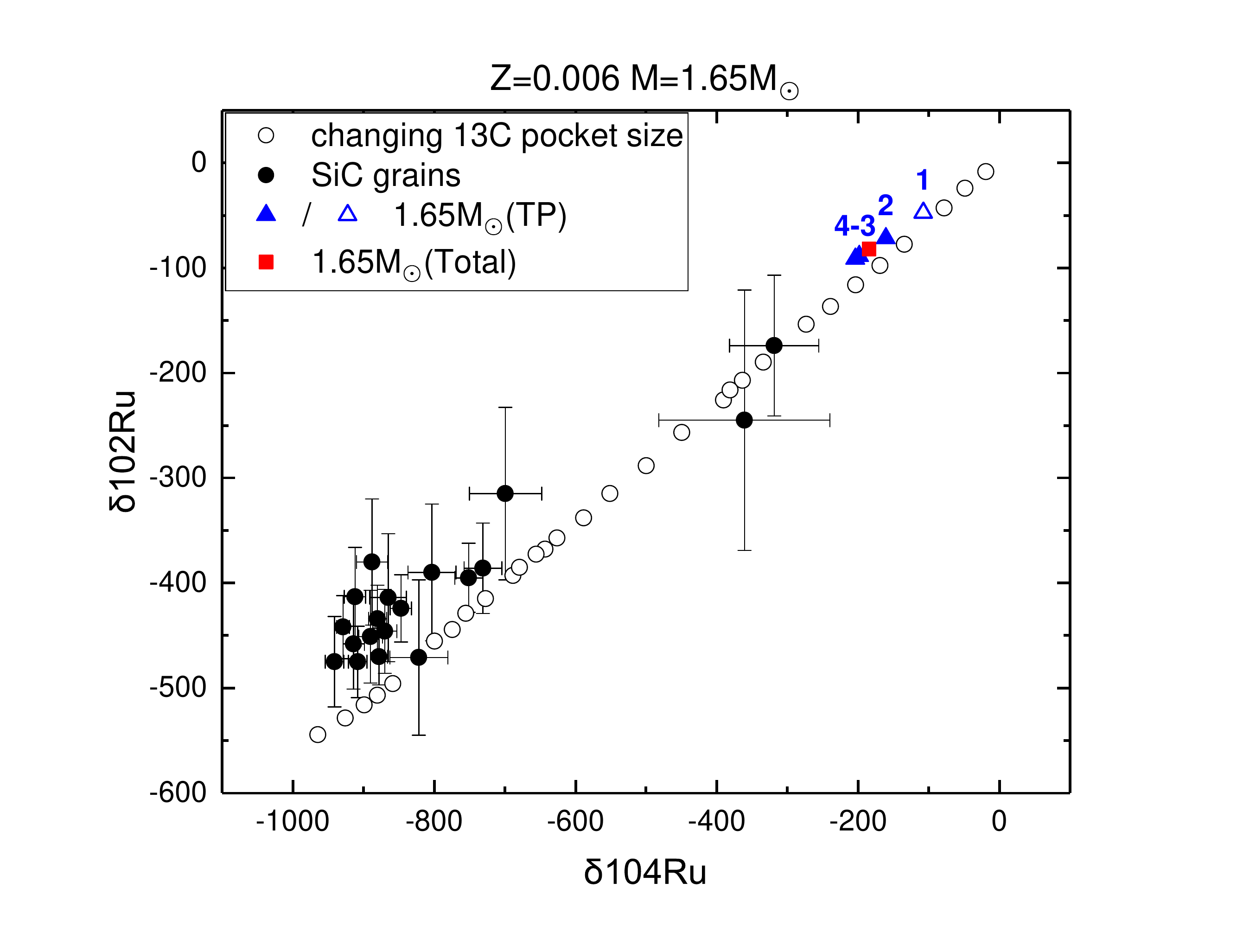}{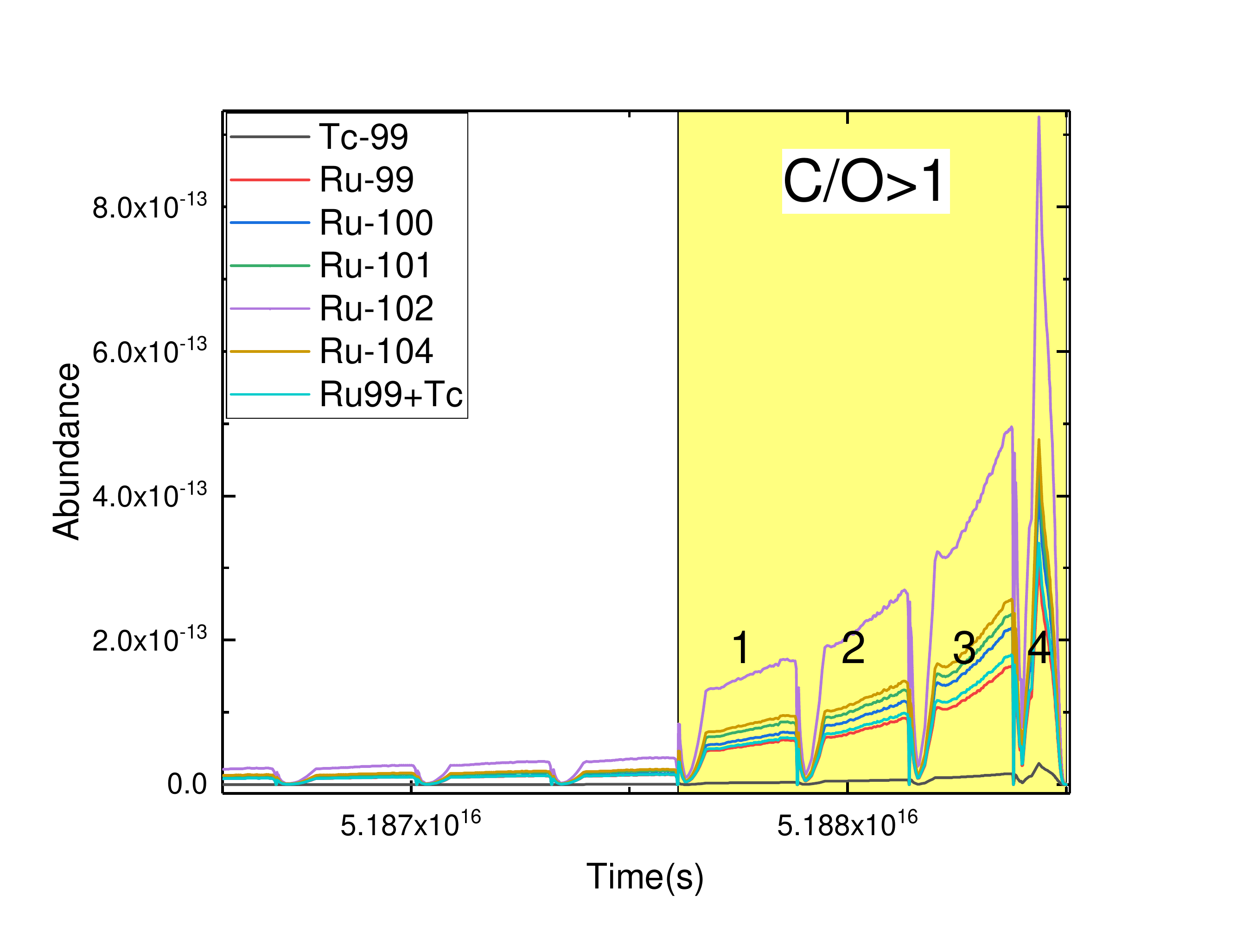}
\caption{A low-mass star with $Z=0.006$ and $M=1.65~M_{\sun}$} 
\end{figure*}

\begin{figure*}
\figurenum{A8}
\epsscale{1.17}
\plottwo{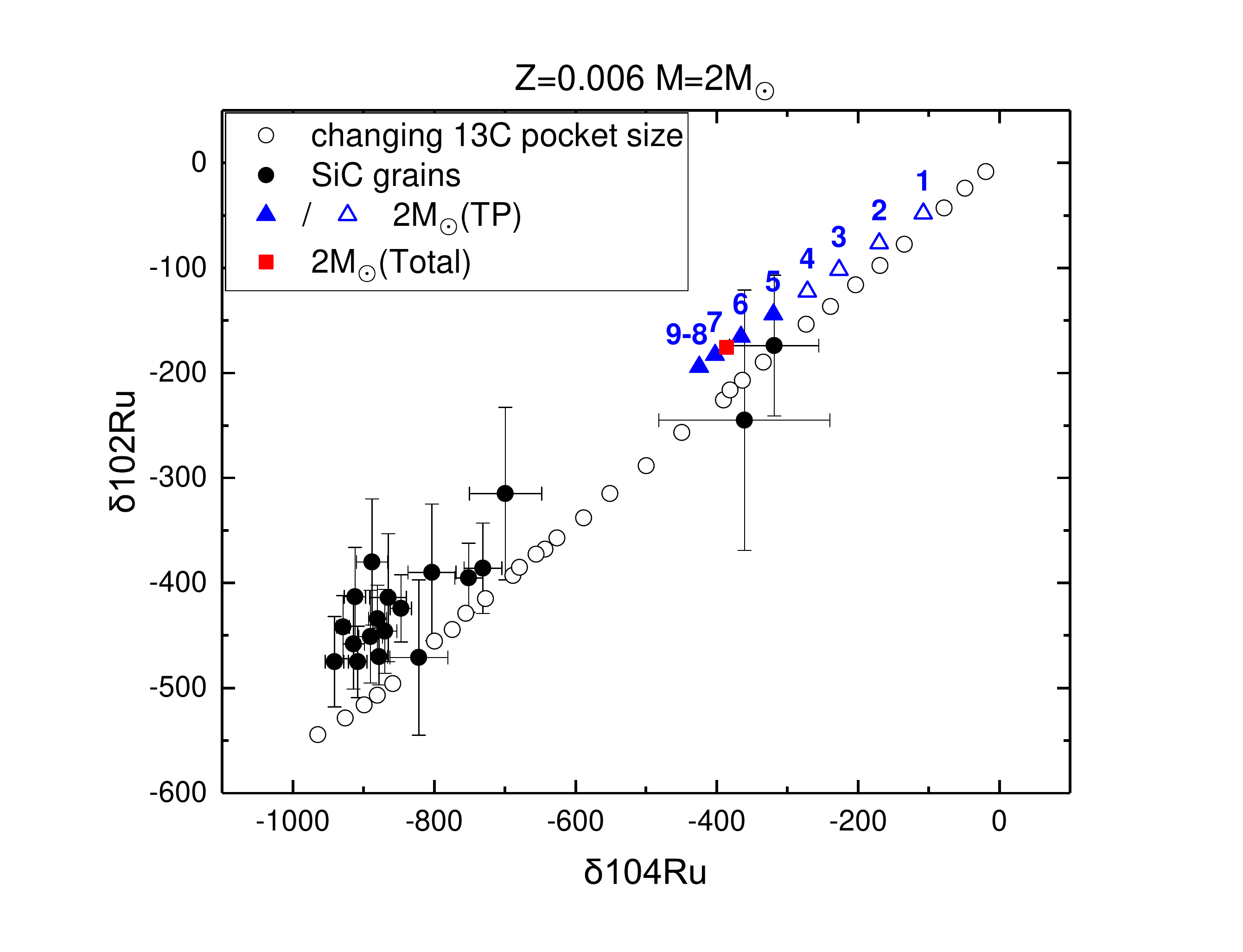}{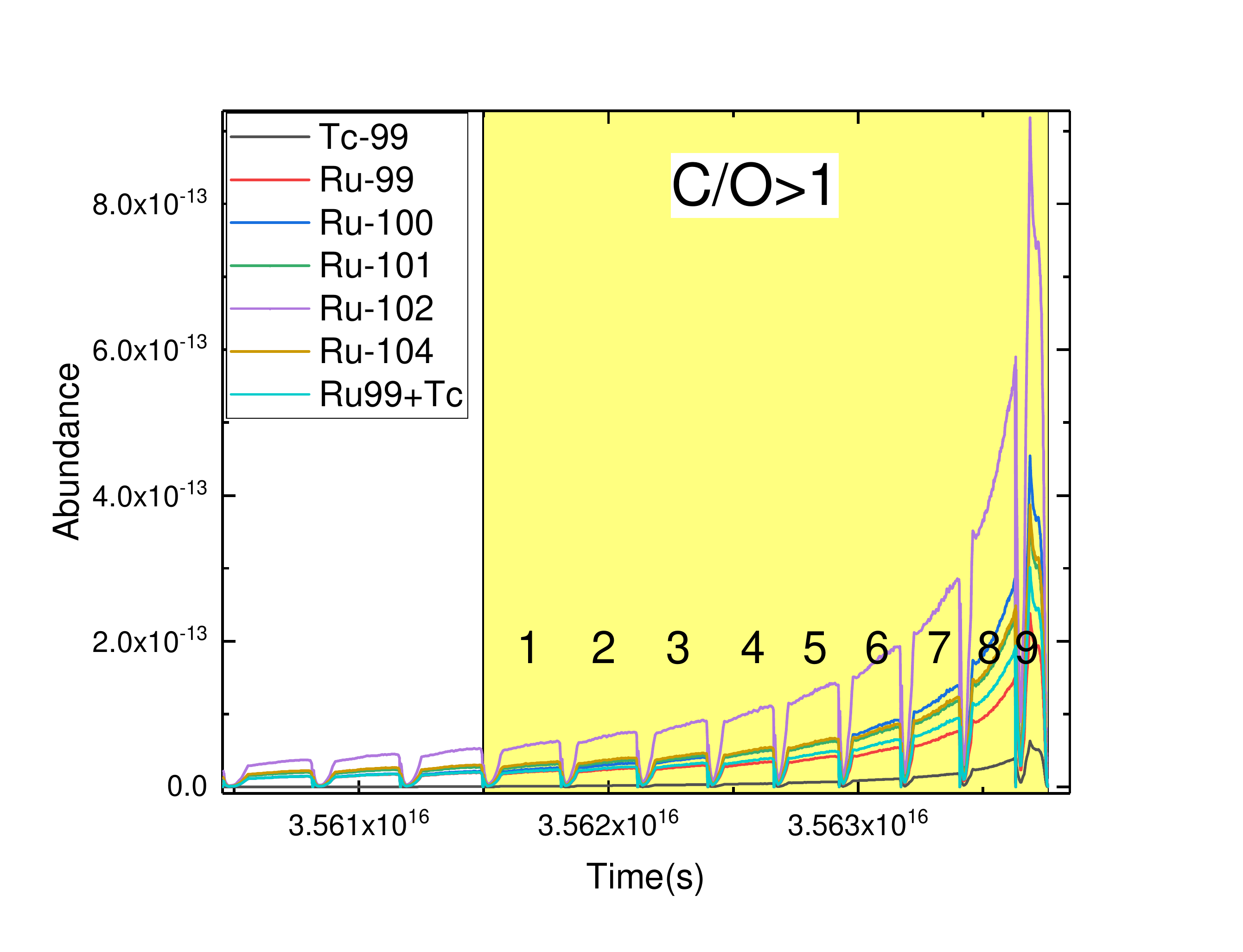}
\caption{A low-mass star with $Z=0.006$ and $M=2~M_{\sun}$} 
\end{figure*}

\begin{figure*}
\figurenum{A9}
\epsscale{1.17}
\plottwo{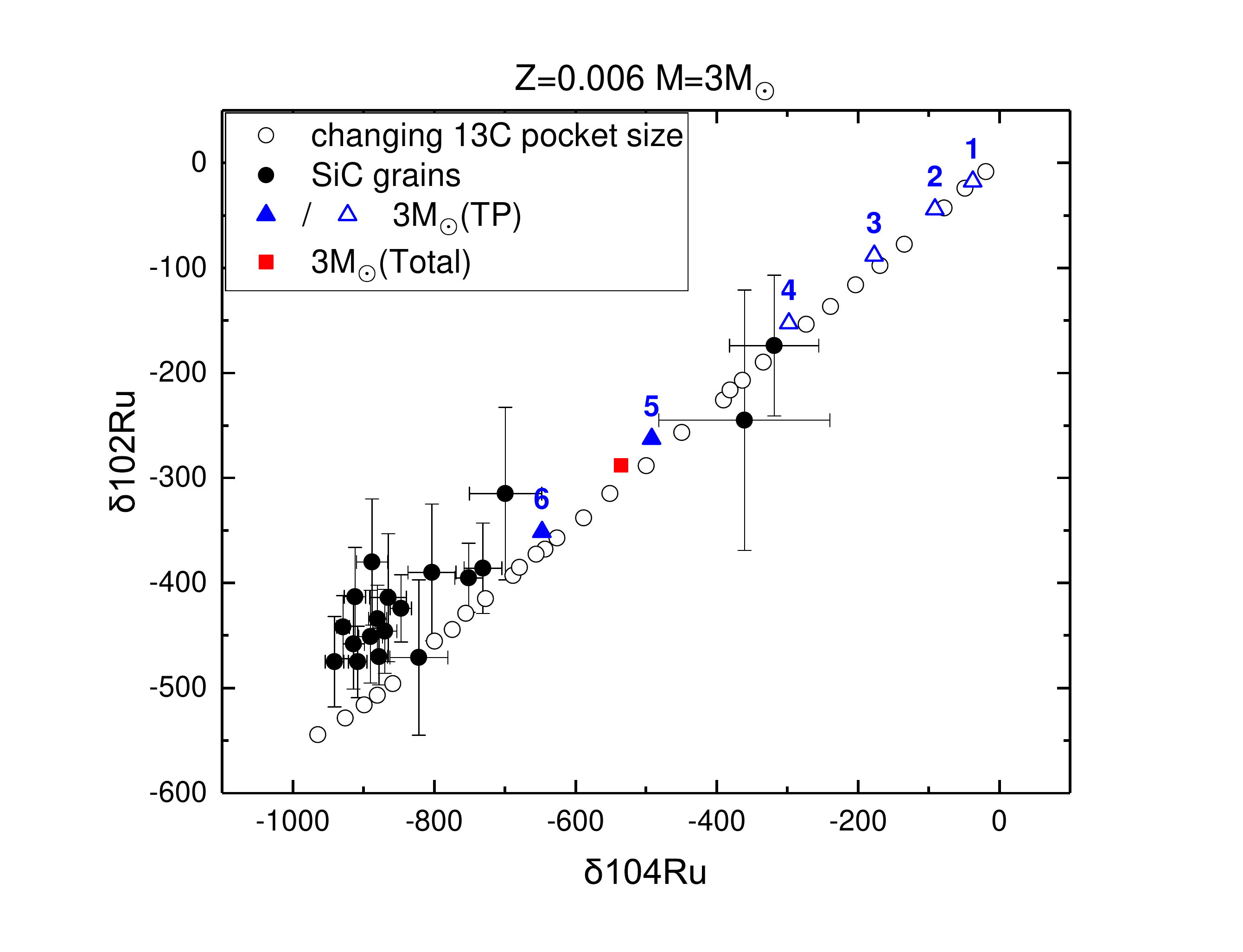}{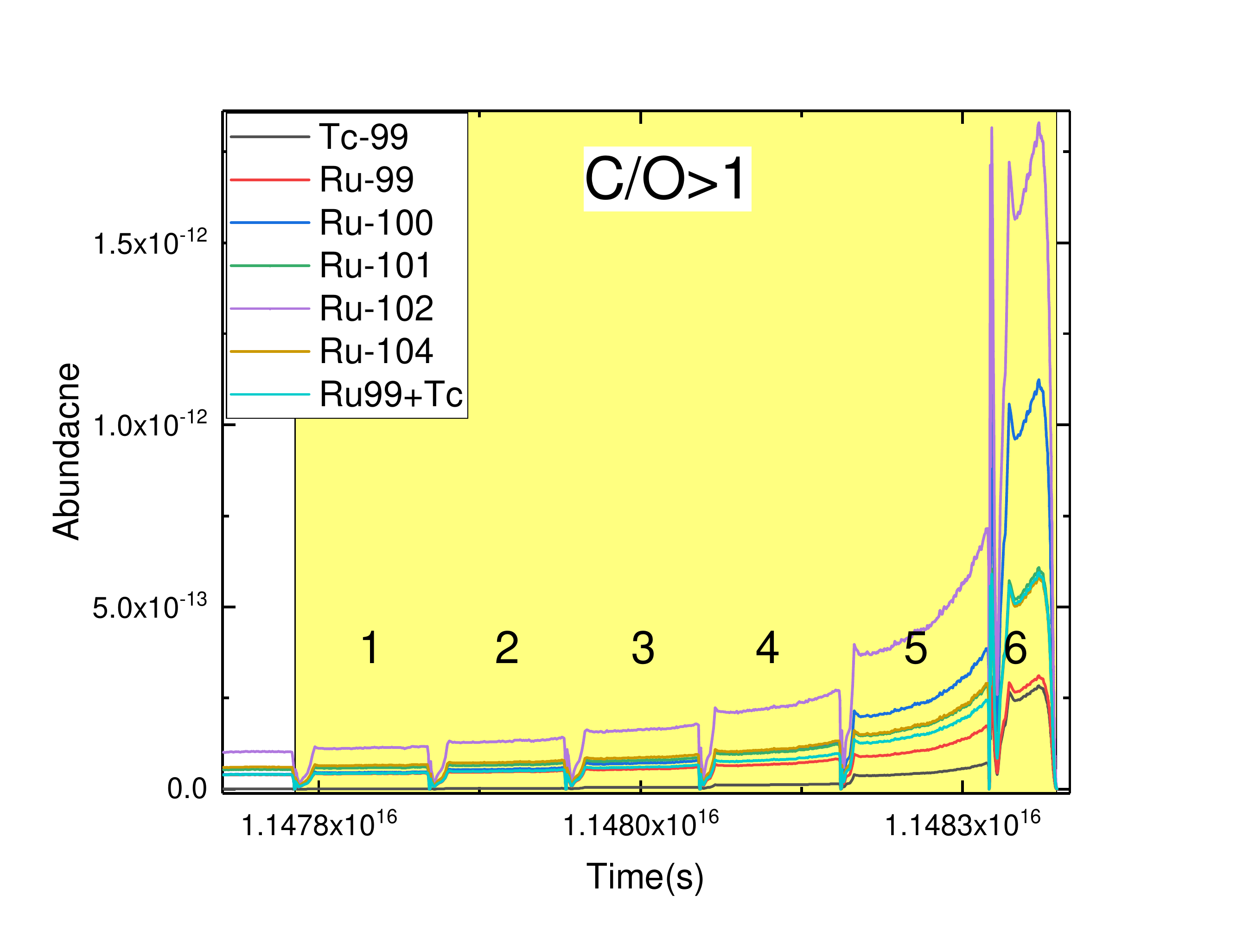}
\caption{A low-mass star with $Z=0.006$ and $M=3~M_{\sun}$} 
\end{figure*}

\begin{figure*}
\figurenum{A10}
\epsscale{1.17}
\plottwo{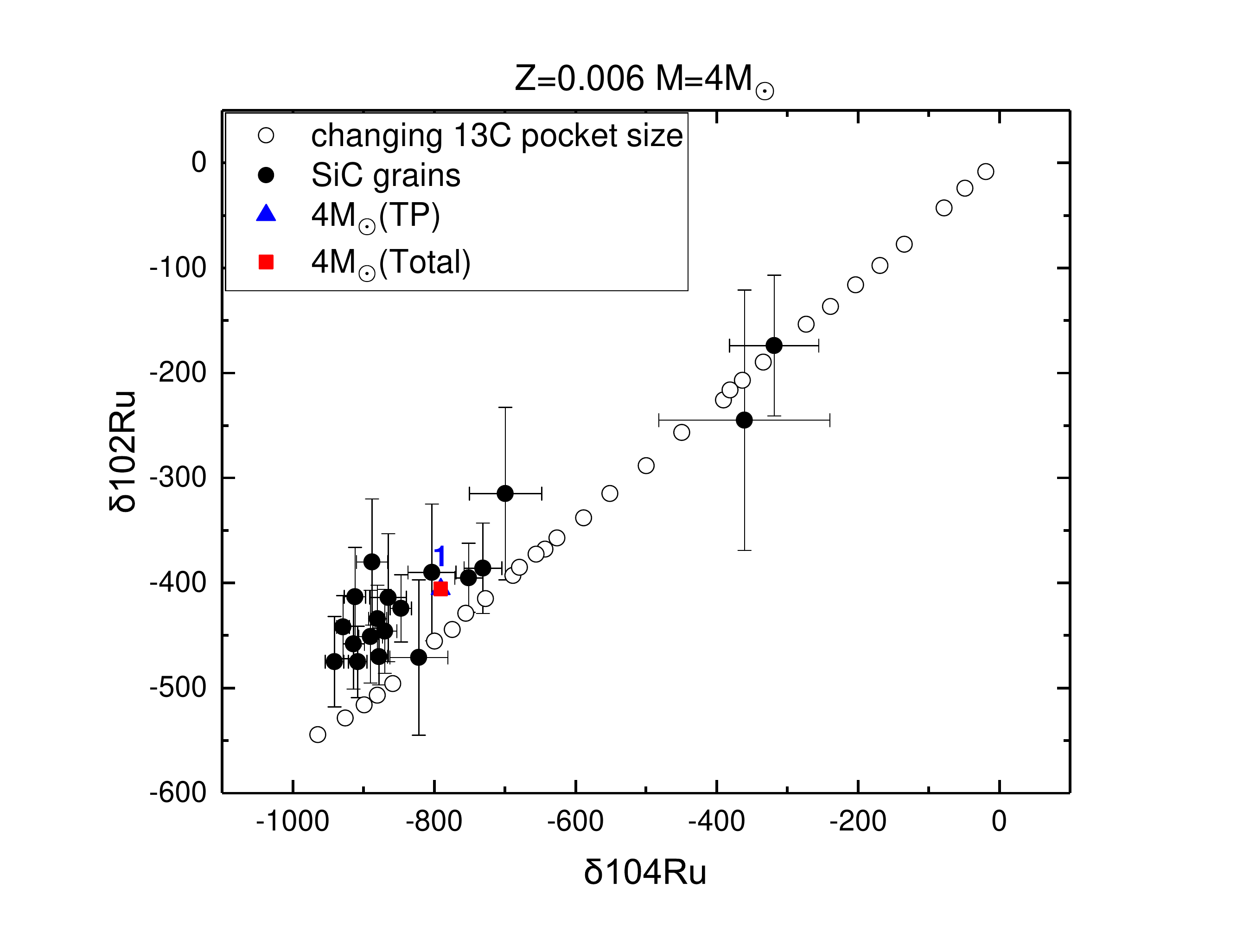}{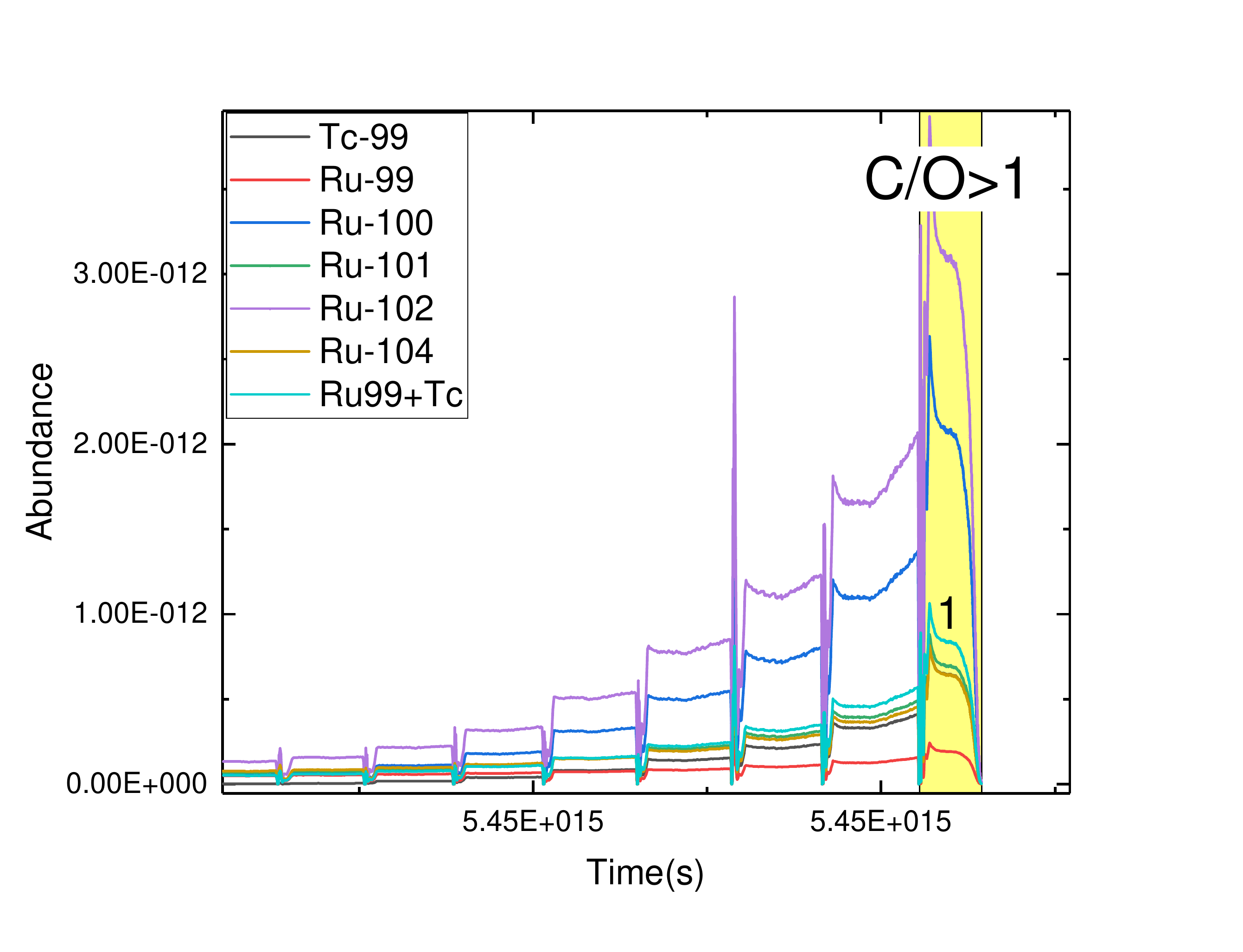}
\caption{A low-mass star with $Z=0.006$ and $M=4~M_{\sun}$} 
\end{figure*}

\begin{figure*}
\figurenum{A11}
\epsscale{1.17}
\plottwo{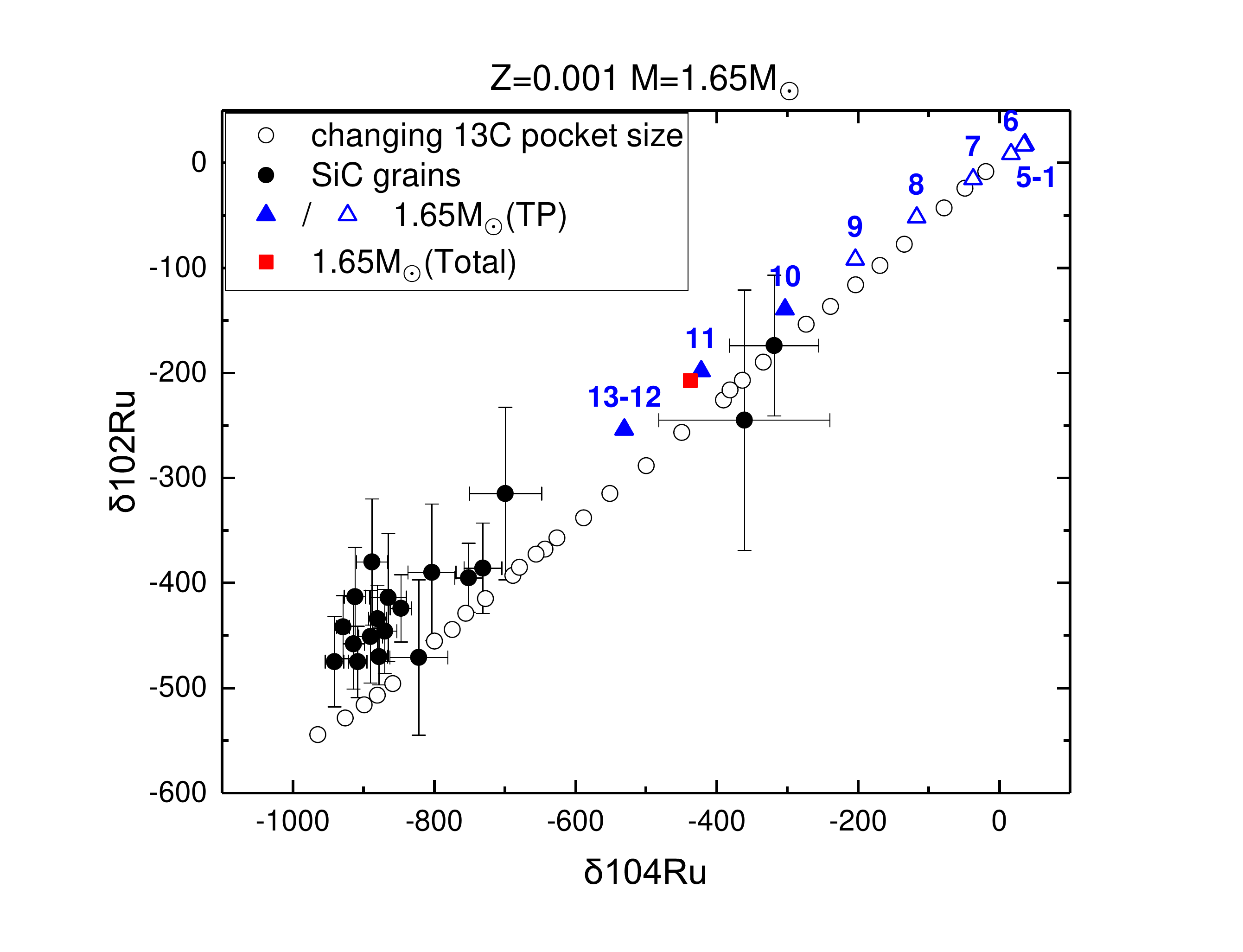}{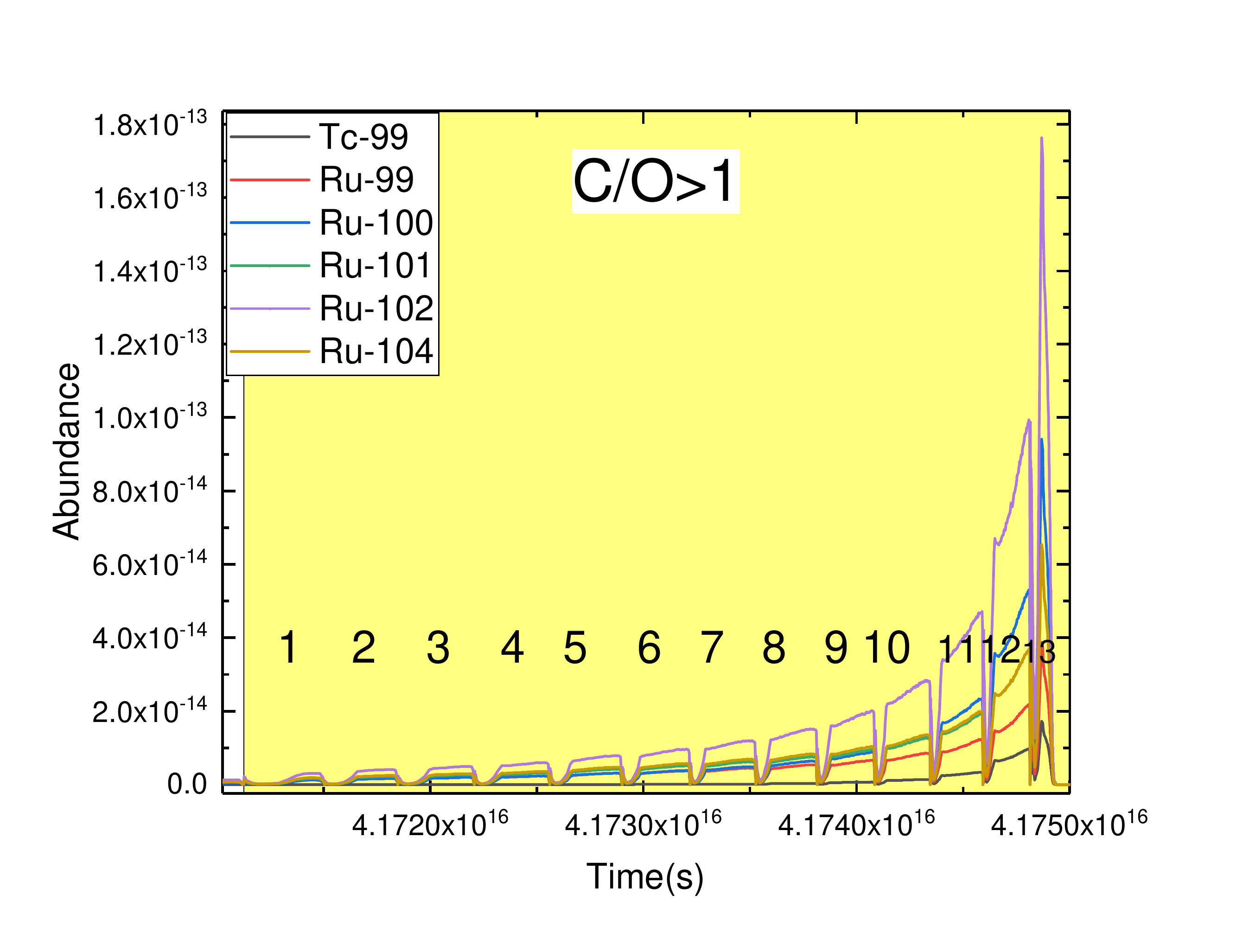}
\caption{A low-mass star with $Z=0.001$ and $M=1.65~M_{\sun}$} 
\end{figure*}

\begin{figure*}
\figurenum{A12}
\epsscale{1.17}
\plottwo{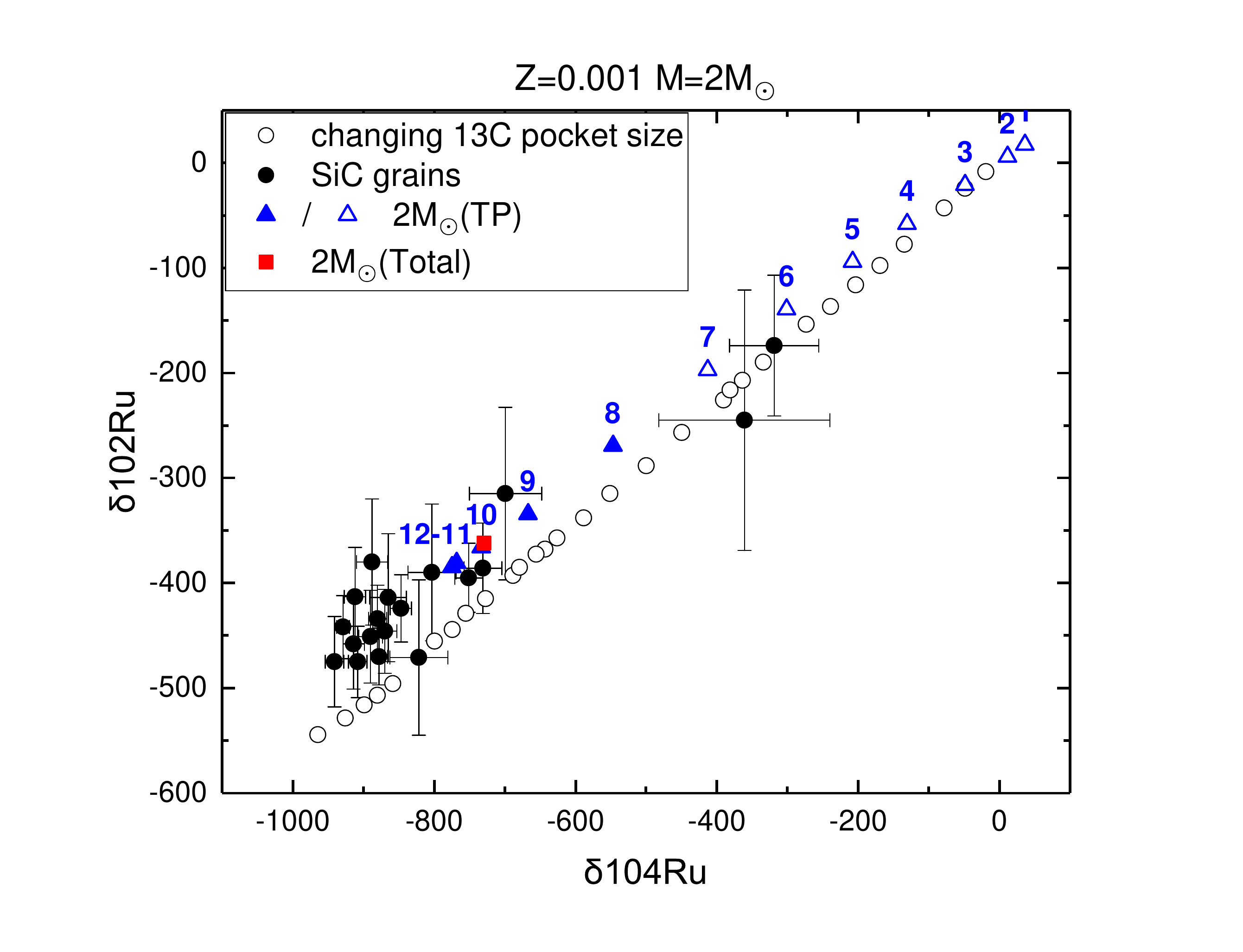}{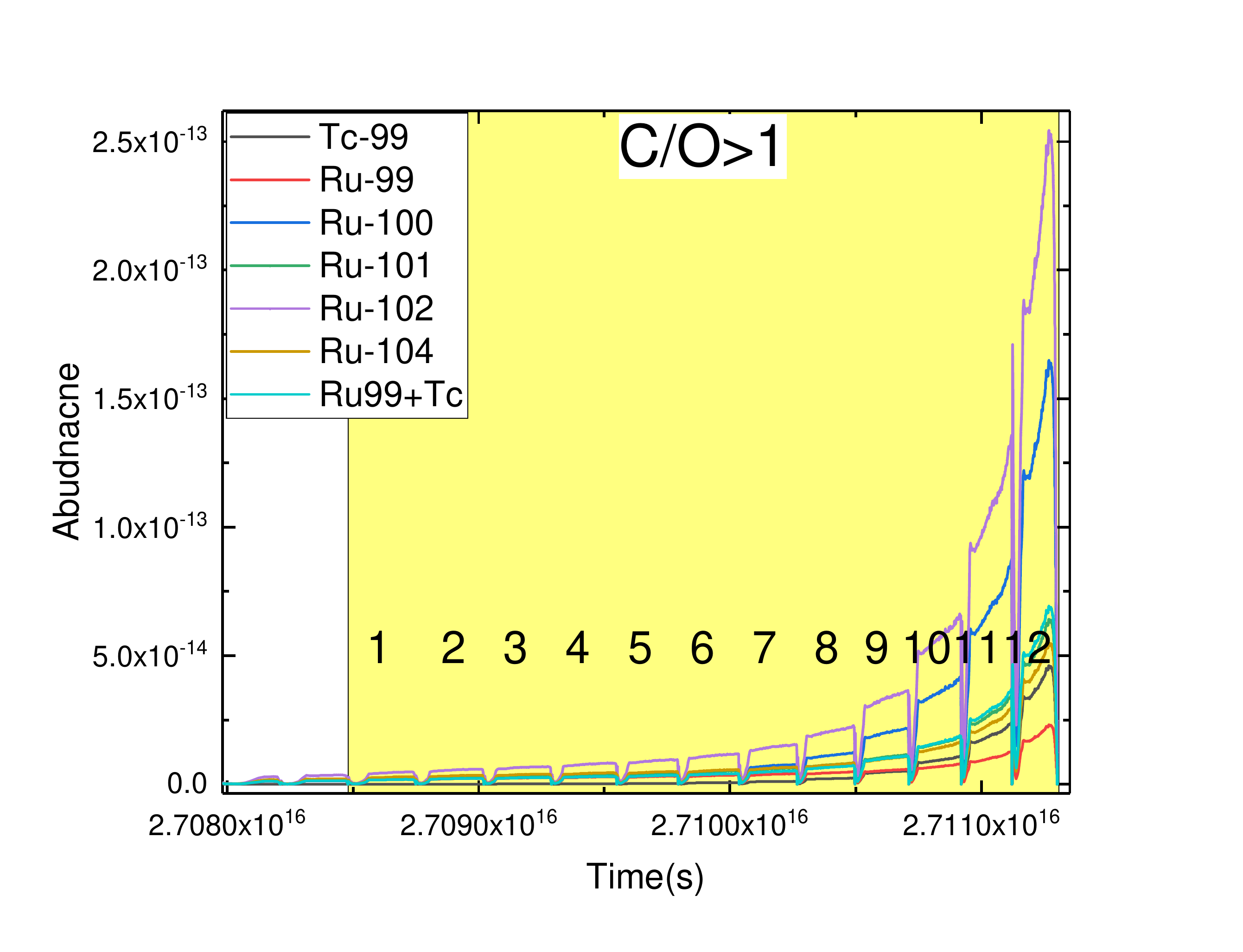}
\caption{A low-mass star with $Z=0.001$ and $M=2~M_{\sun}$}
\end{figure*}

\begin{figure*}
\figurenum{A13}
\epsscale{1.17}
\plottwo{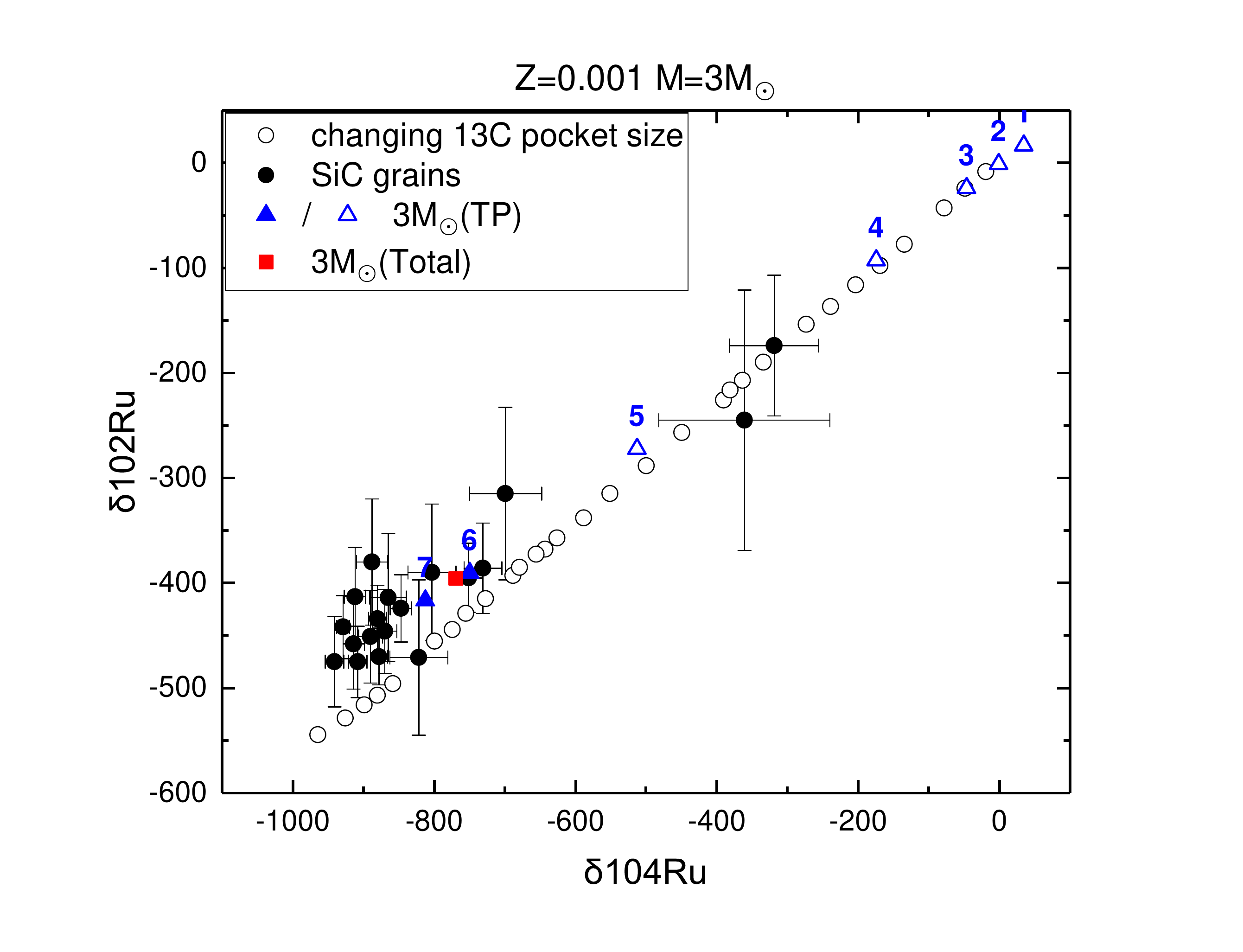}{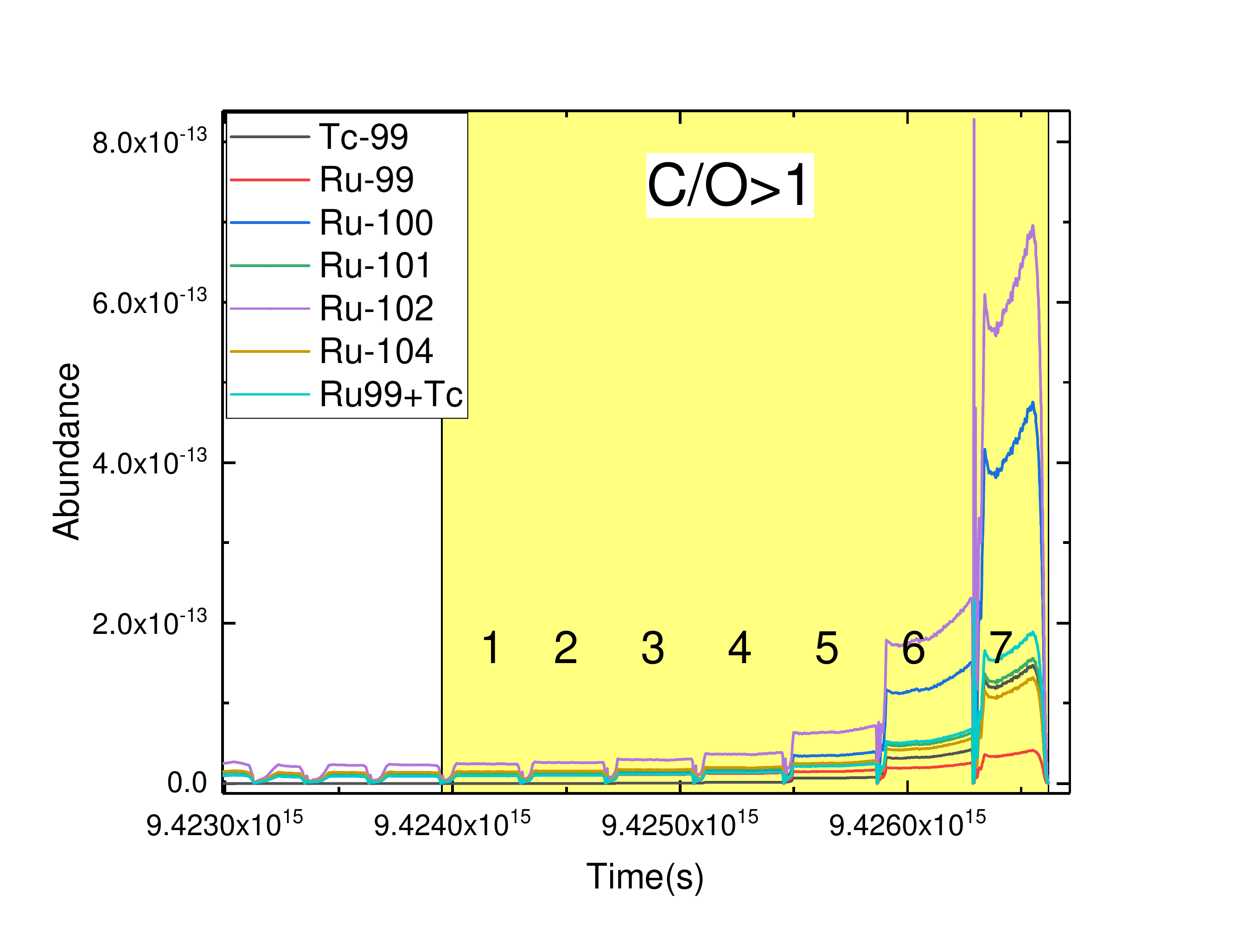}
\caption{A low-mass star with $Z=0.001$ and $M=3~M_{\sun}$} 
\end{figure*}

\begin{figure*}
\figurenum{A14}
\epsscale{1.17}
\plottwo{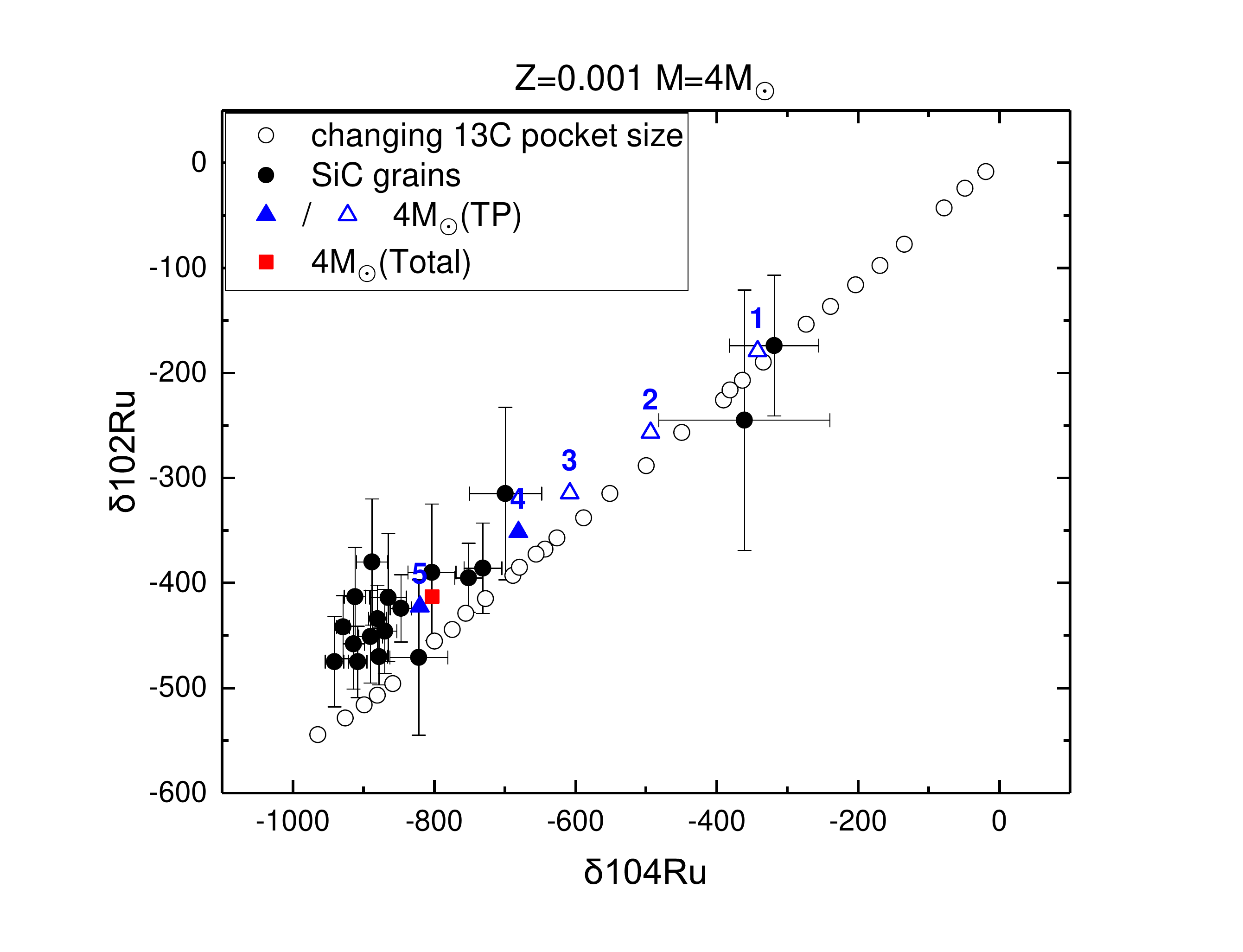}{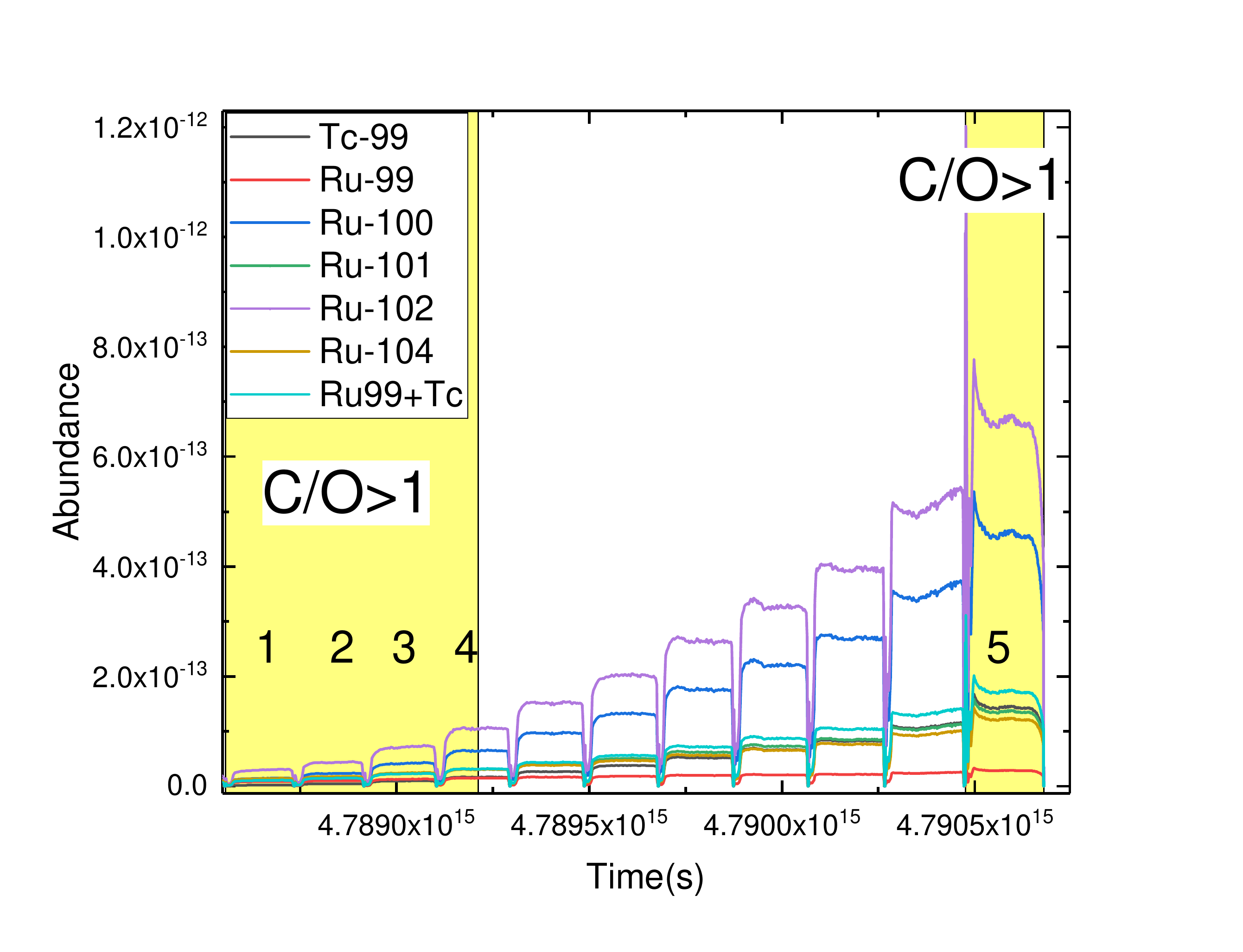}
\caption{A low-mass star with $Z=0.001$ and $M=4~M_{\sun}$} 
\end{figure*}

\begin{figure*}
\figurenum{A15}
\epsscale{1.17}
\plottwo{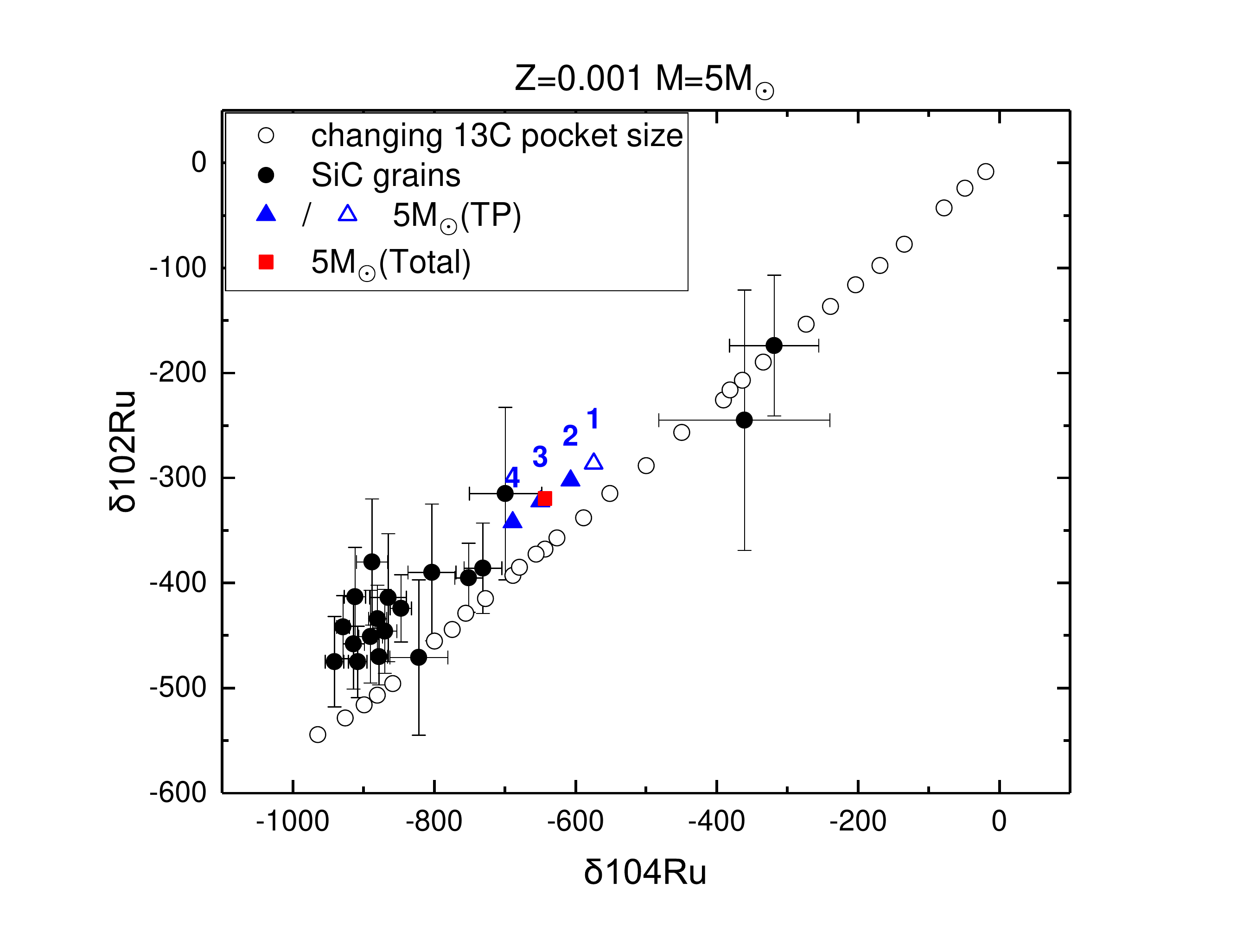}{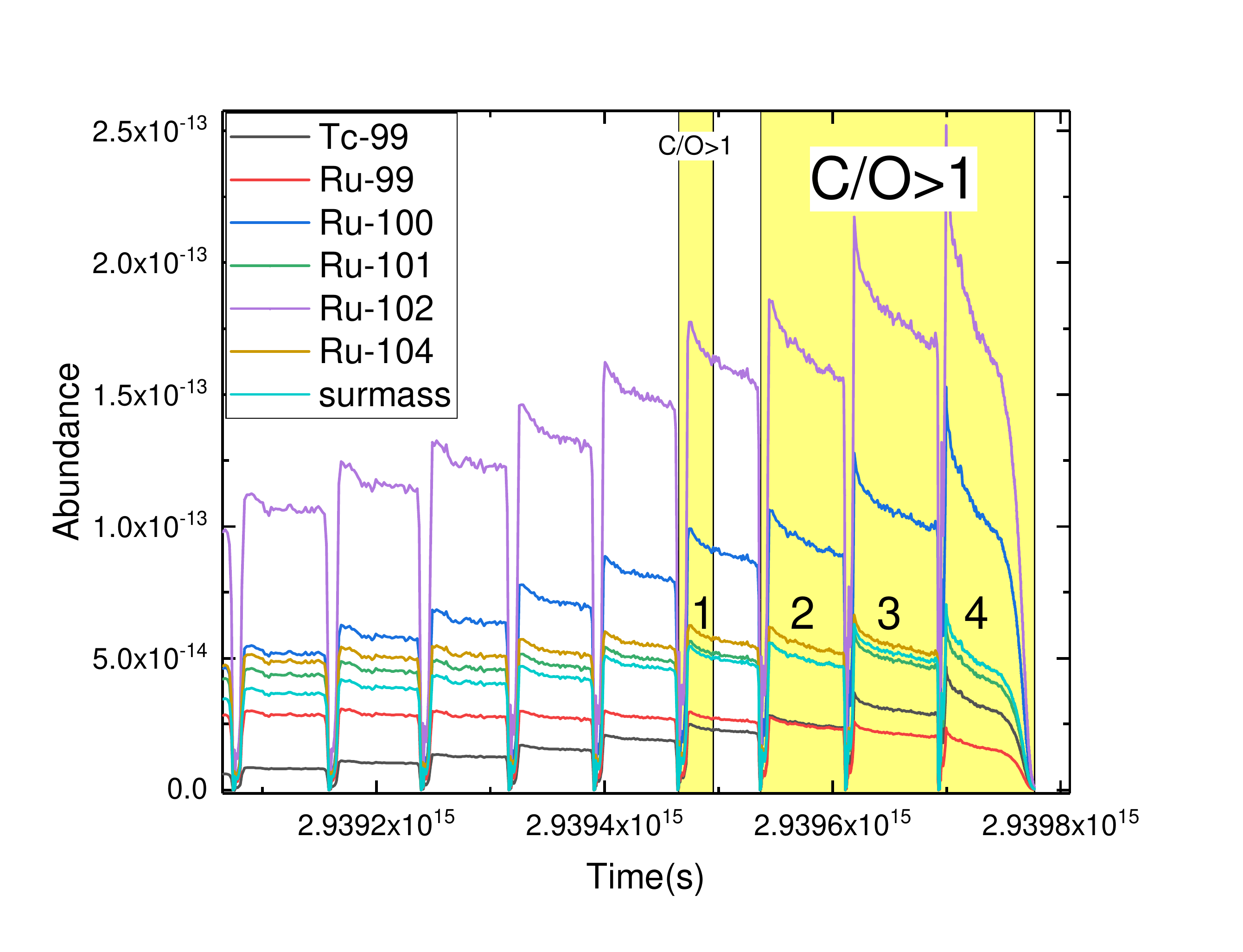}
\caption{A low-mass star with $Z=0.001$ and $M=5~M_{\sun}$} 
\end{figure*}


\begin{figure*}
\figurenum{A16}
\epsscale{1.17}
\plottwo{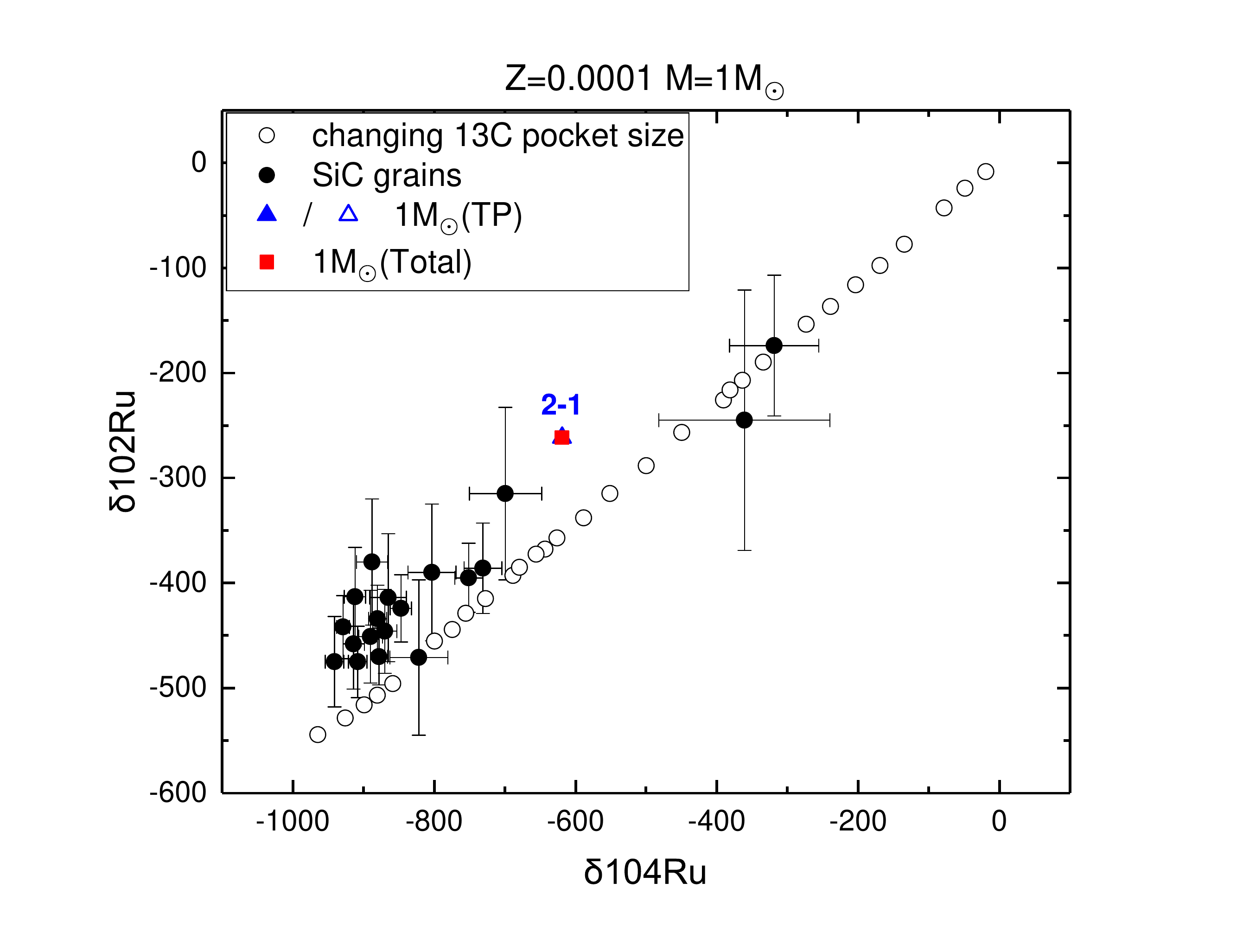}{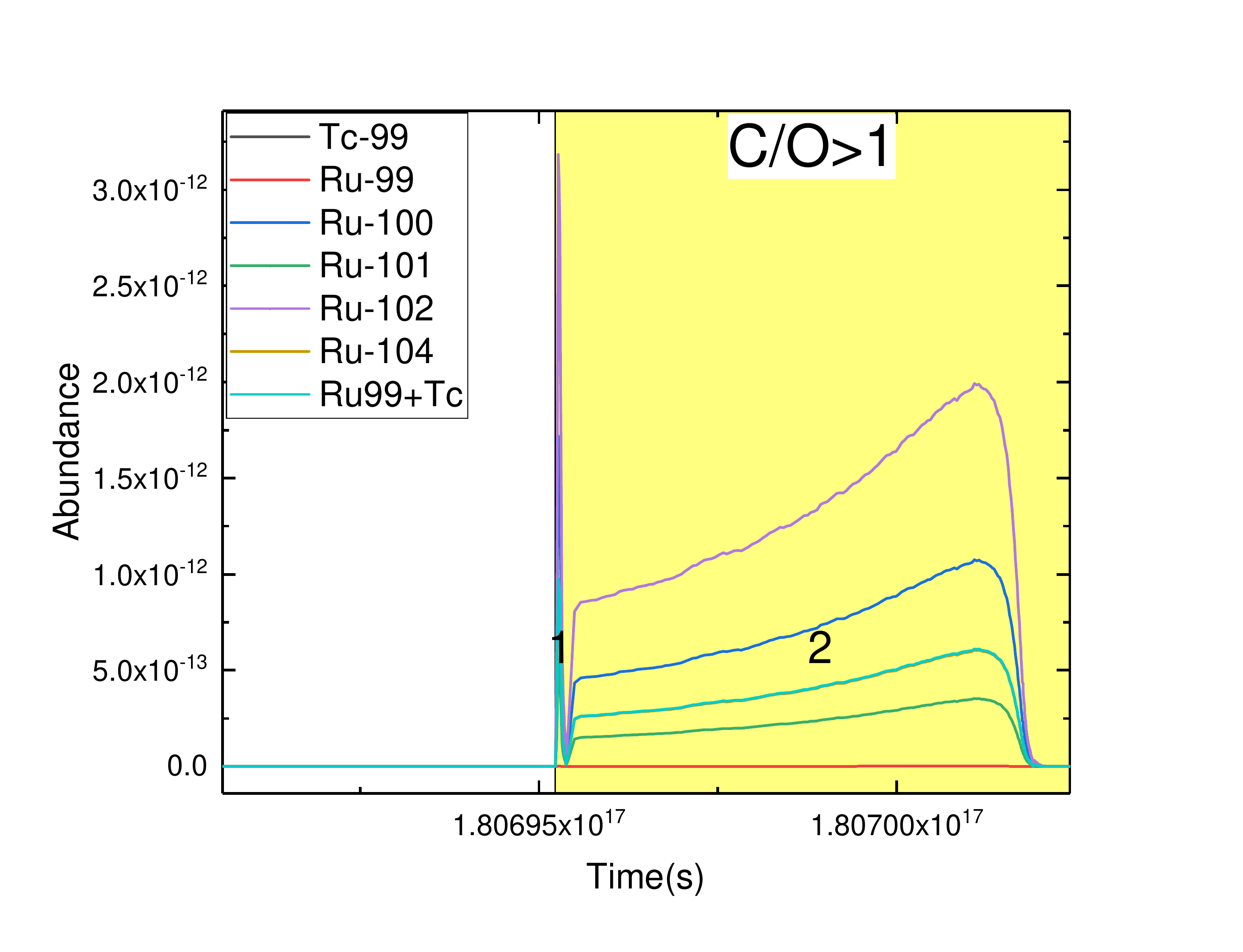}
\caption{A low-mass star with $Z=0.0001$ and $M=1~M_{\sun}$} 
\end{figure*}

\begin{figure*}
\figurenum{A17}
\epsscale{1.17}
\plottwo{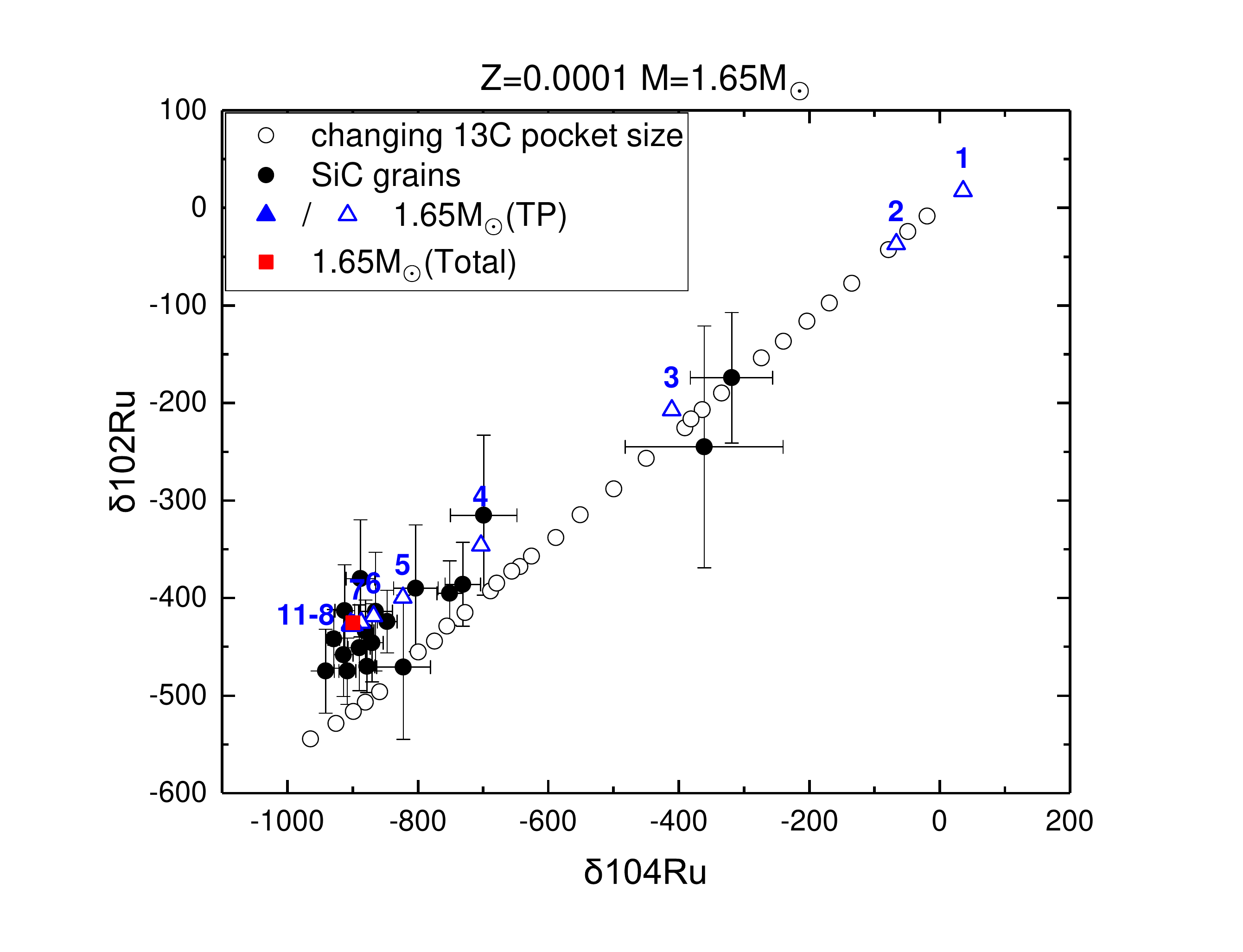}{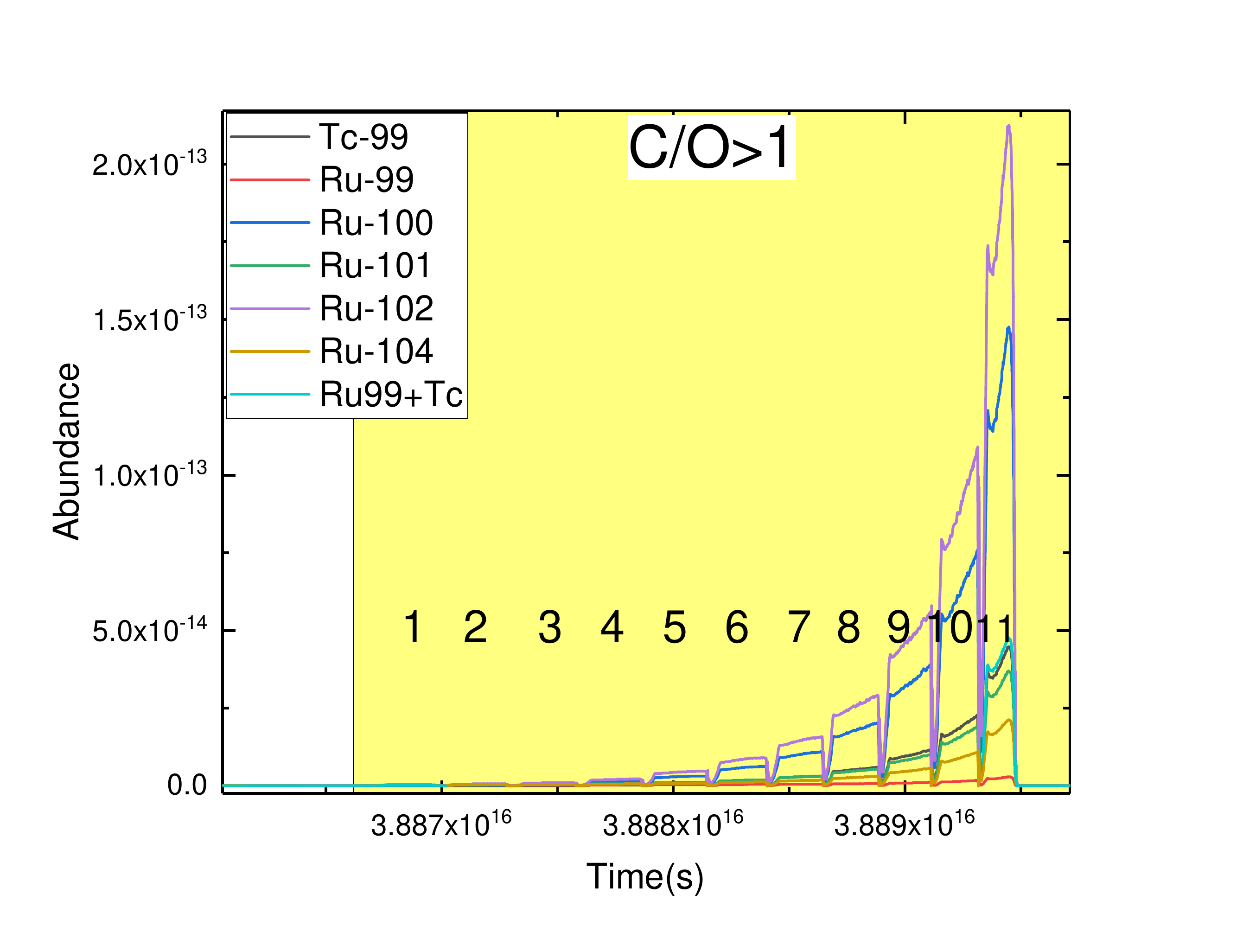}
\caption{A low-mass star with $Z=0.0001$ and $M=1.65~M_{\sun}$} 
\label{special1}
\end{figure*}

\begin{figure*}
\figurenum{A18}
\epsscale{1.17}
\plottwo{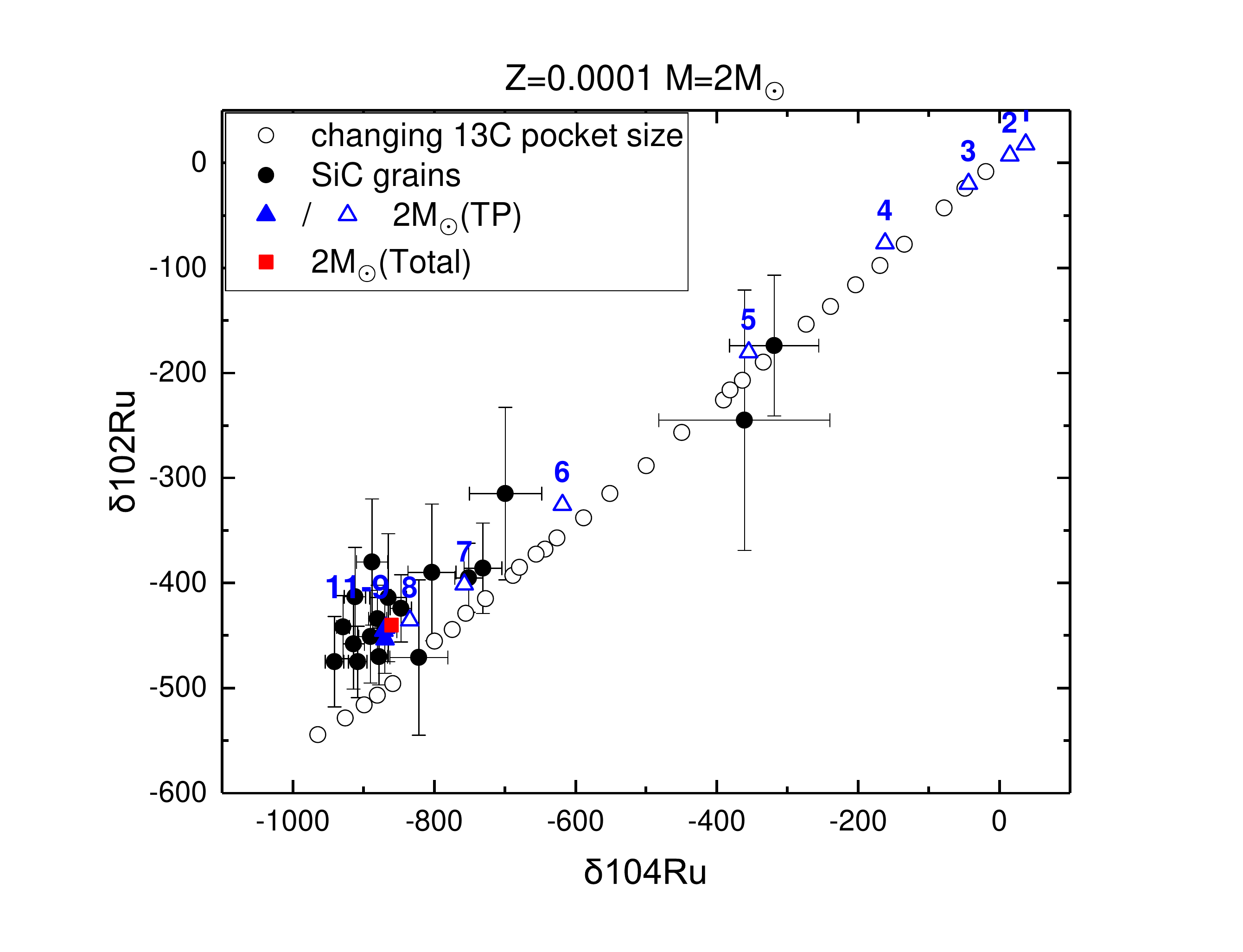}{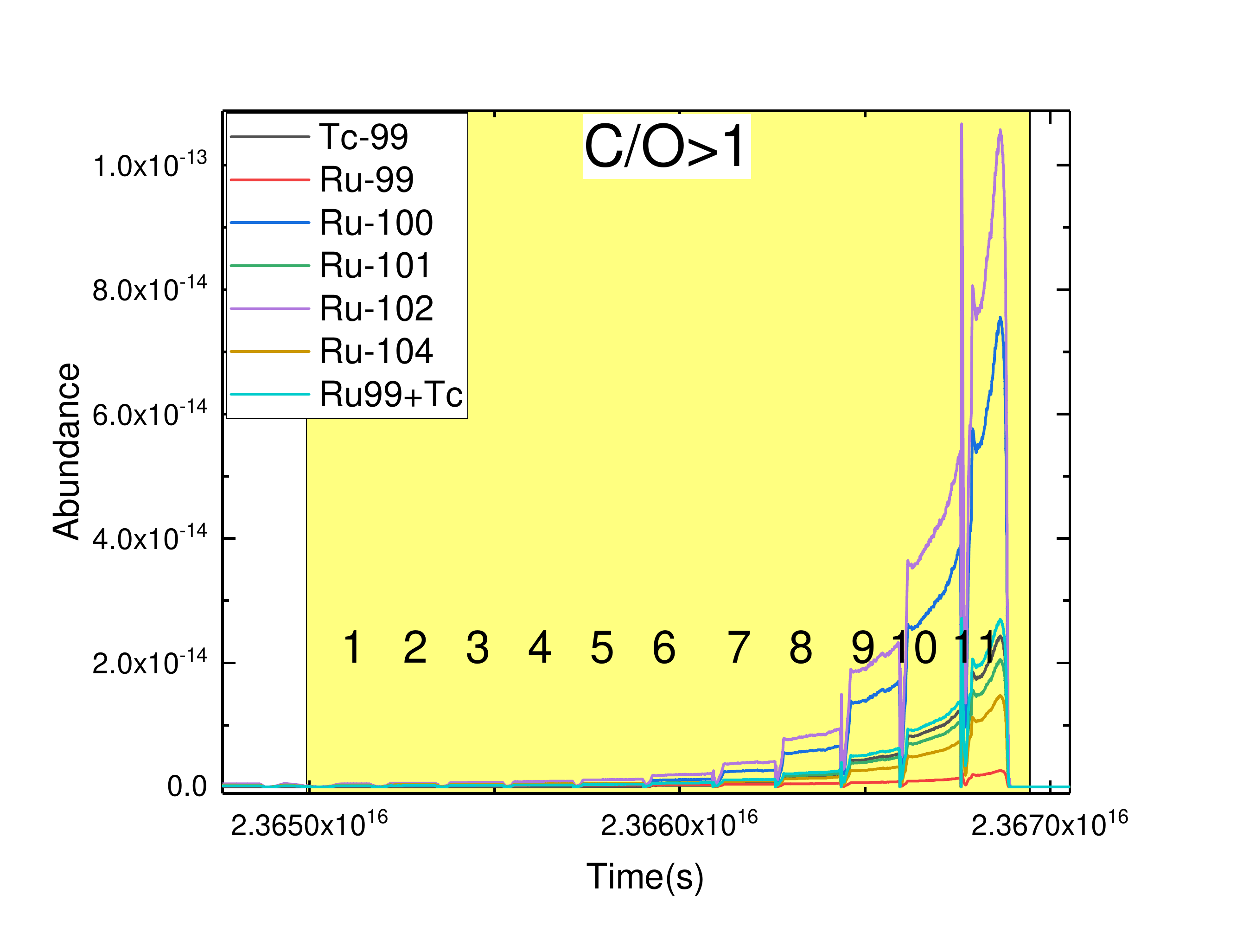}
\caption{A low-mass star with $Z=0.0001$ and $M=2~M_{\sun}$} 
\label{special2}
\end{figure*}

\begin{figure*}
\figurenum{A19}
\epsscale{1.17}
\plottwo{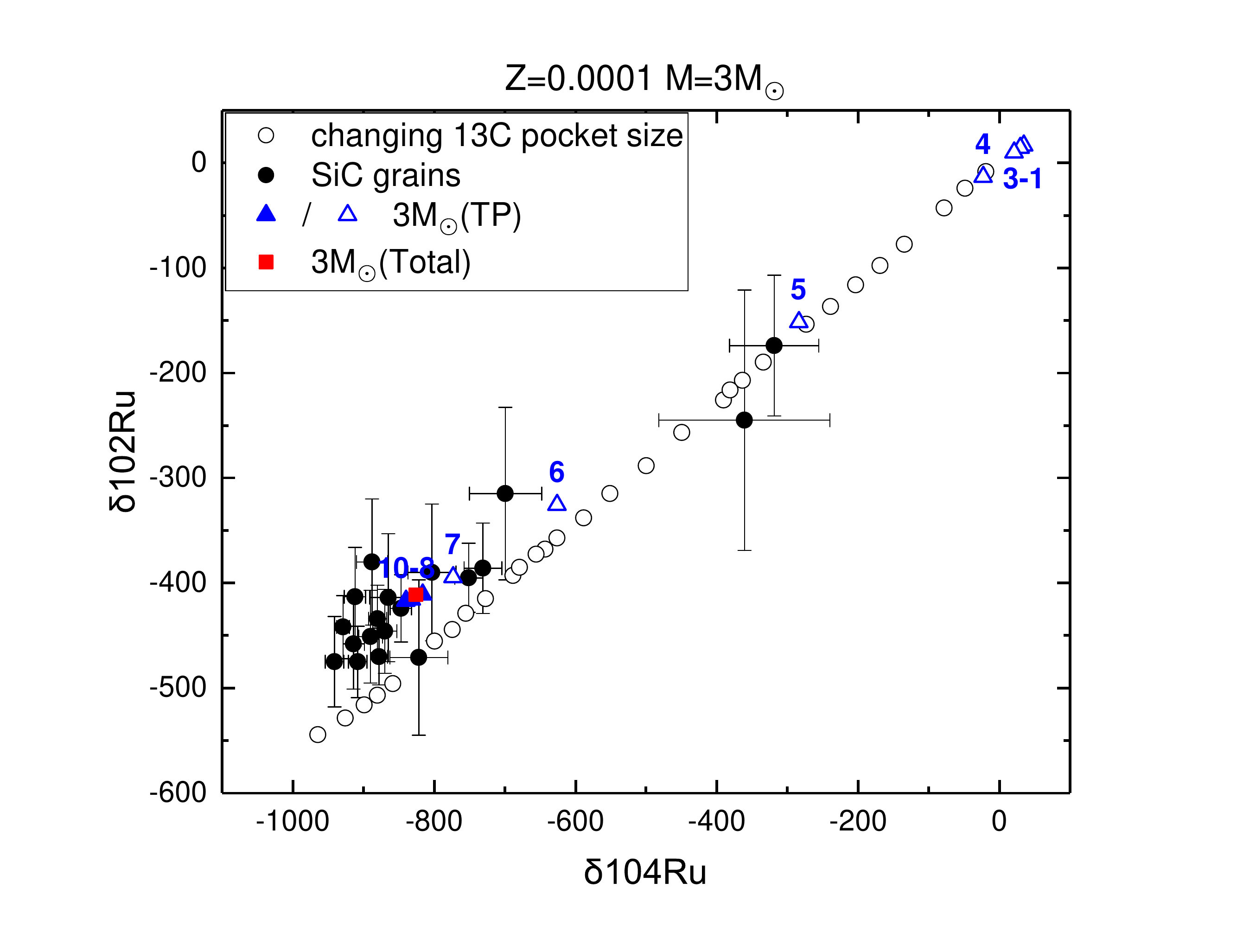}{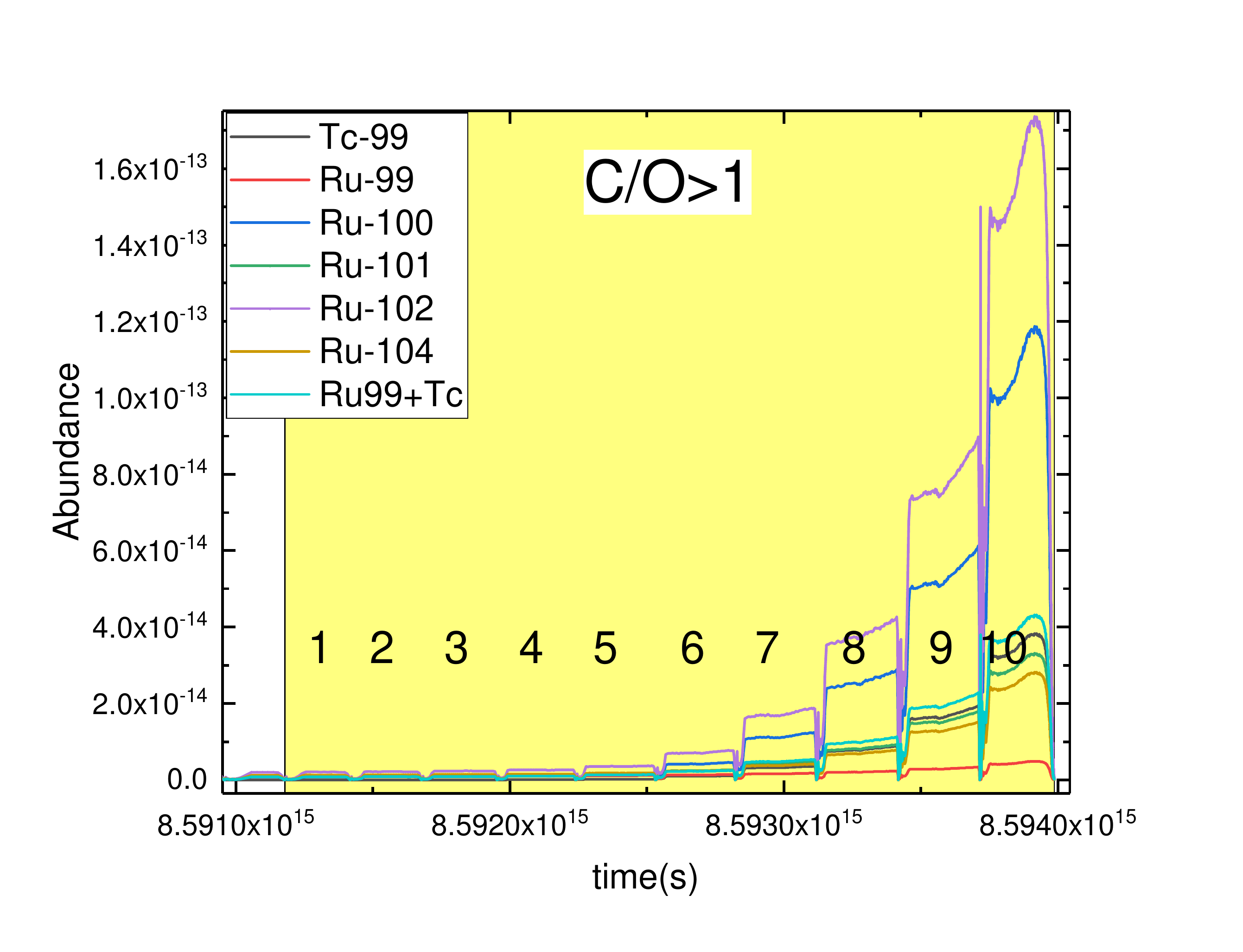}
\caption{A low-mass star with $Z=0.0001$ and $M=3~M_{\sun}$} 
\end{figure*}

\begin{figure*}
\figurenum{A20}
\epsscale{1.17}
\plottwo{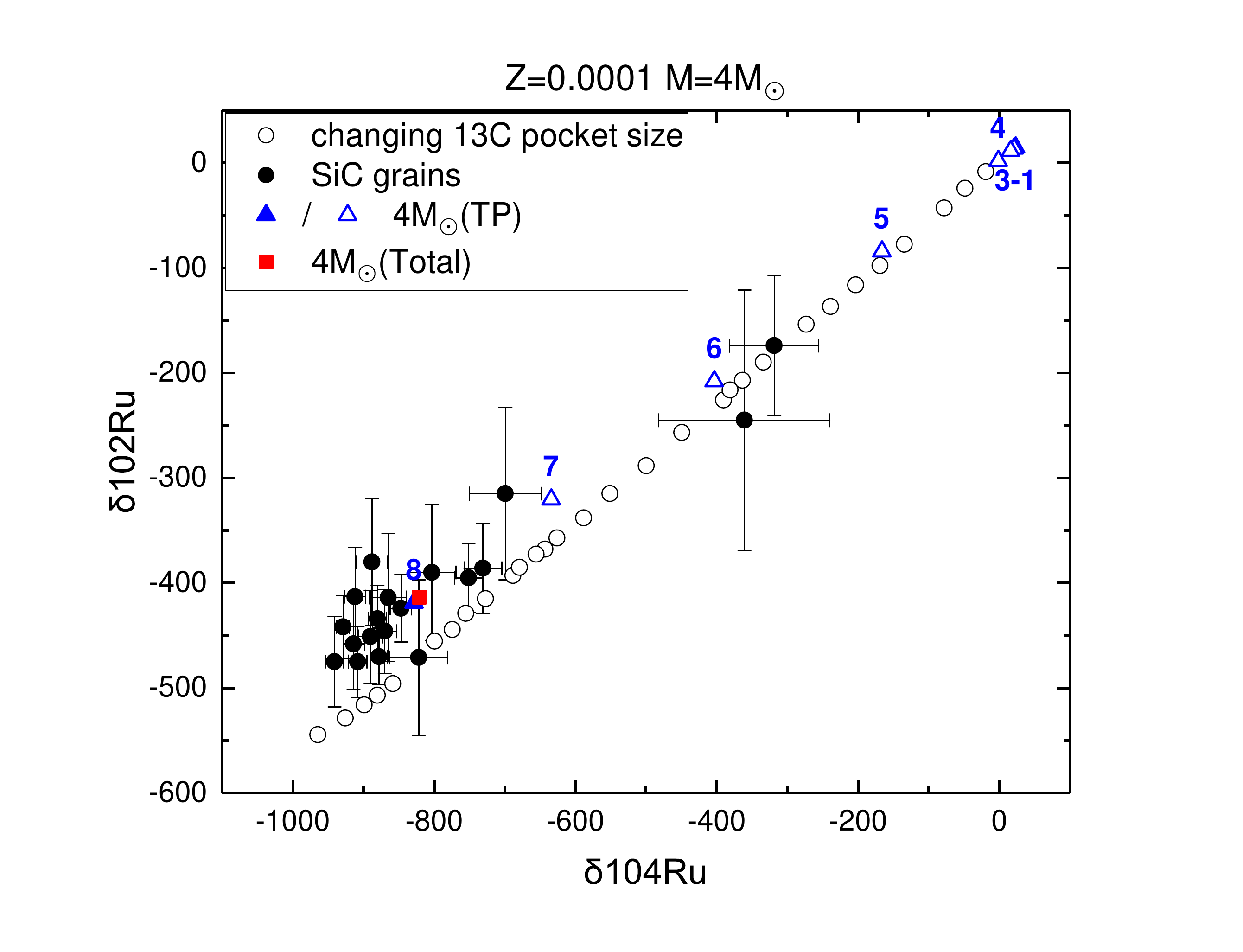}{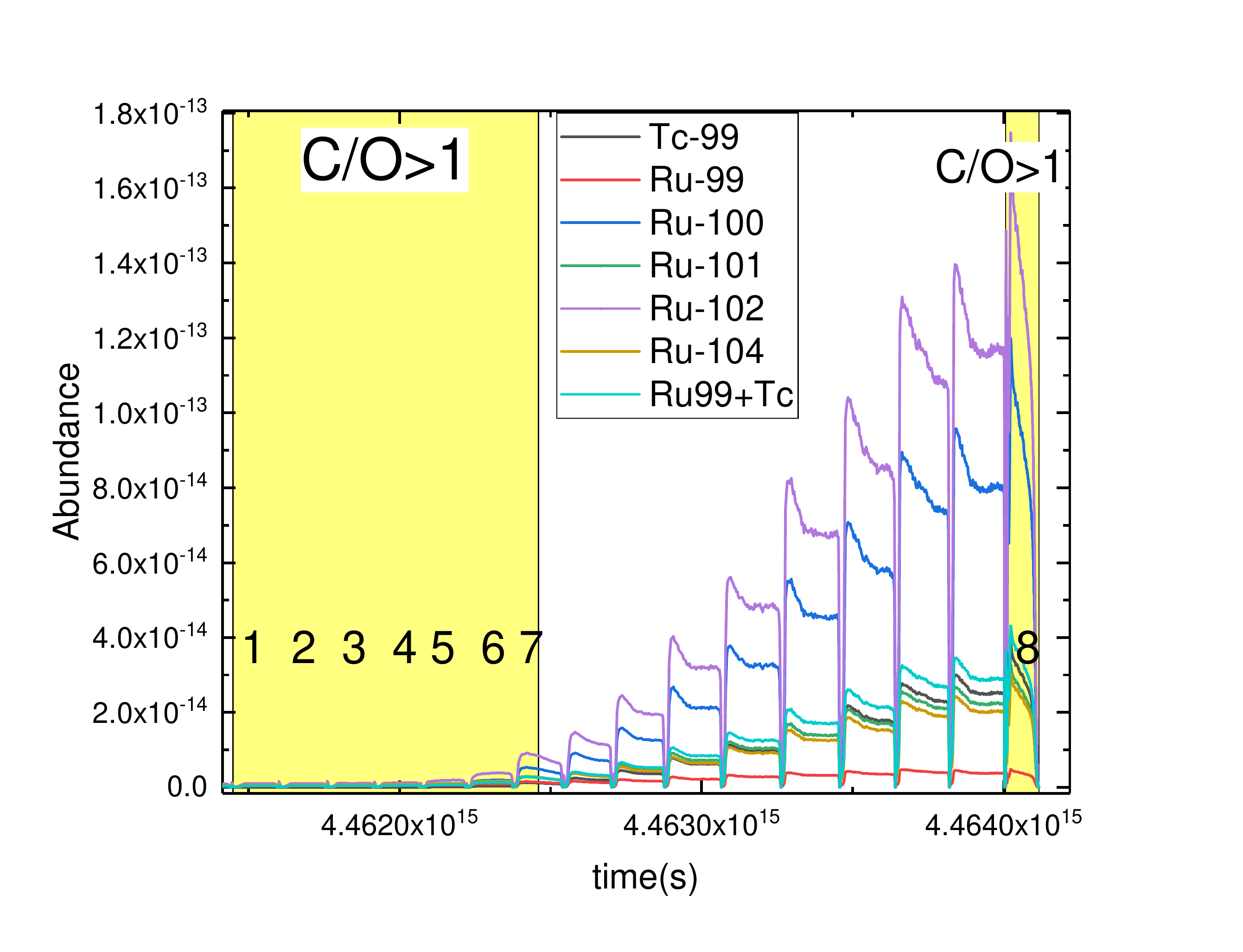}
\caption{A low-mass star with $Z=0.0001$ and $M=4~M_{\sun}$} 
\end{figure*}

\begin{figure*}
\figurenum{A21}
\epsscale{1.17}
\plottwo{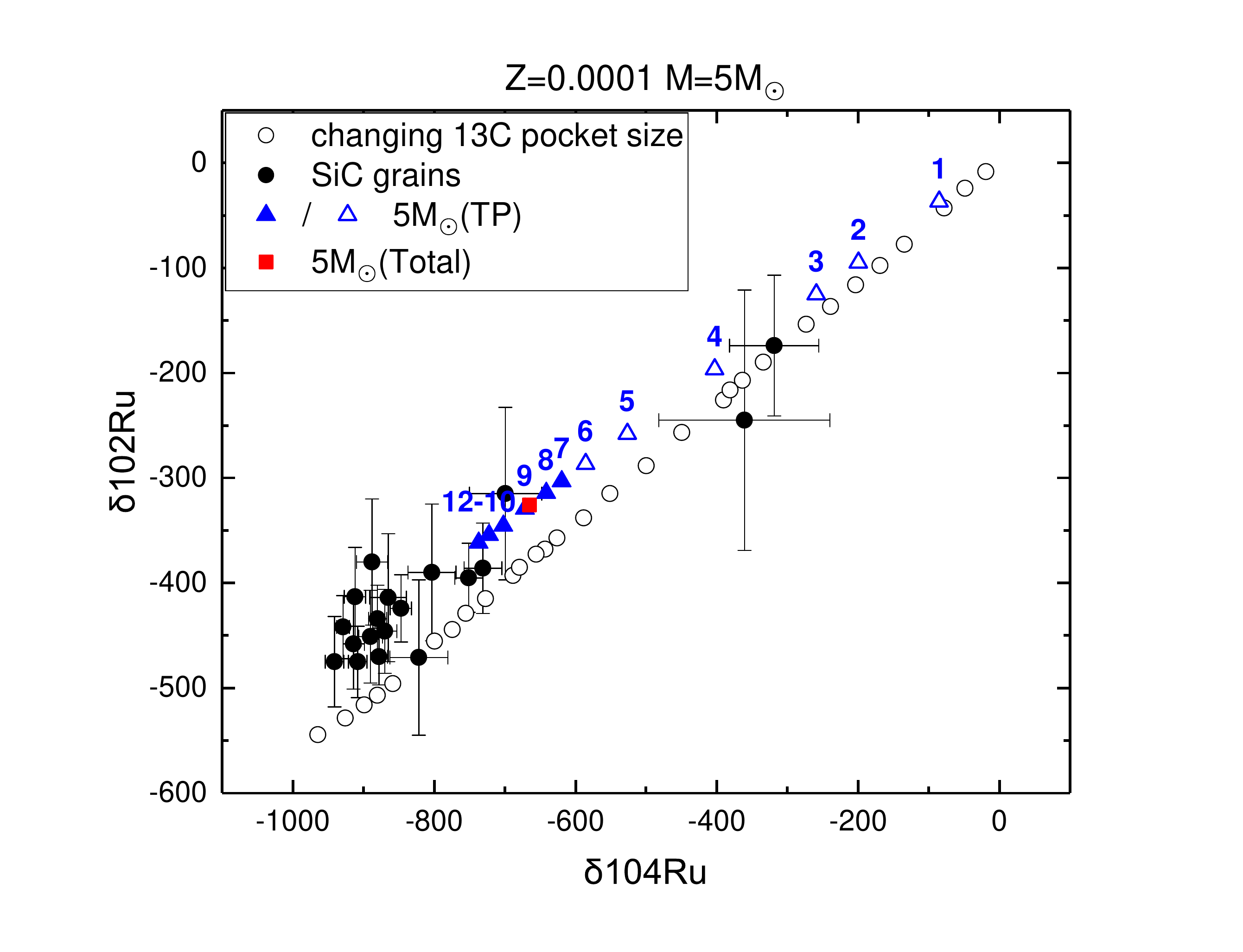}{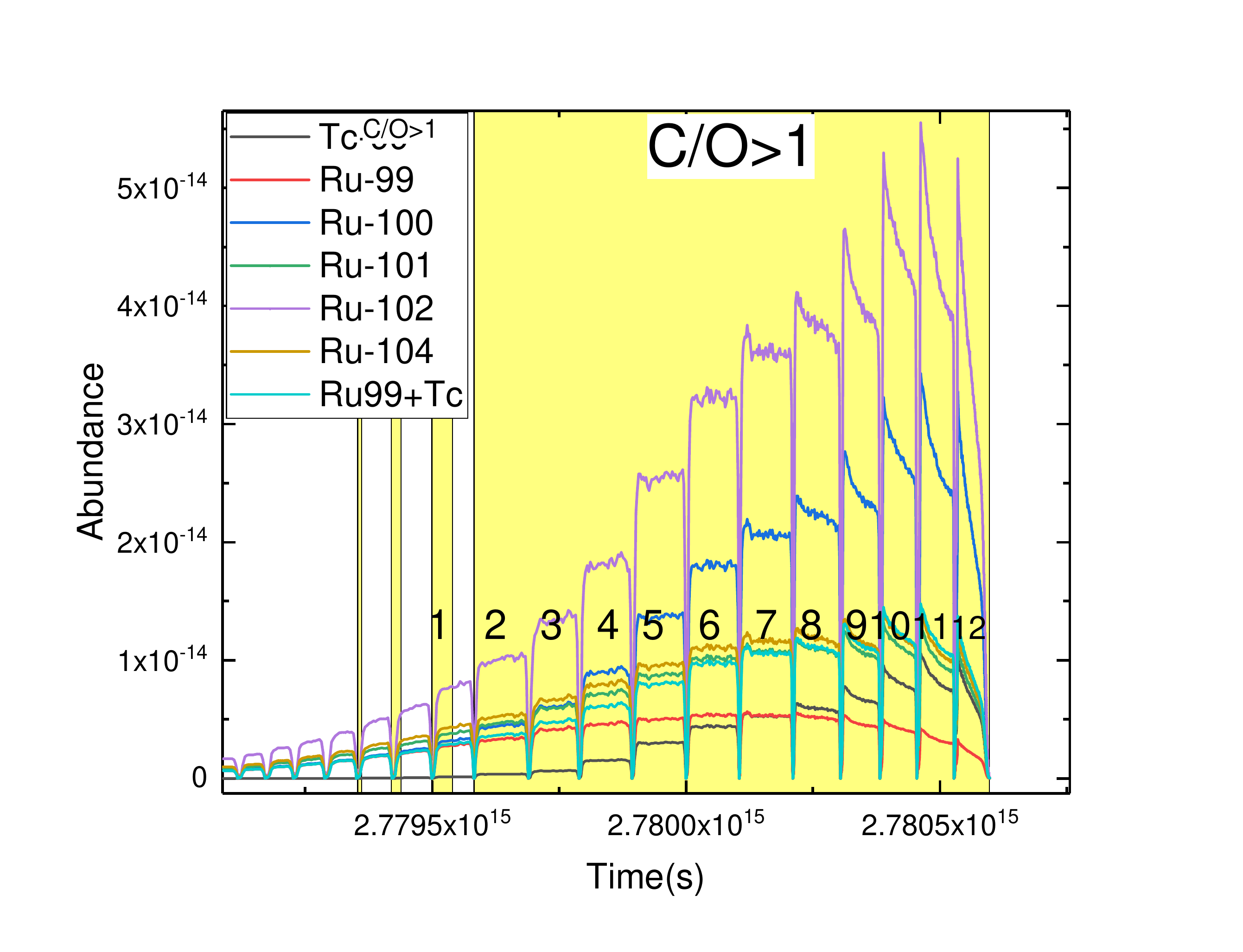}
\caption{A low-mass star with $Z=0.0001$ and $M=5~M_{\sun}$} 
\end{figure*}

\begin{figure*}
\figurenum{A22}
\epsscale{1.17}
\plottwo{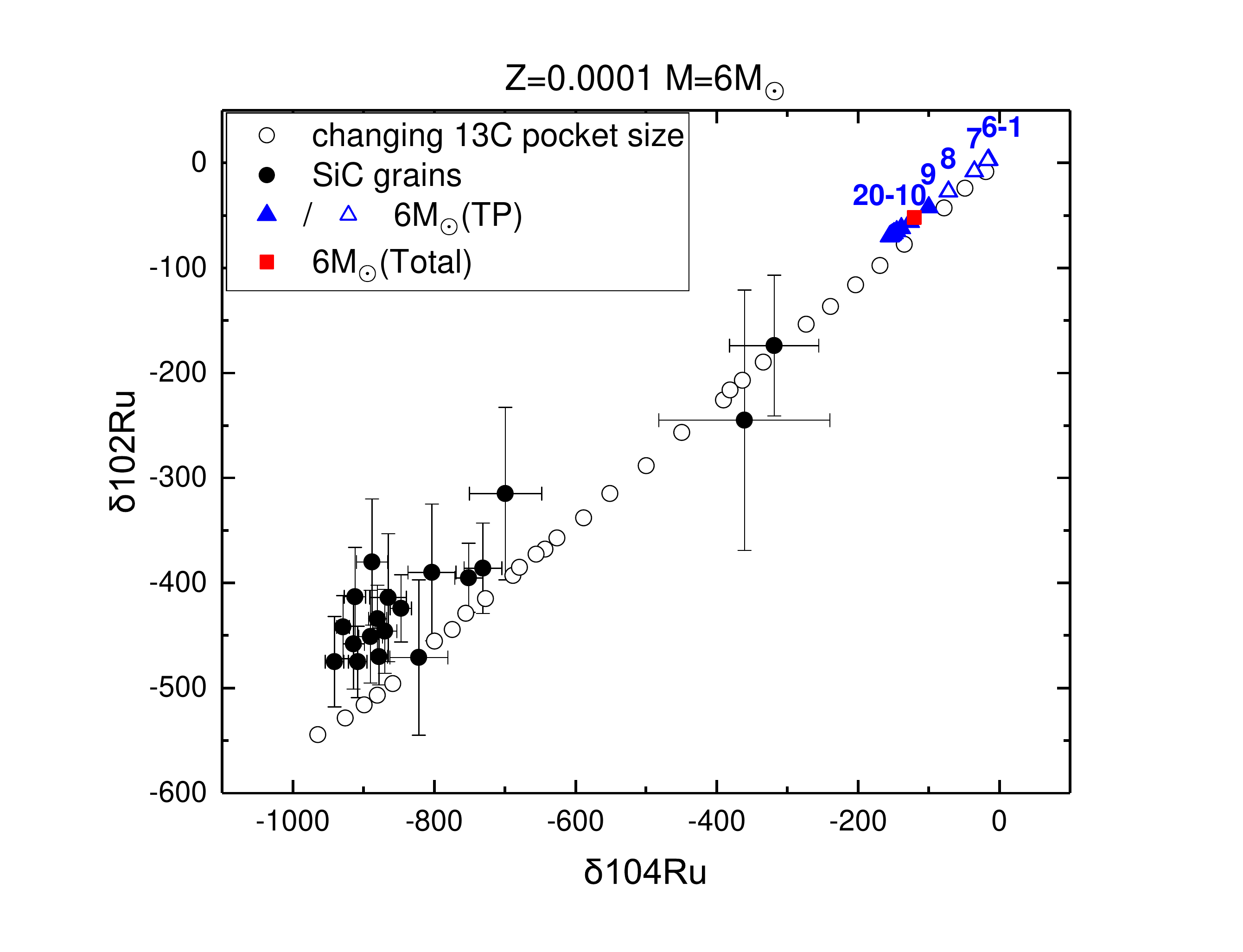}{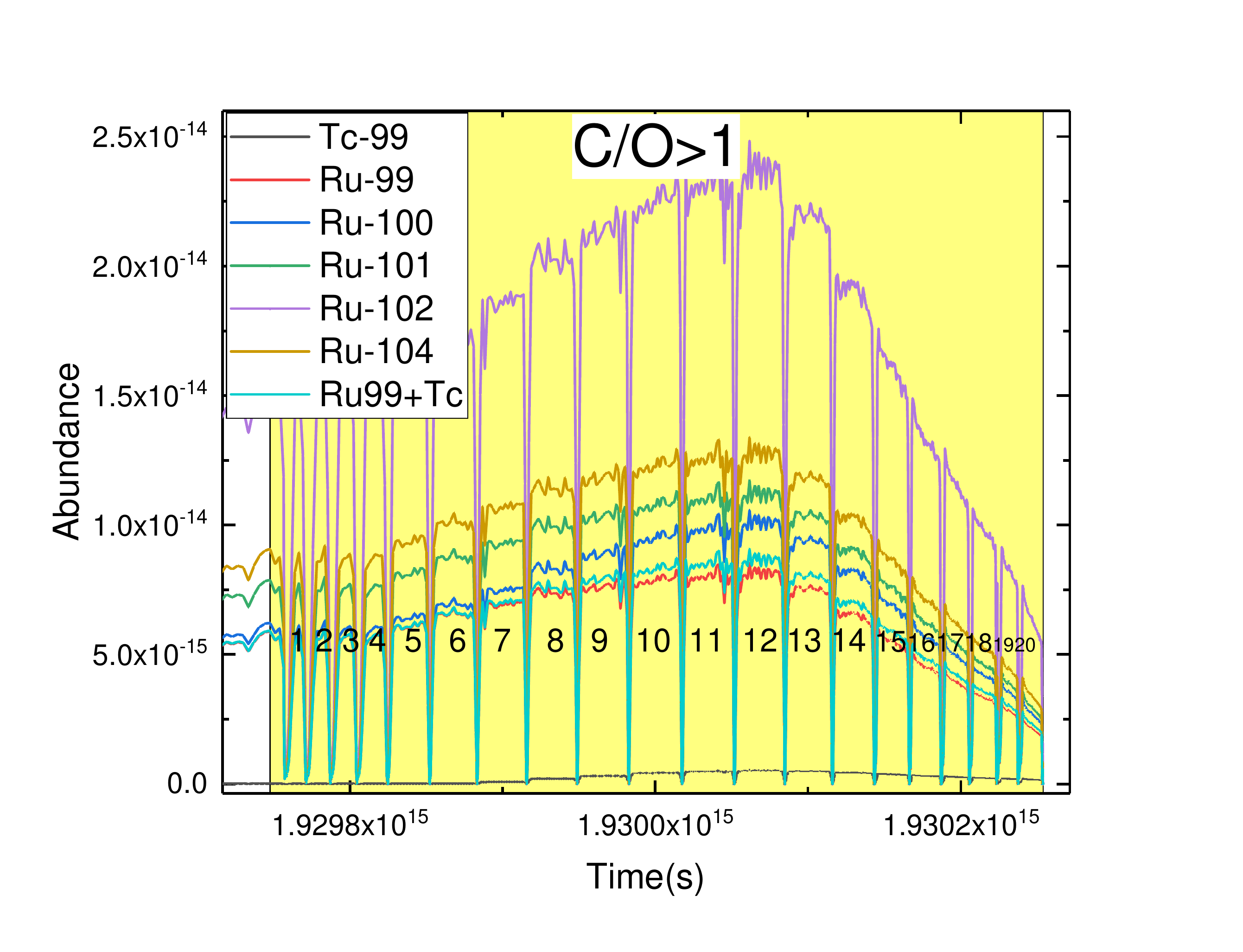}
\caption{A low-mass star with $Z=0.0001$ and $M=6~M_{\sun}$.} 
\end{figure*}

\begin{figure*}
\figurenum{A23}
\epsscale{1.17}
\plottwo{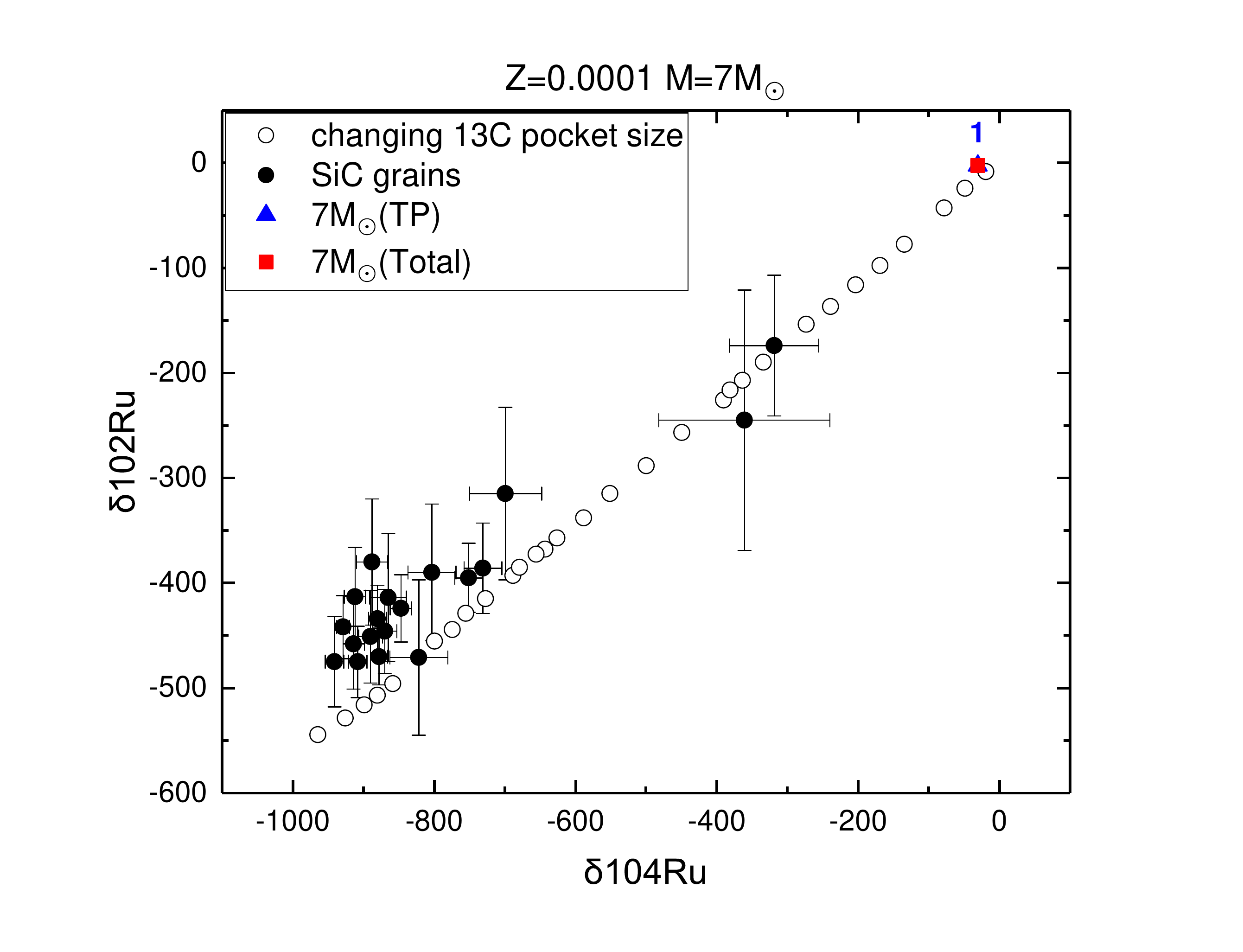}{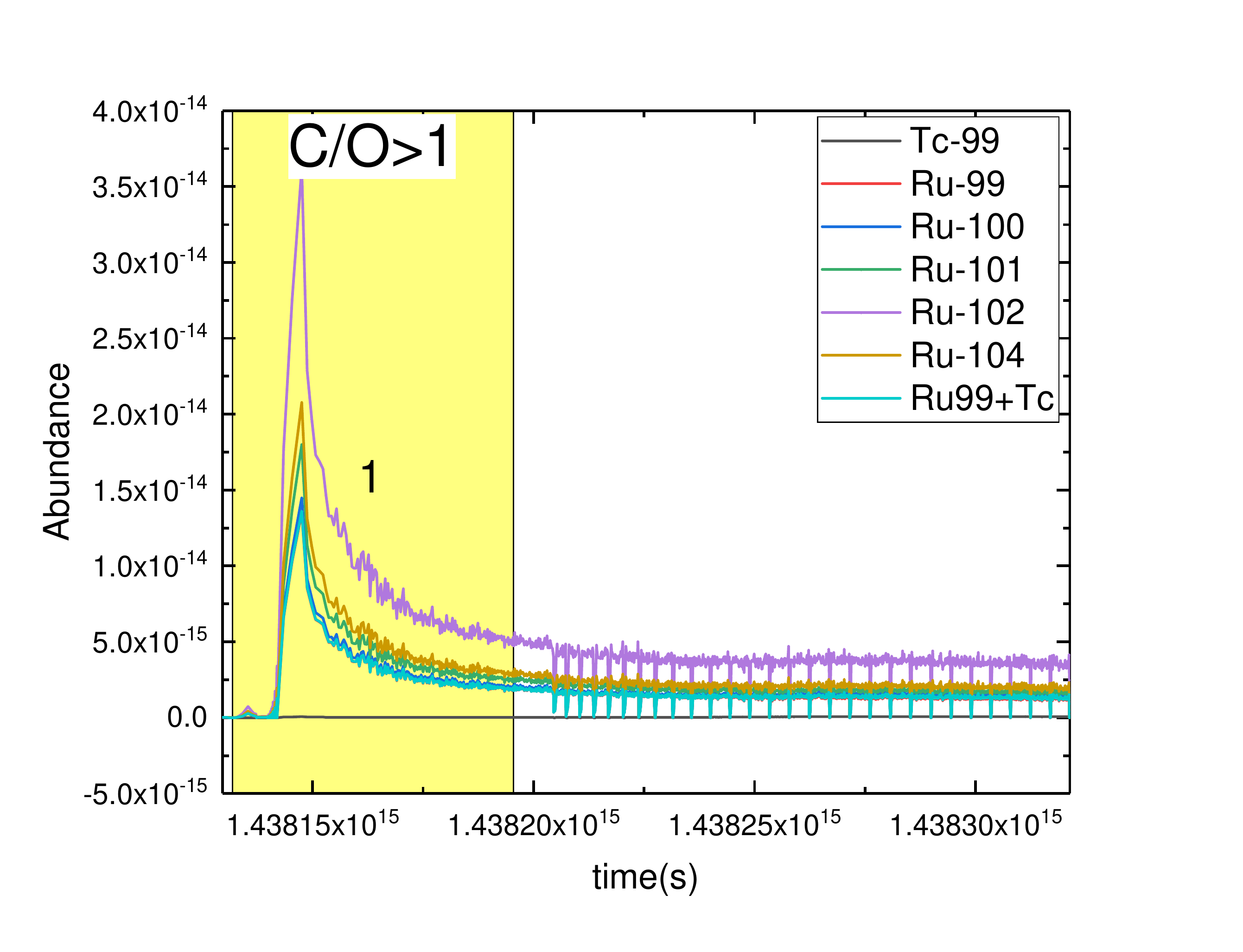}
\caption{A low-mass star with $Z=0.0001$ and $M=7~M_{\sun}$. This set belongs to Type II.} 
\end{figure*}



\end{document}